\documentclass[superscriptaddress,prx,amsmath,amssymb,aps,preprint,floatfix]{revtex4-1}
\usepackage{graphicx}
\usepackage{amsmath}
\usepackage{amssymb}
\usepackage{dcolumn}
\usepackage{color}

\begin{document}
\title{The quantum nature of hydrogen}
\author{Wei Fang}
\affiliation{School of Physics and Collaborative Innovation Centre of Quantum Matter, Peking University, Beijing 100871, P. R. China}
\affiliation{Thomas Young Centre, London Centre for Nanotechnology, and Department of Physics and Astronomy, University College London, London WC1E 6BT, United Kingdom}
\affiliation{Laboratory of Physical Chemistry, ETH Zurich, CH-8093 Zurich, Switzerland}
\author{Ji Chen}
\affiliation{Department of Electronic Structure Theory, Max Plank Institute for Solid State Research, Heisenbergstrasse 1, 70569 Stuttgart, Germany}
\author{Yexin Feng}
\affiliation{School of Physics and Electronics, Hunan University, Changsha, 410082, P. R. China}
\author{Xin-Zheng Li}
\email{xzli@pku.edu.cn}
\affiliation{School of Physics and Collaborative Innovation Centre of Quantum Matter, Peking University, Beijing 100871, P. R. China}
\affiliation{State Key Laboratory for Artificial Microstructure and Mesoscopic Physics, Peking University, Beijing 100871, P. R. China}
\author{Angelos Michaelides}
\email{angelos.michaelides@ucl.ac.uk}
\affiliation{Thomas Young Centre, London Centre for Nanotechnology, and Department of Physics and Astronomy, University College London, London WC1E 6BT, United Kingdom}
\date{\today}

\begin{abstract}
Hydrogen is the most abundant element. It is also the most quantum, in the sense that quantum tunnelling, quantum delocalisation, and zero-point motion can be important. For practical reasons most computer simulations of materials have not taken such effects into account, rather they have treated nuclei as classical particles. However, thanks to methodological developments over the last few decades, nuclear quantum effects can now be treated in complex materials. Here we discuss our studies on the role nuclear quantum effects play in hydrogen containing systems. We give examples of how the quantum nature of the nuclei has a significant impact on the location of the boundaries between phases in high pressure condensed hydrogen. We show how nuclear quantum effects facilitate the dissociative adsorption of molecular hydrogen on solid surfaces and the diffusion of atomic hydrogen across surfaces. Finally, we discuss how nuclear quantum effects alter the strength and structure of hydrogen bonds, including those in DNA. Overall these studies demonstrate that nuclear quantum effects can manifest in different, interesting, and non-intuitive ways. Whilst historically it has been difficult to know in advance what influence nuclear quantum effects will have, some of the important conceptual foundations have now started to emerge.
\end{abstract}

\maketitle
\clearpage

\section{Introduction}
The conventional picture of the atom is simple: a central nucleus clouded by orbiting electrons. Arranging the atoms in various ways and at specific distances from each other leads to the ``structure'' of molecules, solids, and liquids: the tetrahedral structure of methane; the rock-salt structure of salts; and so on. In this conventional picture of the atom it is implicitly assumed that the atomic nuclei are point-like classical particles, whereas the only quantum objects are the orbiting electrons. This is an approximation. In reality the atomic nuclei are not point-like classical particles, but are themselves quantum objects, and like the electrons, distributed in terms of wave functions.

For most elements the nuclear wave function is sufficiently localised so that the nuclei effectively behave as classical particles, which is why the conventional picture of atoms serves us so well. Quantum mechanics tells us, however, that the approximation of atomic nuclei as point-like classical particles gets progressively worse as the temperature drops or as the elements get lighter. 
For the lightest element, hydrogen, nuclear quantum effects (NQEs) can be significant even at room temperature, with the nucleus of the H atom, a proton, being localised in space and able to tunnel through classically forbidden potential energy barriers.

If hydrogen was some obscure element trailing off the end of the periodic table perhaps one would not care much about NQEs, but for the very reason that hydrogen is the most ``quantum'' of the elements, i.e., its lightness, it is also the most abundant: A colossal 90\% of the universe by weight is hydrogen; it is present in all organic compounds, the largest class of which are literally called hydrocarbons; it is essential for life by, for example, being the marriage partner of O in H$_2$O. 
Hydrogen is also of paramount importance to the global economy with practically all industrial catalytic processes having hydrogen implicated either as a reactant, product, or intermediate, with the making and breaking of H-H, C-H, O-H, and N-H bonds at surfaces the bread and butter of heterogeneous catalysis.

Real world manifestations of the importance of NQEs are plentiful. 
For example, the heat capacity of ``classical'' water would be $\sim$40\% \cite{Vega_2010} larger without NQEs; 
thus in a classical world
tea drinkers would have to wait much longer for their water to boil.
Similarly many biological reactions (notably in enzyme catalysis) rely on the tunnelling of protons \cite{SHS_bio_tunel_rev,bio_rev}.
In addition, in the pharmaceutical industry, novel classes of deuterated drugs, which are expected to have greater biochemical potency, are under development

Various experimental approaches are available for understanding NQEs at a fundamental level. 
Experiments involving isotopic substitution, in which hydrogen is replaced by deuterium (or, less often, tritium), are a powerful and widely used approach. 
With this strategy one can then explore how e.g. the structure and properties of materials 
change with isotope substitution \cite{doi:10.1021/cr60292a004,Ubbelohde}.
Experiments have shown that deuteration can sometimes lead to dramatic changes in physical properties, e.g. a $>$100 K change in the ferroelectric to paraelectric transition temperature in H-bonded ferroelectrics
\cite{PhysRevLett.81.5924,doi:10.1063/1.4862740}.
Kinetic measurements in which the rate change upon isotopic substitution is examined (so-called kinetic isotope effects) are also widely employed in chemistry and biology \cite{SHS_bio_tunel_rev,bio_rev,doi:10.1021/cr100436k}.
In terms of more direct experimental probes, deep inelastic neutron scattering (DINS) has emerged as an approach that can directly measure the momentum distribution of protons \cite{DINS_1}. 
It can provide insights into the local environment of H atoms, which can help elucidate NQEs in hydrogen bond (HB) networks, such as in structures of crystalline and amorphous ice \cite{DINS_1,Romanelli_2013,DINS_2,DINS_3}.
On surfaces, the invention of Scanning Tunnelling Microscopy (STM) enabled atomic resolution imaging and atom manipulation \cite{STM_1986}, and STM has also been used to probe NQEs. 
Notably measurements of H and D diffusion on various metal surfaces have been performed and provide strong evidence of H tunnelling at cryogenic temperatures \cite{lauhon_direct_2000,PhysRevB.81.045402,Sykes_Quantum_2012,Davidson_2014_2}.
In addition, isotope-dependent switching of HBs in adsorbed water clusters has been visualised on Cu and NaCl \cite{PhysRevLett.100.166101,tetramer_2015}.
%
Fast diffusion of H atoms on surfaces beyond the time-resolution of STM can also be measured indirectly with helium spin echo (HeSE) experiments, a robust and sophisticated 
scattering experiment \cite{Jardine_HeSE_2009,jardine_determination_2010}.

From the theoretical perspective a variety of schemes can be employed to treat NQEs \cite{LSCIVR,MCTDH,SHS_bio_tunel_rev,bio_rev,Marx-Parr_1994}. 
A particularly elegant approach and the one we focus on in this review is Feynman's path-integral representation of quantum mechanics \cite{Feynman}.
Detailed accounts of the path integral representation, its implementation into computer codes, and applications in chemistry, physics and materials science can be found elsewhere \cite{Tuckerman_book,Tuckerman-Marx_1996,doi:10.1063/1.441588,Berne_1982,PhysRevB.30.2555,Marx-Parr_1994,Cao_Voth_1994,PhysRevLett.105.110602,ipi,PI_Ramirez,PILE,Mass_3,habershon_ring-polymer_2013,markland_nuclear_2018}. 
Thus we do not go in to the details of the theory here except to note that the path-integral framework formulates quantum mechanics as a summation of paths rather than through the wavefunction view of Schr\"odinger.
In doing so it provides a classical analogy for quantum mechanics, which is widely applicable for sampling quantum ensembles and approximating quantum dynamics.
In the 1980s, pioneering path-integral simulations of materials were performed with either path-integral molecular dynamics (PIMD) \cite{doi:10.1063/1.446740,Gillan_H_qtst} or path-integral Monte Carlo (PIMC) \cite{doi:10.1063/1.441588,Berne_1982,PhysRevB.30.2555,PhysRevB.31.4234,PhysRevLett.58.1648}. 
In these early studies the interatomic interactions were described with empirical potentials (otherwise known as forcefields). 
However, with the emergence of density functional theory (DFT), \textit{ab initio} based path integral approaches became possible \cite{Marx-Parr_1994,Tuckerman-Marx_1996}.
Early DFT-based PIMD studies 
appeared in the 1990s with exciting applications on e.g. water and ice \cite{Marx1998,Tuckerman817,Parrinello_2003}.
Over the following 20 years or so numerous algorithmic and computational advances served to make the path integral methodology robust and computationally tractable. 
This has included important work on quantum dynamics within the path-integral framework, with e.g. centroid molecular dynamics (CMD) \cite{CMD,voth_rigorous_1989,MARX1999166},
ring-polymer molecular dynamics (RPMD) \cite{craig_refined_2005,habershon_ring-polymer_2013,trpmd}, 
and the combination of path integral methods with other electronic structure methods beyond DFT \cite{PI_QMC_1,doi:10.1063/1.4941091,Tachikawa_2013,C4CP05192K}. 
Developments have also been made in combining the path integral framework with other quantum theories, for example the instanton theory \cite{miller_semiclassical_1975,richardson_ring-polymer_2009,Jonsson_2009,Kastner_2014,Inst_persp}.
Indeed, over the years increasingly complex systems have been explored \cite{Klein2003,XZLi_2011,tuckerman_nature_2002,doi:10.1021/jp810590c,Wei_rev_2016,doi:10.1021/acs.jpclett.7b00979,PhysRevLett.120.225901,rossi_nuclear_2016,markland_nuclear_2018}
and in some respects it is now a ``golden era" for investigating NQEs with path integral methods.
In this brief review, we highlight some key recent findings on the role and importance of NQEs in H containing systems. 
Although work from other groups is discussed, this review focuses heavily on work carried out by the authors. 
As such it is not intended as a comprehensive overview of the field; for such reviews the interested reader is referred to refs. \cite{Marx_Tuckerman_2010,Wei_rev_2016,sremarks,markland_nuclear_2018}.
Three prominent topics where NQEs manifest in interesting manners under experimental or real world conditions are covered: the phase diagram of hydrogen; the adsorption and diffusion of hydrogen on surfaces and 2D materials; and the structure and stability of H-bonded systems.

\section{Quantum nature of condensed phase hydrogen}
\begin{figure}[!ht]
\includegraphics[width=10cm]{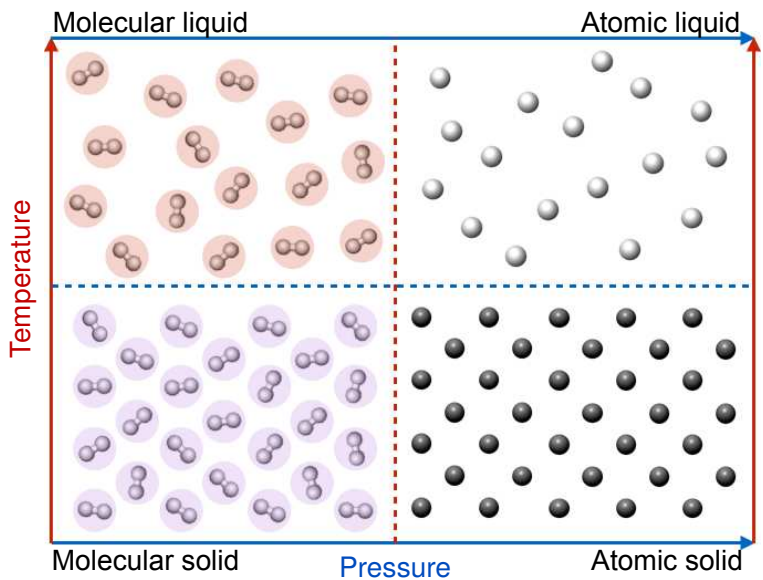}
\centering
\caption{Schematic cartoon illustration of four broad types of phase for condensed phase hydrogen, namely a molecular solid, a molecular liquid, an atomic solid and an atomic liquid.  
The molecular solid is composed of $\text{H}_2$ molecules on crystal lattice sites, and depending on the solid phase the molecules are either orientationally ordered (e.g. phase II and III) or disordered (e.g. phase I).
Upon heating, the crystalline molecular solid melts into a molecular liquid.
The $\text{H}_2$ molecules in both the molecular solid and the liquid have been predicted to dissociate when compressed, and could transform into either an atomic solid or an atomic liquid.
}
\label{figure_phase-diagram}
\end{figure}

%
Hydrogen, when condensed under megabar pressures, exhibits extremely complex solid and liquid phase transitions, accompanied by intriguing physics such as metalisation, superconductivity, and superfluidity.
Exploring the phase diagram of hydrogen, therefore, is one of the major (and most heavily debated) topics in condensed matter physics \cite{dias_observation_2017, silvera_response_2017, liu_comment_2017, goncharov_comment_2017}.
Fig. \ref{figure_phase-diagram} illustrates schematically that condensed phase hydrogen can broadly be classified into four regimes, entailing molecular solid(s), molecular liquid(s), atomic solid(s), and atomic liquid(s).
The boundaries between the various condensed phases depend sensitively on the relative free energies of the different phases, on which NQEs such as zero point energy (ZPE) and quantum delocalisation may have a significant impact.
For example, it has been estimated that the difference in the ZPE 
between different phases can be as large as $\sim$10 meV per atom \cite{pickard_structure_2007,drummond_quantum_2015}.
This is enough to alter the relative thermodynamic stability of competing structures and can significantly shift the pressure-temperature phase boundaries by hundreds of kelvin or tens of gigapascal \cite{pickard_structure_2007,drummond_quantum_2015}.

In experiments in this area NQEs are often indirectly probed by examining H/D isotope effects with e.g. vibrational spectroscopies in diamond anvil cells.
H/D isotopic effects have been discussed in the four crystal structures of the molecular solid observed so far, namely phases I, II, III and IV.
These studies have shown that
the phase I/II boundary has a strong isotope dependence, whereas the phase II/III boundary is almost identical for H 
and D \cite{silvera_new_1981, lorenzana_evidence_1989, lorenzana_orientational_1990, cui_megabar_1994, mazin_quantum_1997, goncharov_invariant_1995, goncharov_raman_1996,PhysRevLett.119.065301}.
In phase I, the $\text{H}_2$ molecules rotate freely, following the quantum rotational partition function \cite{mcmahon_properties_2012}, whereas in phase II the hydrogen molecules are preferentially aligned on their crystalline lattice sites \cite{goncharenko_neutron_2005}.
With the discovery of phase IV of solid hydrogen, isotope effects have now been explored up to the 300 GPa regime \cite{eremets_conductive_2011,howie_mixed_2012,howie_proton_2012, zha_synchrotron_2012, loubeyre_hydrogen_2013, eremets_infrared_2013,zha_high-pressure_2013, zha_raman_2014, howie_raman_2015, dalladay-simpson_evidence_2016}.
Phase IV consists of alternating layers of an orientationally disordered molecular layer and an atomic layer on a honeycomb lattice.
Hence NQEs are also expected to have an impact 
on the $\text{H}_2$ molecules in phase IV \cite{ackland_bearing_2015}.
Room temperature proton tunnelling was also suggested in phase IV to explain measured Raman data \cite{howie_proton_2012}.
Overall, the studies mentioned above indicate that in certain regimes NQEs can be important.
However, a comprehensive picture of NQEs in condensed phase hydrogen is yet to be established.
In what follows we review certain aspects of condensed phase hydrogen with a focus on understanding the role of NQEs on: (i) the solid phase boundaries in the molecular solid regime; and (ii) the melting of the atomic solid.
NQEs have also been revealed in other regimes of the hydrogen phase diagram \cite{doi:10.1063/1.1893956, morales_nuclear_2013, kang_revealing_2013, kang_nuclear_2014}.
Comprehensive reviews and previous studies to explore the pressure-temperature phase diagram of hydrogen can be found elsewhere, and the interested reader is refereed to e.g. refs. \onlinecite{mao_ultrahigh-pressure_1994, mcmahon_properties_2012, silvera_new_1981, lorenzana_evidence_1989, lorenzana_orientational_1990, cui_megabar_1994, mazin_quantum_1997, goncharov_invariant_1995, goncharov_raman_1996,PhysRevLett.119.065301, eremets_conductive_2011,howie_mixed_2012,howie_proton_2012, zha_synchrotron_2012, loubeyre_hydrogen_2013, eremets_infrared_2013,zha_high-pressure_2013, zha_raman_2014, howie_raman_2015, dalladay-simpson_evidence_2016}.

\subsection{The quantum nature of solid molecular hydrogen}
\begin{figure}[!ht]
\includegraphics[width=8cm]{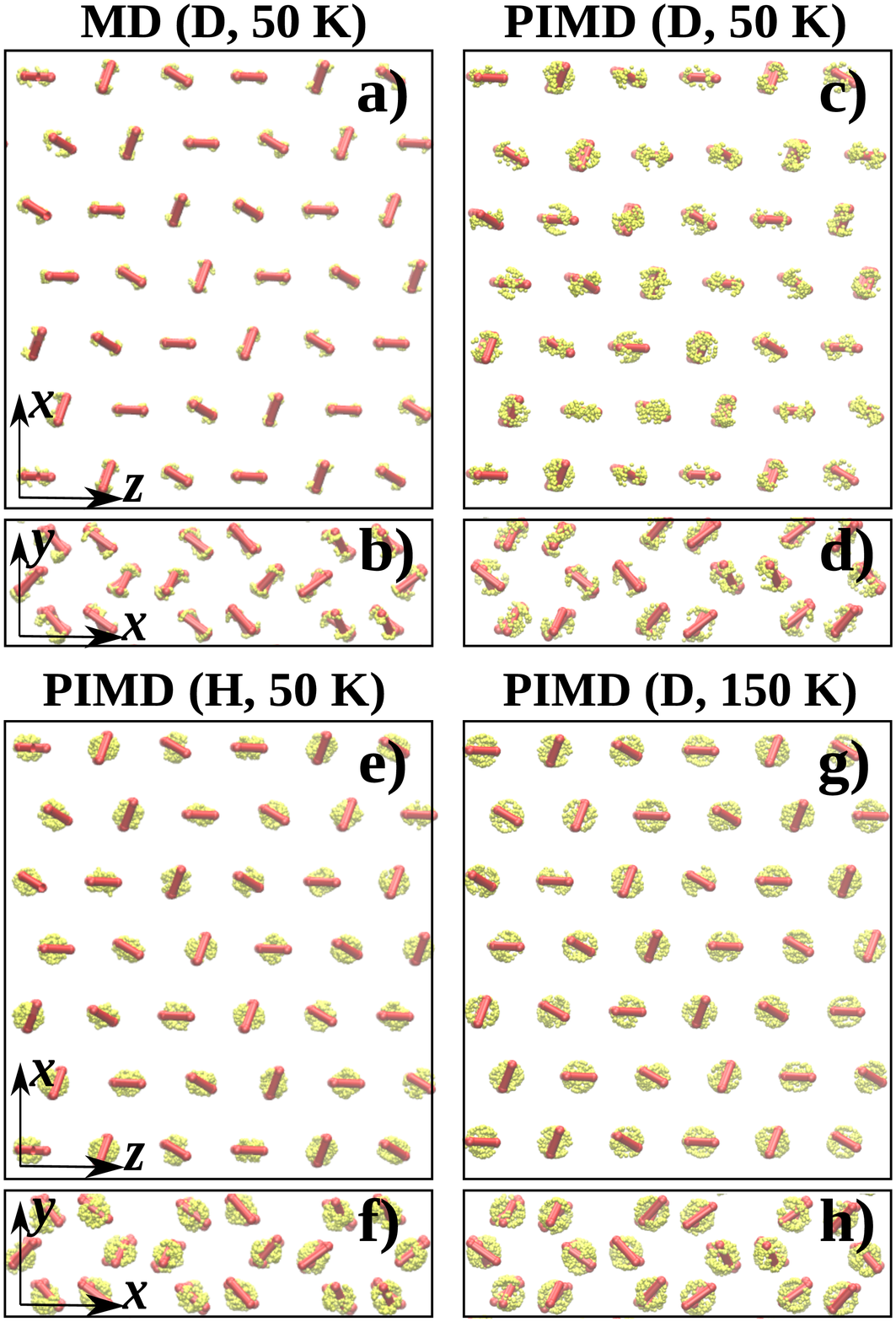}
\centering
\caption{
NQEs have a significant impact on the phase transition between molecular phases I and II of hydrogen.
Trajectories of structures obtained from simulations with classical and quantum nuclei at 80 GPa starting from the phase II (P21/c-24) structure. Yellow balls show the representative configurations of the centroids throughout the course of the simulation. The red rods show the static (geometry-optimized) structure. A conventional hexagonal cell containing 144 atoms was used. Panels (a), (c), (e), and (g) show the z-x plane and panels (b), (d), (f), and (h) show the x-y plane of the hcp lattice. The four simulations are: (1) MD with classical nuclei at 50 K (panels (a) and (b)), (2) PIMD for deuterium at 50 K (panels (c) and (d)), (3) PIMD for hydrogen at 50 K (panels (e) and (f)), and (4) PIMD for deuterium at 150 K (panels (g) and (h)). In the MD simulation, the anisotropic inter-molecular interaction outweighs the thermal and quantum nuclear fluctuations. Therefore, the molecular rotation is highly restricted. The thermal plus quantum nuclear fluctuations outweigh the anisotropic inter-molecular interactions in the PIMD simulations of hydrogen at 50 K and deuterium at 150 K.
Reprinted with permission from ref \cite{li_classical_2013}. Copyright 2013 IOP Publishing.
}
\label{figure_XZLi-JPCM-Fig2}
\end{figure}

State of the art \textit{ab initio} path integral molecular dynamics simulations have helped to
reveal the role of NQEs on dense hydrogen in the last two decades.
(For studies with complementary approaches see e.g. ref. \cite{drummond_quantum_2015}.)
Notably Biermann \textit{et al}. simulated solid hydrogen at 50 K with \textit{ab initio} path integral molecular dynamics. 
They found that the lattice structure of hydrogen above a pressure of 350 GPa was very diffuse, due to quantum fluctuations of the hydrogens far from their equilibrium lattice positions \cite{biermann_proton_1998,biermann_quantum_1998}.
This unexpected fluxional structure suggests a structure qualitatively different from the classical picture.
Later Kitamura \textit{et al}. reported different rotational order due to NQEs for some of the solid phases (specifically phases I, II and III) \cite{kitamura_quantum_2000}.
They found that in phase I the hydrogen molecules rotate easily on their hcp lattice sites and that the orientational order of the molecules in the crystal is smeared out. 
For phase II the simulations suggested that the hydrogen molecules had a preferred orientational order with $\text{Cmc}2_1$ symmetry.
For phase III the rotational order was also suppressed but the averaged position of the hydrogen molecules suggests phase III has Cmca symmetry.
Later Li \textit{et al}. carried out a comprehensive study using \textit{ab initio} PIMD simulations on the same three low temperature solid hydrogen phases.
%
In these studies van der Waals dispersion forces were also taken into account through the application of a van der Waals inclusive DFT functional \cite{PhysRevB.83.195131,doi:10.1063/1.4754130}. 
This study showed that the classically orientated $\text{H}_2$ molecules in phase II also lose orientational order
when NQEs are accounted for, whereas for $\text{D}_2$ the orientational order was maintained (Fig. \ref{figure_XZLi-JPCM-Fig2}).
%
This study also explained the large isotopic effect in the I-II phase transition due to a less corrugated potential energy landscape, where quantum fluctuations play a more significant role.
Overall, these simulations significantly improved the agreement between experiment and theory on the phase boundaries of solid hydrogen. They also highlighted the need to account for both van der Waals interactions and NQEs simultaneously, when treating the low temperature region of the hydrogen phase diagram.
We note that at a similar time Geneste \textit{et al}. also studied the role NQEs in solid hydrogen, reaching similar conclusions to Li \textit{et al}.

\subsection{The role of NQEs on the melting of hydrogen}
\begin{figure}[!ht]
\includegraphics[width=8cm]{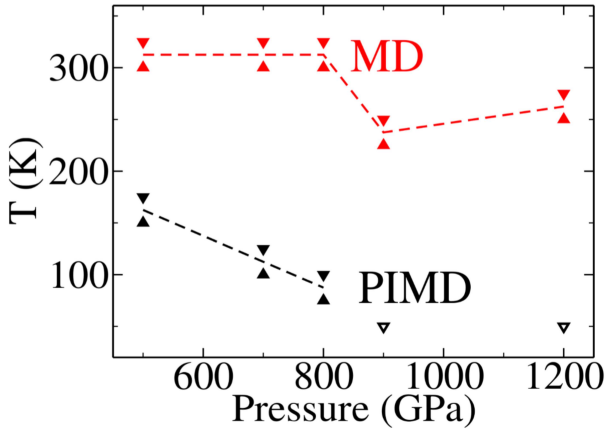}
\centering
\caption{
NQEs significantly lower the melting temperature of high pressure atomic hydrogen. MD (red) and PIMD (black) label the results considering the H atoms as classical or quantum particles, respectively. Triangles show the upper and lower limit of the melting temperature estimates, with the melting temperature taken as the average (dashed lines). The lowest temperature simulations were performed at 50 K. Therefore, only the upper bounds from PIMD simulations are shown with dashed triangles at 900 GPa and 1200 GPa.
Reprinted with permission from ref \cite{chen_quantum_2013}. Copyright 2013 Nature Publishing Group, under the creative commons license 3.0.
}
\label{figure_Chen-Ncomm-Fig2}
\end{figure}

The melting of hydrogen is another aspect of interest, intimately connected with hydrogen superconductivity and superfluidity at high pressures and low temperatures \cite{babaev_superconductor_2004}. 
It has been suggested that NQEs affect melting, most likely lowering the melting temperature because of quantum fluctuations \cite{bonev_quantum_2004}. 
It is also expected that such an effect would be stronger for atomic phases of hydrogen than
for molecular phases because of the heavier mass and higher melting temperatures of the $\text{H}_2$ solids 
 \cite{bonev_quantum_2004, deemyad_melting_2008}.
Since the atomic phases only exist at pressures on the limits of what can be reached experimentally, computer simulation 
can play a crucial role. 
With this in mind Chen \textit{et al}. studied the melting of atomic hydrogen in the 500 GPa to 1.2 TPa regime with DFT-based PIMD simulations \cite{chen_quantum_2013}. 
Interestingly it was found that NQEs significantly reduced the melting temperature from about 300 K to less than 200 K at 500 GPa (Fig. \ref{figure_Chen-Ncomm-Fig2}).
At higher pressures the effect is even more pronounced and 
at 900 GPa and above, the melting temperature drops below 50 K. 
This suggests that in this pressure regime a 
low temperature quantum metallic liquid state of hydrogen is possible. 
%
%
Later Geng \textit{et al}. also studied the role of NQEs on the melting of high pressure hydrogen, and predicted slightly smaller NQEs, that lower the melting temperature by 50 to 100 K \cite{geng_lattice_2015, geng_predicted_2016}.
Although not yet confirmed in experiments or substantiated with higher level electronic structure theories, 
these studies reveal that NQEs can have profound consequences on the melting and physics of hydrogen at high pressures.

Overall, at lower pressure, the phase diagram of condensed hydrogen is relatively well resolved experimentally.
Further experimental studies into the vibrations and rotations of hydrogen molecules could shed light on the explicit quantum nature of the nuclei.
However, upon increasing pressure, the phase behaviour of hydrogen is under debate and recent experimental progress suggests that metallic hydrogen at low temperatures is not far away if not already detected
\cite{dias_observation_2017, silvera_response_2017, liu_comment_2017, goncharov_comment_2017}.
The question of NQEs in the metallic transition of hydrogen at low temperatures also
remains open.
Entering the TPa regime of the phase diagram at low temperatures is a great challenge experimentally, but it is important to get there since many interesting properties are predicted in this regime.
At high temperatures evidence of non-negligible NQEs has been obtained, and it is also of interest to further investigate the impact of NQEs on phase transition phenomenon near the critical point \cite{li_supercritical_2015}.
Theoretically, disagreements have been shown in different studies because of e.g. the choice of exchange correlation functional in density functional theory, therefore improving the understanding of NQEs based on electronic structure calculations of higher level, such as quantum Monte Carlo, is desired.

\section{Hydrogen at surfaces}
The adsorption of hydrogen on, and diffusion of hydrogen across surfaces is of central importance to disciplines such as surface science, astrophysics and astrochemistry, and heterogeneous catalysis.
%
%
NQEs can in principle be important to these processes, particularly at the low temperatures of astrochemistry \cite{Interstellar_2013,Interstellar_2}.
%
%
%
Here we discuss four different systems we have worked on, with each one illustrating a different aspect of NQEs at surfaces. 
%

\subsection*{A. Adsorption of atomic hydrogen on graphene}
Hydrogen adsorption on sp$^2$-bonded carbon materials is relevant to hydrogen storage, graphene based electronic and spintronic devices, and $\text{H}_{2}$ formation in the interstellar medium \cite{doi:10.1021/la051659r,PhysRevLett.103.016806,Elias610,doi:10.1021/ja804409f,Hgra_PRL,Interstellar_2013,Kastner_2010}.
Upon H atom adsorption on sp$^2$-bonded carbon materials the carbon atom the hydrogen bonds to transforms to sp$^3$ hybridisation. 
As a result there is expected to be an energy barrier for the adsorption process. 
The nature and height of this energy barrier has been extensively studied, using graphene, graphite, and polycyclic aromatic hydrocarbons as model systems \cite{JELOAICA1999157,doi:10.1063/1.1463399,EPJB_Hadgra,Hornekaer1943,Hgra_PRL,Davidson_2014_1,doi:10.1063/1.4931117,doi:10.1021/acs.jpca.5b12761}.
Experimentally, the barrier for H atom adsorption on graphite was placed within a broad range of 25 to 250 meV, by means of adsorption
and abstraction experiments using H atoms with varying kinetic energies \cite{doi:10.1063/1.3518981}.
%
Early theoretical studies estimated the barrier to be 
$\sim$ 0.2 eV \cite{JELOAICA1999157,doi:10.1063/1.1463399,EPJB_Hadgra}; a barrier of this height would mean e.g. at the low temperatures of the interstellar medium H atoms would have insufficient thermal energy to adsorb on sp$^2$ carbon substrates. 
This, in turn, has implications for the mechanism of H$_2$ formation in the interstellar medium.
However, the earlier computational studies neglected NQEs and also neglected vdW dispersion forces; both of these effects are likely to influence the chemisorption process. %
Especially, quantum tunnelling have been known to be crucial to the formation of an extraordinary rich variety of molecules in the universe \cite{sims_low-temperature_2013,Interstellar_2013,doi:10.1146/annurev.pc.46.100195.000545,Kastner_2010}. 
%

\begin{figure}[!ht]
\includegraphics[width=10cm]{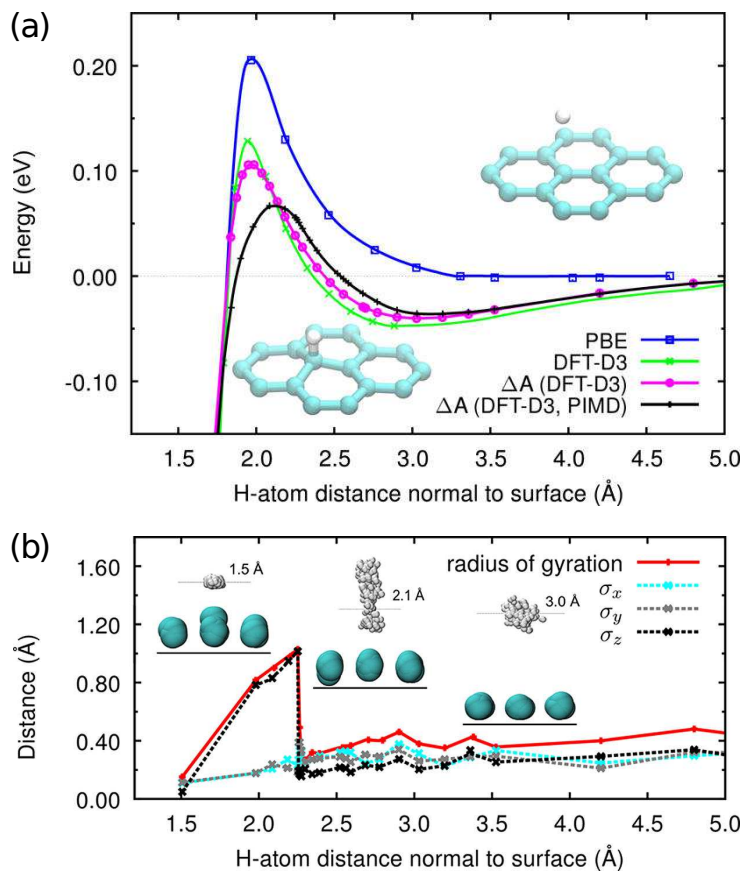}
\caption{\label{H-ads}
NQEs on low temperature H atom adsorption on graphene.
(a) Energy profiles calculated with different approaches for the adsorption of a single H atom on graphene.  
The highest barrier (blue) is the PBE potential energy barrier to chemisorption.
Upon including vdW dispersion forces with PBE-D3 (green) the potential energy barrier gets lower and narrower. 
The PBE-D3 free energy barrier (pink) computed using \textit{ab initio} MD at 50 K does not differ significantly from the underlying potential energy barrier. 
However, accounting for NQEs with \textit{ab initio} PIMD (black) significantly lowers the free energy barrier. 
The H atom height above the surface is measured from the surface plane of the graphene sheet prior to chemisorption. 
(b) Radius of gyration for the path-integral ring-polymer as a function of the H atom distance from the surface. 
This is decomposed into lateral (x,y) and normal (z) components relative to the surface plane. 
Snapshots are also shown from calculations with the ring-polymer constrained close to the physisorption well at 3.0 \AA, at the transition state at 2.1 \AA, 
and unconstrained in the chemisorbed state at 1.5 \AA~ 
above the graphene sheet. 
The snapshots are an aggregation of bead positions for several hundred molecular dynamics steps.
The ring polymer is significantly more delocalised in the transition state region (2.1 \AA), a signature of quantum mechanical tunnelling through the barrier.
Reprinted with permission from ref. \cite{Davidson_2014_1}. Copyright 2014 American Chemical Society.
}
\end{figure}

In a recent study \cite{Davidson_2014_1}, some of us investigated in detail 
the role of both vdW interactions and NQEs in the H adsorption process on graphene at cryogenic temperatures to understand this process under interstellar conditions. 
The key findings are 
summarised in Fig.~\ref{H-ads}.
To cut a long series of simulations short, the
main conclusions were: (i) vdW interactions make the chemisorption barrier slightly smaller and narrower; reducing it from 200 meV to 175 meV with the specific vdW-inclusive DFT treatment employed; and (ii)
%
%
%
%
%
Thermal, but mostly NQEs, reduce the barrier further to $ca.$ 100 meV at 50 K. 
The quantum free energy barrier was calculated with a constrained-centroid PIMD approach \cite{Gillan_H_qtst}
and the reduction in the barrier compared to the classical one was attributed to tunnelling of the H atom near the transition state.
This can be seen in Fig. \ref{H-ads}b by 
the delocalisation of the H atom in the vicinity of the transition state. 
%
Overall, this study shows that both vdW interactions and NQEs work together in a cooperative manner to dramatically reduce the barrier to the formation of a covalent (C-H) bond; an effect that is likely to apply broadly to many chemical processes.
%
For this specific system it means that low temperature H atom chemisorption on sp$^2$ bonded materials is likely to be much more facile than previously anticipated, implying that chemisorbed H could be much more prevalent under interstellar medium conditions.
%
%

%

\subsection*{B. Dissociative adsorption of H$_2$ at metal surfaces}
\begin{figure}[!ht]
\includegraphics[width=12cm]{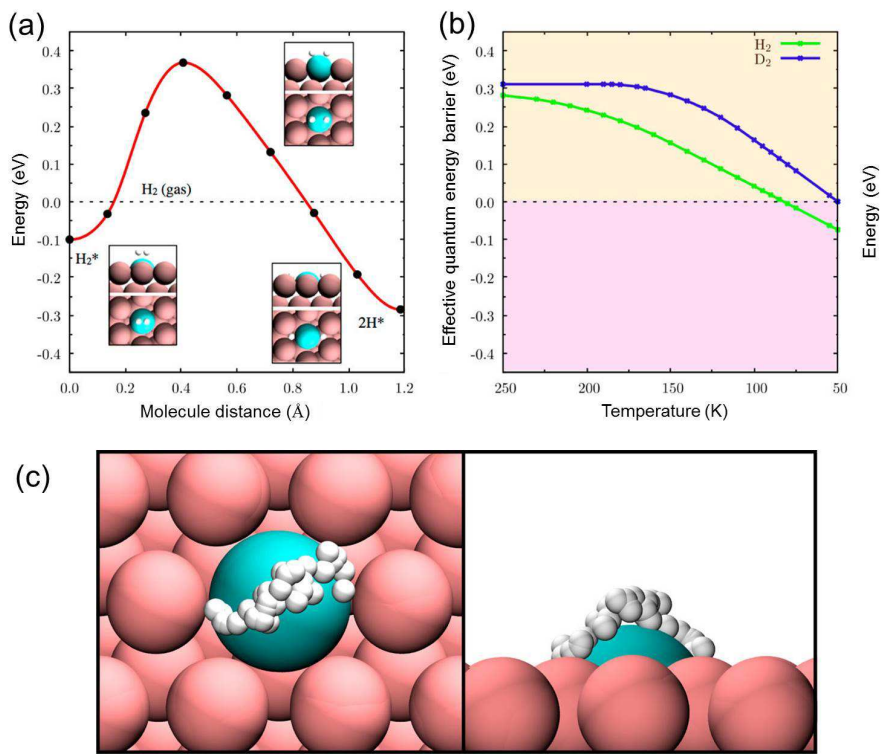}
\caption{\label{H2_ad_metal}
Quantum tunnelling facilitates H$_2$ dissociative adsorption on metal surfaces at low temperatures.
(a) Potential energy surface for H$_2$ dissociation from the physisorbed state on the Pd/Cu(111) substrate. 
The total energy of the clean surface and the H$_2$ in the gas phase is used as the energy zero. 
Insets are top and side views of the initial physisorbed state (H$_2$), the transition state (TS), and the final state (2H). 
White, pink, and cyan spheres indicate H, Cu, and Pd atoms, respectively. 
(b) Temperature dependence of the effective quantum energy barrier (relative to the gas phase) obtained from harmonic quantum transition state theory calculations that take into account tunnelling through the chemisorption barrier and zero-point energies. 
(c) Top and side view snapshots taken from an 85 K \textit{ab initio} path integral molecular dynamics simulation of a single H$_2$ at the classical saddle point for dissociative chemisorption on the Pd/Cu(111) surface.
The large spread of the beads along the dissociation reaction coordinate indicates that at this temperature the H$_2$ molecule can tunnel through the dissociation barrier.
Reprinted with permission from ref \cite{Davidson_2014_2}. Copyright 2014 American Chemical Society.
}
\end{figure}

Dissociative adsorption of molecular hydrogen plays an important role in a wide variety of chemical and physical processes, such as heterogeneous catalysis, energy storage, and sensing \cite{B718842K,lopez_when_2004,teschner_roles_2008,SAKINTUNA20071121}. 
In particular H$_2$ at metal surfaces is central to many processes in heterogeneous catalysis and has been widely studied on well-defined atomically flat metal surfaces for the last 30-40 years. 
Recently, highly dilute metal alloys (so called ``single atom alloys'') have emerged as a promising class of materials with unique catalytic functionality \cite{PhysRevLett.103.246102,doi:10.1021/acs.jpclett.8b01888,doi:10.1021/acscatal.8b00881,doi:10.1038/nchem.2915}.
In particular by doping a reactive transition metal (e.g. Pd, Pt, Ni) into a relatively unreactive (Cu, Ag, Au) host they offer great potential for highly selective hydrogenation reactions. 
With this in mind a detailed experimental and computational study of 
H$_2$ dissociation on a single atom alloy surface of Pd/Cu was recently performed \cite{Davidson_2014_2}.
Interestingly it was found experimentally that as the temperature was lowered to cryogenic temperatures the rate of H$_2$ dissociative adsorption \textit{increased}.
Complementary measurements of D$_2$ dissociative adsorption showed more conventional behaviour with the rate of adsorption decreasing as the temperature was lowered. 
The application of DFT and path integral based techniques proved to 
be very helpful in understanding this system. 
The key results of which are shown in Fig.~\ref{H2_ad_metal}, 
from where it can be seen that at the classical level the barrier to H$_2$ dissociative adsorption is fairly large (0.4 eV). 
This is too large to enable facile H$_2$ adsorption at low temperatures. 
However, when NQEs are taken into consideration the effective quantum free energy barrier for H$_2$ dissociation drops significantly as the temperature is lowered below 250 K. 
Although the D$_2$ barrier also drops with temperature, the effect kicks in at a much lower temperature ($ca.$ 150 K). 
Analysis again reveals that the origin of the barrier reduction is quantum mechanical tunnelling (Fig.~\ref{H2_ad_metal}c).
Interestingly, the calculations also suggest that tunnelling enables a different dissociation mechanism at very low temperatures (below 80 K for H$_2$, 50 K for D$_2$), where incident molecules at the Pd site can undergo barrierless dissociation avoiding a physisorbed state into which they would otherwise be trapped.
Note that at the higher temperatures of many existing catalytic processes we do not expect NQEs to play a significant role. 
However, examining these effects in other systems may uncover temperature regimes at room temperature and below where quantum effects can be harnessed, yielding better control of bond-breaking processes at surfaces and uncovering useful chemical properties such as selective bond activation or isotope separation.

\subsection*{C. Diffusion of atomic hydrogen}
Once H atoms are adsorbed onto a surface, NQEs can also be important to how they diffuse across it. 
The role of NQEs in H atom diffusion has been extensively studied on a wide variety of surfaces \cite{Interstellar_2013,martinazzo_hydrogen_2013,DURR201361,Hgra_PRL,JPCC_Hdiff_gra,PhysRevB.79.115429,doi:10.1063/1.5029329,Skulason20121400}. 
Here, we focus on work that we have recently been involved in to understand H atom diffusion across atomically-flat metal surfaces. 
Such systems have been widely studied thanks to the 
development of surface sensitive experimental techniques (see e.g. \cite{lin_diffusion_1991,lauhon_direct_2000,jardine_determination_2010,Ru_2013}).
Interestingly, experimental measurements of diffusion rates yield qualitatively different temperature dependencies upon moving from one metal surface to the next. 
%
Relatively straightforward behaviour is seen on e.g. Pt(111) where according to helium spin echo measurements \cite{jardine_determination_2010}, the rate drops as temperature is lowered. 
On Ru(0001), a gradual transition from Arrhenius behaviour to a temperature independent (i.e. quantum) regime has been reported \cite{Ru_2013}. 
However on Ni(100) \cite{lin_diffusion_1991} and Cu(100) \cite{lauhon_direct_2000}, diffusion rates suddenly become T-independent below a certain temperature, indicating a sharp classical to quantum transition.
%

\begin{figure}[!ht]
    \centering
    \includegraphics[width=8cm]{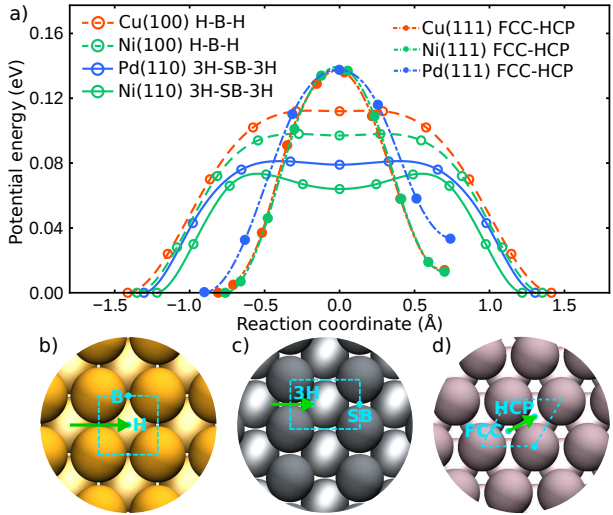}
    \caption{
    a) Potential energy barriers for H atom diffusion on metal surfaces can vary considerably from a conventional parabolic shape to an unconventional broad-topped shape. A variety of H diffusion barriers is shown, obtained from nudged elastic band calculations using DFT, for several transition metal surfaces.
    The filled symbols show data for the conventional barriers that are parabolic near the top, and the open symbols are the data points for broad-top barriers.
    In (b)-(d) the structures of the various metal surfaces considered are reported: b) Top view of the (100) surface;
    c) Top view of the (110) surface;
    d) Top view of the (111) surface.
    Green arrows show the diffusion paths.
    Reprinted with permission from ref \cite{Wei_HD_2017}. Copyright 2017 American Physical Society.
    }
    \label{barrier}
\end{figure}

\begin{figure}[!ht]
\includegraphics[width=9cm]{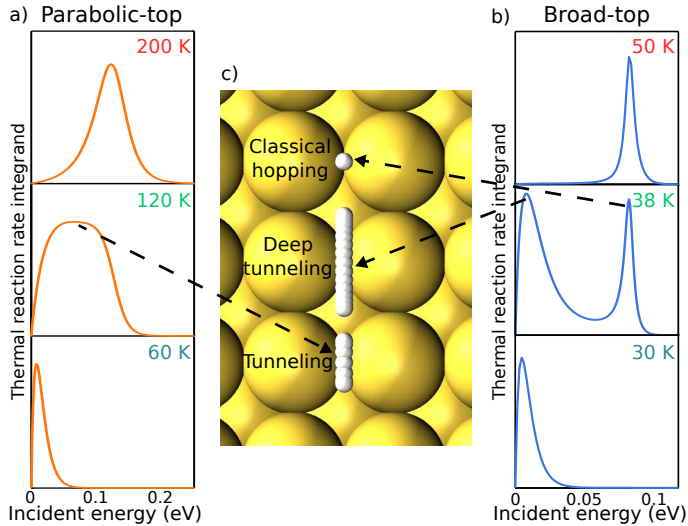}
\caption{\label{diffu_cmp}
    The shape of the potential energy barrier strongly influences the transition from classical hopping to quantum tunnelling for H diffusion on metals.
    In 
    %
    a) and b), the thermal rate integrand is plotted against the incident energy $E$ for examples of the conventional parabolic-top barrier (Ni(111)) and the broad-top barrier (Pd(110)) respectively.
    c) Illustration of the tunnelling behaviour the different peak positions of the rate integrand represent, using tunnelling paths represented by the Feynman path-integral.
    Adapted with permission from ref \cite{Wei_HD_2017}. Copyright 2017 American Physical Society.
}
\end{figure}

Theoretical studies of H diffusion on metal surfaces has been useful in helping to understand this behaviour for certain specific experiments \cite{lin_diffusion_1991,lauhon_direct_2000,doi:10.1063/1.479392,ZHANG2011689}. 
For example, the sharp transition on Ni(100) and Cu(100) was attributed to the particular shape of the diffusion barrier \cite{mattsson_h_1993,mattsson_isotope_1997,sundell_quantum_2004,sundell_hydrogen_2005,Sundell_2004_2,Skulason20121400}.
In a recent study we set about rationalising this behaviour in general and performed a systematic DFT-based instanton study of H atom diffusion across a range of metal surfaces \cite{Wei_HD_2017}. 
The key finding was that H atom diffusion on metal surfaces could be 
categorised into systems possessing either conventional 
parabolic shaped barriers (“parabolic-tops”) or into systems possessing unusually broad barriers resembling a top hat (“broad-tops”) (Fig.~\ref{barrier}).
The unusually broad barriers are found on surfaces (see also refs.~\cite{Skulason20121400,doi:10.1063/1.5029329}) partly because of the very large mismatch in size between the small H atom and the relatively large surface atoms of the lattice. 
Such barriers have also been seen for magnetic transitions \cite{C6FD00136J}.
Comparing the thermal rate integrand (integration of which gives the rate at a given temperature) of the two types of barriers (Fig.~\ref{diffu_cmp}), one finds a qualitative difference.
With the conventional parabolic-top barriers, hydrogen diffusion evolves gradually from classical hopping to shallow tunnelling to deep tunnelling as the temperature decreases, and noticeable quantum effects persist at moderate temperature.  
In contrast, with broad-top barriers quantum effects become important only at the lowest temperatures and the classical to quantum transition is sharp, at which point classical hopping and deep tunnelling both occur (Fig.~\ref{diffu_cmp}b).
The unusual behaviour revealed by the simulations has a number of far reaching implications \cite{Wei_HD_2017}, including 
%
a new way of defining the classical to quantum crossover temperature (T$_\text{W}$) and a prediction of the sudden emergence of strong isotope effects around T$_\text{W}$ \cite{Wei_HD_2017}. 
%
%
These insights are likely to help and guide the interpretation of existing and future experiments for H diffusion on metals and quantum tunnelling in chemical systems in general.

\subsection*{D. Proton penetration of 2D materials}
Another topic recently brought into focus is the direct transfer of H atoms or protons through 2D materials such as graphene and h-BN.
Generally it was assumed that pristine graphene and h-BN were impermeable to H atoms and protons.
Indeed DFT calculations have shown that the barrier for a chemisorbed H or proton to penetrate a pristine graphene sheet is 3.5 eV or more \cite{C3CP52318G,doi:10.1021/acs.jpclett.6b01507,C6CP08923B}.
However, recent experiments provide strong evidence that protons can, in fact, penetrate pristine layers of graphene and h-BN \cite{hu_proton_2014}.
Based on temperature-dependent proton conductivity measurements, the barriers for the proton penetration process were estimated to be only 0.8 and 0.3 eV for single-layer graphene and h-BN, respectively.
In a subsequent study, focusing on the role of isotope effects, 
the penetration rate reduced by an order of magnitude when the protons (H$^+$) were replaced by deuterons (D$^+$) and the barrier estimated to increase by about 60 meV \cite{Lozada-Hidalgo68}.

Partly because of the fascinating measurements, considerable theoretical effort have gone into understanding the microscopic details of how protons penetrate
graphene and h-BN and to understand the role of NQEs in the process \cite{doi:10.1021/acs.jpclett.6b01507,doi:10.1021/acs.jpcc.7b08152,Tka_tunneling_2016,doi:10.1021/acs.jpclett.7b02820}.
In work that we were involved in, we used DFT-based PIMD to
examine the proton penetration process \cite{doi:10.1021/acs.jpclett.7b02820}.
With the specific DFT functional used and model system employed we found that the classical $\text{H}^+$  penetration barrier was approximately 3.5 eV, in line with earlier studies. 
When NQEs were accounted for the 
%
penetration barrier for $\text{H}^+$ on graphene was reduced by 0.46 eV ($12\%$) at 300K and for $\text{D}^+$ the reduction was less.
The reduction in the free-energy barrier is due to ZPE effects and quantum tunnelling near the transition state (as shown in the insets, the beads are more delocalised near the TS compared with the initial state).
Although the role of NQEs is far from negligible, NQEs alone do not describe the enormous discrepancy between the computed and experimental proton penetration barriers. 
%
%
Instead we suggested in ref. \cite{doi:10.1021/acs.jpclett.7b02820} that
hydrogenation of the graphene (and h-BN) plays a critical role. 
In particular hydrogenation can destabilise the low-lying chemisorbed initial state and slightly expands the hexagonal lattice through which the proton penetrates. 
\begin{figure}[!ht]
\includegraphics[width=12cm]{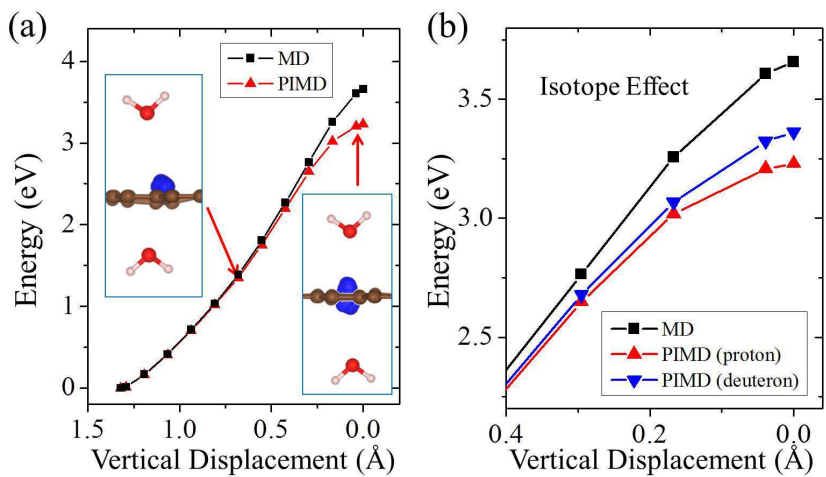}
\caption{\label{proton_pene}
Quantum contributions and isotope effects for proton penetration of pristine graphene.
(a) Free-energy profiles at 300 K obtained with \textit{ab initio} constrained MD and PIMD simulations for proton transfer across a graphene sheet in the presence of water molecules. 
PIMD simulation snapshots for the initial state and transition state are also shown. 
Blue (red and pink) balls represent the beads of protons (O and H atoms) for one snapshot in a PIMD simulation. 
The centroids of the C atoms are shown as brown balls. 
(b) Zoom-in view of the free energy profiles shown in (a) close to the transition state, for both proton and deuteron penetration.
Reprinted with permission from ref \cite{doi:10.1021/acs.jpclett.7b02820}. Copyright 2017 American Chemical Society.
}
\end{figure}

\section{The role of NQEs on hydrogen bonds}
Hydrogen is obviously also present in hydrogen-bonded systems and such systems are essential for life, 
being an essential component of the binding in the condensed phases of water, the structure of many biomolecules, and molecular crystals.
In fact, several early studies on NQEs with PIMD were on H-bonded systems such as water and ice.
See refs. \cite{Marx_waterrev_2006,doi:10.1021/jp810590c,Marx_Tuckerman_2010,Wei_rev_2016} for reviews of this work.
Here,  we go beyond aqueous systems towards bio-molecules, organic crystals, and HBs at interfaces. 
We focus on two specific aspects: (A) quantum delocalisation in HBs; and (B) how NQEs impact upon the strength of HBs. 
%

%
\begin{figure}[!ht]
\includegraphics[width=8cm]{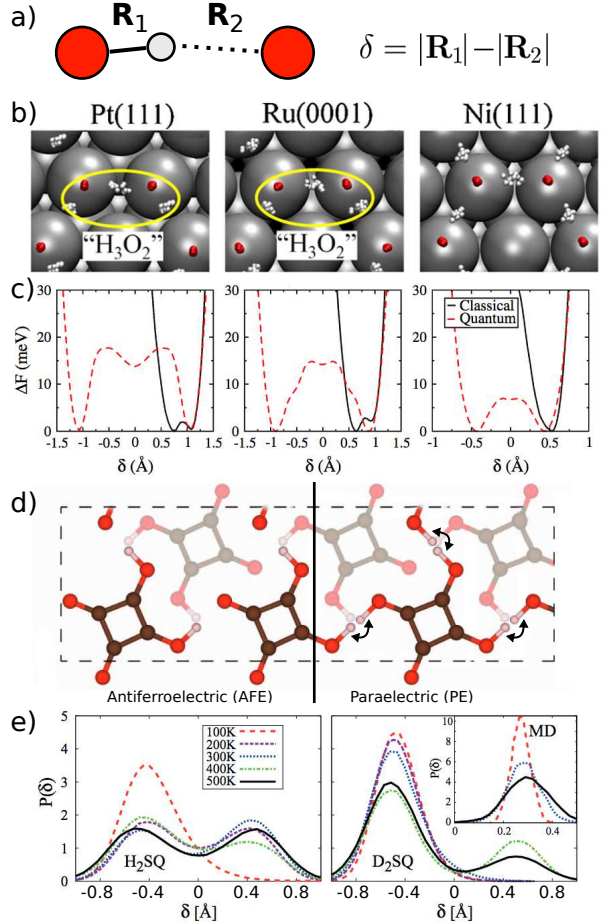}
\caption{\label{Qdelocal}
NQEs can strongly influence H-bonded systems, as shown here for two distinct types of system. 
(a) Illustration of a HB and the order parameter $\delta$ for describing the H position in a HB.
(b) Snapshots for typical spatial configurations of an overlayer of water and hydroxyl molecules on Pt (left), Ru (middle), and Ni (right) obtained from PIMD.
(c) Free energy profiles ($\Delta F$) for H atom transfer from water to hydroxyl along the intermolecular axes for the systems shown in (b) on Pt (left), Ru (middle), and Ni (right) at 160 K, obtained from MD and PIMD.
Reprinted with permission from ref \cite{water_metal_NQEs}. Copyright 2010 American Physical Society.
(d) Crystal structure of squaric acid, illustrating the difference between AFE (left) and PE (right) ordering.
(e) Probability distributions of the $\delta$ parameter for H$_2$SQ (left) and D$_2$SQ (right) from PIMD, and (inset in right) from MD.
In the MD simulations the nuclei are treated as classical particles, in the PIMD simulations they are quantum. 
All simulations have been performed at the DFT level of theory.
Adapted with permission from ref \cite{doi:10.1063/1.4862740}. Copyright 2014 AIP Publishing.
}
\end{figure}

\subsection*{A. Quantum Delocalisation in hydrogen bonds}
%
As noted in the introduction, the quantum nature of hydrogen means that it does not behave like a point-like particle. 
A key question has been to what extent hydrogens are delocalised in space when incorporated in HBs and how this varies from system to system. 
In a traditional HB the hydrogen is covalently bonded to one electronegative element and H-bonded to a second, such as the OH-O bond that holds two water molecules together. 
A HB such as this is asymmetric and conveniently characterised with an order parameter  $\delta=|\textbf{R}_1|-|\textbf{R}_2|$ as illustrated in Fig. \ref{Qdelocal}a. 
Applying pressure to the HB, as done in high pressure studies of e.g. ice, can leave the proton shared symmetrically between the two oxygens it is bonded to  \cite{Marx1998,doi:10.1063/1.4818875,doi:10.1063/1.465467,doi:10.1063/1.1677221,Goncharov218}.
For ice the intermolecular separation was controlled by pressure.
However water on metals represents another class of systems 
where variable intermolecular separations can be found. 
In these systems the intermolecular separation between the adsorbed molecules is dictated to a large extent by the spacing between the substrate lattice sites \cite{Morgenstern_2007_1,Carrasco_2012_1,Salmeron_2015,doi:10.1021/acs.chemrev.6b00045}. 
One particular class of adsorption structures, comprised of a mixture of water and hydroxyl forms across many metal surfaces. 
Upon moving from one surface to the next, the key difference in the adsorption structure is that the intermolecular separation changes and is tuned by the lattice constant of the underlying substrate \cite{doi:10.1021/ja003576x,PhysRevB.69.113404}. 
Some years ago we performed a series of DFT-based PIMD simulations on such systems and identified surprisingly pronounced NQEs \cite{water_metal_NQEs}.
Specifically we found that metal substrates, by templating the water overlayer, could shorten the corresponding HBs (in analogy to ice under high pressure). 
Depending on the substrate and the intermolecular separations it imposes, the traditional distinction between covalent and HBs is partially or almost entirely lost (Fig. \ref{Qdelocal}b).
Such a picture only emerges with a quantum treatment of the system; treating the nuclei classically preserves the traditional picture because of large classical free energy barriers for proton hopping (Fig. \ref{Qdelocal}c).   

Soon after the theoretical prediction of strong quantum delocalisation in such systems, direct experimental evidence was obtained from STM for a closely related system \cite{PhysRevB.81.045402}.
However, more experiments are needed on a broader range of systems to test the theoretical predictions of ref. \cite{water_metal_NQEs}. 
Indeed it seems to us that interfacial water is an excellent model system for probing the delicate connection between HB length and quantum delocalsation and studying systems such as these is likely to be a fruitful area of future research.

Quantum delocalisation is also key to order-disorder transitions in hydrogen-bonded crystals. 
This includes high pressure phases of ice \cite{Marx1998,HERRERO2015125,doi:10.1063/1.4818875}
and certain organic crystals where the order-disorder transition is connected with an (anti-)ferroelectric to paraelectric transition.
One system of particular interest that we have looked at recently is squaric acid \cite{doi:10.1063/1.4862740}.
In particular we focussed on understanding the antiferroelectric to paraelectric transition and its isotope dependence (Fig.~\ref{Qdelocal}d).
Interestingly, in agreement with experiment, we were able to show that 
quantum delocalisation and concerted tunnelling resulted in a $ca.$ 200 K difference in the order-disorder transition temperature between the hydrogenated and deuterated crystals (Fig.~\ref{Qdelocal}e).
Specifically, as shown in Fig.~\ref{Qdelocal}e, the HBs become disordered (paraelectric) in H$_2$SQ at 200 K and above, while in D$_2$SQ the HBs are ordered up to 400 K.
NQEs also explain the superconductivity transition in a recently discovered high temperature superconductor, solid H$_3$S.
Calculations accounting for NQEs correctly predict the phase transition from an asymmetric to a symmetric proton phase, reproducing experimental H/D isotope effects in the superconducting transition temperature \cite{errea_quantum_2016}.

\subsection*{B. NQEs on the structure and strength of hydrogen bonds}
%
Regardless of the position of hydrogen within a HB, NQEs can influence the structure of H-bonded systems.
This geometric effect is sometimes known as 
the Ubbelohde effect \cite{Ubbelohde}, where replacing H with D can change the heavy atom separation in a HB, and in H-bonded molecular crystals consequently alter their lattice constants.
Model studies of the Ubbelohde effect have been performed  \cite{McKenzie2012,McKenzie_2014}), and a few years ago some of us carried out an \textit{ab initio} PIMD study of NQEs on a range of H-bonded clusters and crystals \cite{XZLi_2011}. 
The key conclusion of our work is that NQEs make strong HBs shorter and weak HBs longer. 
This observation was explained by generalising earlier observations of Manolopoulos and co-workers  \cite{Habershon_2009,C1CP21520E,Markland_2012} of an effect that is is now known as the competing quantum effects (CQEs) picture. 
The CQEs picture has been very successful in understanding and reproducing experimental H/D isotope fractionation ratios between liquid water and its vapour \cite{Markland_2012,Ceriotti_2013_1,LWang_2014}, and has been reviewed in detail recently \cite{Wei_rev_2016}.
In short, it arises out of a balance of effects: quantum nuclear motion along the HB direction that strengthens the bond, versus quantum fluctuations out of the plane of the HB that weakens it. 
Thanks to this competition, for certain systems (such as liquid water), simulations with classical nuclei can give comparable results to those obtained with quantum nuclei \cite{Wei_rev_2016,doi:10.1063/1.4907554,rossi_nuclear_2016,doi:10.1021/acs.jpclett.7b00979}.
In fact, the balance of the CQEs in liquid water is so delicate, that even the choice of the electronic structure approach (or force field) can lead to qualitatively opposite results on which of the two CQEs dominates \cite{Parrinello_2003,Morrone_2008,water_ordf}.
Temperature can also change the balance between the competing effects \cite{Markland_2012,Wei_BP_BFE}.
%

The existence of CQEs has been supported by several experimental investigations \cite{Romanelli_2013,DINS_3}, specifically 
DINS experiments on water \cite{Romanelli_2013,DINS_3,Cheng_2016} and  
STM experiments for water on NaCl \cite{Guo321}.
Nonetheless, most previous work has focused on geometrical properties, while direct information on how and to what extent NQEs influence the strengths of HBs has been lacking.
Recent developments allow one to estimate this quantitatively in computer simulations for materials and molecular systems of moderate size, through a combination of PIMD and thermodynamic integration using mass as the order parameter \cite{Mass_2,Wei_BP_BFE,Rossi_2015}.
Note that there are different ways to combine thermodynamic integration with PIMD \cite{Morales_Singer,DaanFrenkel_book,doi:10.1063/1.2966006,doi:10.1002/jcc.21070,scaled_coordinate}, with early developments dating back to 1991 \cite{Morales_Singer}.

\begin{figure}[!ht]
\includegraphics[width=11cm]{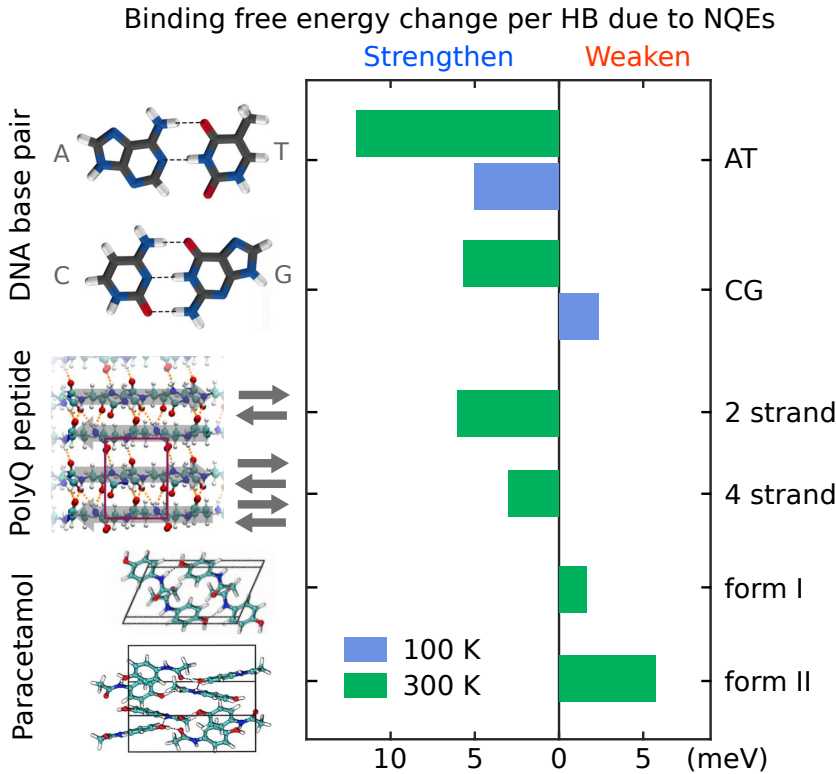}
\caption{\label{BFE}
NQEs can either strengthen or weaken HBs, as illustrated here for the binding free energy change of several important H-bonded organic systems: Watson-Crick base pairs \cite{Wei_BP_BFE}, polyglutamine(polyQ) strands \cite{Rossi_2015}, and paracetamol crystals \cite{Rossi_2016_2}, calculated from \textit{ab initio} PIMD simulations.
Structural images are adapted with permission from refs \cite{Wei_BP_BFE,Rossi_2015,Rossi_2016_2}. Copyright: 2016 American Chemical Society; 2015 American Chemical Society; 2016 American Physical Society, respectively.
}
\end{figure}

Fig.~\ref{BFE} summarises recent quantitative determinations of the impact of NQEs on HB strength for several systems important to everyday life, namely DNA base pairs, a peptide, and paracetamol.
It can be seen from Fig.~\ref{BFE} that the absolute influence on the stability of the various materials is small at 10 meV of less per HB.
However, NQEs can act to both strengthen or weaken the HBs and can exhibit interesting temperature dependencies.
Taking the DNA base pairs as an example, these were examined at room temperature and at a cryogenic temperature (100 K) \cite{Wei_BP_BFE}.
%
It was found that NQEs stabilise the base pairs at room temperature, while counterintuitively, the influence of NQEs was smaller at cryogenic temperatures than it was at room temperature. 
This was rationalised, as with other systems, in terms of a competition of NQEs between low-frequency and high-frequency vibrational modes.
Upon forming a HB, certain high-frequency (covalent bond stretching) modes are softened, reducing the quantum kinetic energy hence stabilising the system. 
This stabilisation, however, is offset by the quantum kinetic energy gained when low-frequency modes are hardened or created upon forming the HB. 
%
%
%
Moving on from DNA to proteins, for stacked polyglutamine (polyQ) strands, a peptide often found in amyloid aggregate, NQEs serve to provide an additional degree of stabilisation \cite{Rossi_2015}. 

Another important field where quantitative estimates of NQEs are highly desirable is in the assessment of the stability ranking of different polymorphs of molecular crystals, where a small free energy change of 10 meV can make a difference.
A recent \textit{tour de force} study addressing this found that for paracetamol, $\sim$ 20 \% of the free energy difference between its form I and form II comes from NQEs \cite{Rossi_2016_2}.
This work also suggests that estimates of NQEs on the binding free energy within the harmonic approximation can be reasonable for molecular crystals.

The evaluations of quantum free energy contributions to H-bonded systems is enjoying rapid development with interesting findings in key systems.
Extending these findings, one may ask are there more efficient ways or simple models for estimating the importance of such effects?
Recently we have taken a step in this direction with the presentation of a simple model based on the CQEs picture which predicts the temperature dependence of NQEs on the binding strength of broad range of H-bonded complexes \cite{Wei_BP_BFE}.
%

\section{Outlook and future challenges}
To sum up, we have discussed examples where NQEs are important and interesting.
From solid hydrogen to hydrogen on surfaces to H-bonded systems, NQEs can have a qualitative and quantitative impact on the physiochemical properties of materials.
This is especially true at cryogenic temperatures.
However, we have also shown examples where such effects can be important at room temperatures. 
This review has not covered all types of systems where NQEs are important, but rather gives a flavour of the systems that have piqued our interest over the last few years, with a focus on hydrogen containing systems. 

As noted in the introduction it is an exciting and thriving time for the field due to the development of relatively fast, efficient, and accurate simulation approaches for treating NQEs as well as the development of complementary experimental techniques. 
However, there are, of course, many outstanding challenges for simulation techniques in this area. 
Some of these have been discussed elsewhere \cite{Wei_rev_2016,markland_nuclear_2018,Inst_persp,habershon_ring-polymer_2013},
but the issues we feel that are of particular importance are: 

(i) \textit{Accuracy of the underlying electronic structure theory} -- The results discussed in this brief review have come mainly from DFT-based approaches. 
DFT is the most widely used electronic structure method for the simulation of materials. 
However, it has a number of well-documented shortcomings, many of which are directly relevant to the types of systems discussed here. 
For example, HB strengths, physiorption on surfaces, and covalent bond breaking processes are all sensitive to the choice of the exchange-correlation functional used (see e.g. \cite{Gillan_DFT_water,doi:10.1063/1.4704546,doi:10.1063/1.4869598,PhysRevB.94.115144,doi:10.1063/1.4754130}).
Given that the emergence of NQEs can often depend on a subtle balance of competing effects, qualitatively different behaviour can be observed with different exchange-correlation functionals \cite{XZLi_2011,water_ordf,chen_room-temperature_2014}.
Thus one always has to take care that the exchange-correlation functional used is suitable for the problem at hand. 
More broadly, combining path integral methods with higher level methods such as quantum Monte Carlo and quantum chemistry approaches is clearly an exciting avenue for future research \cite{RevModPhys.73.33,AlaviFCIQMC,mcmahon_properties_2012}. 
Indeed, given the recent progress with quantum Monte Carlo approaches in terms of accuracy, efficiency, and the calculation of forces, QMC-based PIMD simulations look like a promising way of treating particularly challenging systems \cite{PI_QMC_1,doi:10.1063/1.4917171,Zen201715434,PhysRevB.93.241118}. 
Methods for dealing with non-adiabatic effects is another area where excellent progress is being made 
on topics such as understanding NQEs in electron transfer \cite{RICHARDSON2017124,doi:10.1063/1.4932362,doi:10.1021/acs.jpclett.7b01249,doi:10.1063/1.453440,doi:10.1063/1.4863919,doi:10.1002/cphc.201200941}. 
Other non-adiabatic effects, such as electron-phonon coupling, are important for some problems (i.e. dissociative adsorption, dynamics of adsorbates on surfaces) \cite{doi:10.1080/23746149.2017.1381574,PhysRevLett.68.3444,sundell_quantum_2004,doi:10.1063/1.469915,PhysRevLett.116.217601,WANG200266}, but it remains to be fully understood to what extent they impact upon quantum H atom diffusion and the other topics covered in the review.
Indeed coupling friction and quantum nuclear approaches is an interesting area for future research.
%

(ii) \textit{Connection with experiment} -- Much of the work discussed in this review was motivated by or aimed at explaining interesting experimental results. 
And, indeed recent years have seen huge advances in experimental capabilities, from the emergence of deep inelestaic neutron scattering, to scanning tunnelling microscopy, to diamond anvil measurements at ever higher pressures. 
Given the emergence of such techniques there is an ever greater need for more efficient and more robust techniques to compute experimental observables measured in scattering and spectroscopic studies, including time-dependent properties \cite{PhysRevLett.105.110602,trpmd,PIGLET,Romanelli_2013,doi:10.1063/1.3675838,doi:10.1021/acs.jpcb.8b03896}.
Excellent work in this area is ongoing and more is welcomed. 

(iii) \textit{System complexity} -- When connecting with experiment it is challenging to make the simulation model realistic enough that it faithfully mimics the complexity present in the real experimental system. This is especially true when treating defects for which large simulation cells are required, or liquids where large cells and sufficient sampling of phase space is required. 
Although enormous strides have been made in reducing the computational overhead of incorporating NQEs (see review \cite{markland_nuclear_2018}), most studies are still limited by the availability of computational resource. 
Developing higher accuracy interatomic potentials with e.g. machine learning approaches \cite{PhysRevLett.104.136403,doi:10.1002/qua.24890}
and efficiently combining enhanced sampling techniques with path integral methods are two promising avenues for future study \cite{Cheng_2016,Laude_2018}.
With our interest in graphene, as discussed above, we have recently developed a machine learning potential for graphene \cite{PhysRevB.97.054303}
that accurately reproduces DFT for a broad range of properties and are currently using this potential to understand the role of NQEs on the vibrational properties of graphene. 

Addressing these and other computational challenges will provide interesting work for the future and consequently will help to deepen understanding of the role of NQEs in materials yet further. 
Although we believe that great strides have been made in the field in the last decade or so, there is still considerably more to learn about the role of NQEs in the chemistry and physics of hydrogen-rich systems.

\begin{acknowledgements}
The authors thank Ms X.Y. Jiang for helpful suggestions on the review.
J.C. and A.M. are grateful to the Alexander von Humboldt Foundation for a post doctoral research fellowship and a Bessel Research Award, respectively. A.M. is also supported by the European Research Council under the European Union's Seventh Framework Programme (FP/2007-2013)/ERC Grant Agreement number 616121 (HeteroIce project). Y.X.F and X.Z.L are supported by the National Basic Research Programs of China under Grant Nos. 2016YFA0300900, the National Science Foundation of China under Grant Nos.  11774003, 11604092, and 11634001, and the high-performance computing platform of Peking University. The collaboration between UCL and Peking University was supported in part by UCL's Global Engagement Fund.
\end{acknowledgements}

 \bibliography{ref}

\begin{thebibliography}{228}%
\makeatletter
\providecommand \@ifxundefined [1]{%
 \@ifx{#1\undefined}
}%
\providecommand \@ifnum [1]{%
 \ifnum #1\expandafter \@firstoftwo
 \else \expandafter \@secondoftwo
 \fi
}%
\providecommand \@ifx [1]{%
 \ifx #1\expandafter \@firstoftwo
 \else \expandafter \@secondoftwo
 \fi
}%
\providecommand \natexlab [1]{#1}%
\providecommand \enquote  [1]{``#1''}%
\providecommand \bibnamefont  [1]{#1}%
\providecommand \bibfnamefont [1]{#1}%
\providecommand \citenamefont [1]{#1}%
\providecommand \href@noop [0]{\@secondoftwo}%
\providecommand \href [0]{\begingroup \@sanitize@url \@href}%
\providecommand \@href[1]{\@@startlink{#1}\@@href}%
\providecommand \@@href[1]{\endgroup#1\@@endlink}%
\providecommand \@sanitize@url [0]{\catcode `\\12\catcode `\$12\catcode
  `\&12\catcode `\#12\catcode `\^12\catcode `\_12\catcode `\%12\relax}%
\providecommand \@@startlink[1]{}%
\providecommand \@@endlink[0]{}%
\providecommand \url  [0]{\begingroup\@sanitize@url \@url }%
\providecommand \@url [1]{\endgroup\@href {#1}{\urlprefix }}%
\providecommand \urlprefix  [0]{URL }%
\providecommand \Eprint [0]{\href }%
\providecommand \doibase [0]{http://dx.doi.org/}%
\providecommand \selectlanguage [0]{\@gobble}%
\providecommand \bibinfo  [0]{\@secondoftwo}%
\providecommand \bibfield  [0]{\@secondoftwo}%
\providecommand \translation [1]{[#1]}%
\providecommand \BibitemOpen [0]{}%
\providecommand \bibitemStop [0]{}%
\providecommand \bibitemNoStop [0]{.\EOS\space}%
\providecommand \EOS [0]{\spacefactor3000\relax}%
\providecommand \BibitemShut  [1]{\csname bibitem#1\endcsname}%
\let\auto@bib@innerbib\@empty
\bibitem [{\citenamefont {Vega}\ \emph {et~al.}(2010)\citenamefont {Vega},
  \citenamefont {Conde}, \citenamefont {McBride}, \citenamefont {Abascal},
  \citenamefont {Noya}, \citenamefont {Ramirez},\ and\ \citenamefont
  {Ses\'e}}]{Vega_2010}%
  \BibitemOpen
  \bibfield  {author} {\bibinfo {author} {\bibfnamefont {C.}~\bibnamefont
  {Vega}}, \bibinfo {author} {\bibfnamefont {M.~M.}\ \bibnamefont {Conde}},
  \bibinfo {author} {\bibfnamefont {C.}~\bibnamefont {McBride}}, \bibinfo
  {author} {\bibfnamefont {J.~L.~F.}\ \bibnamefont {Abascal}}, \bibinfo
  {author} {\bibfnamefont {E.~G.}\ \bibnamefont {Noya}}, \bibinfo {author}
  {\bibfnamefont {R.}~\bibnamefont {Ramirez}}, \ and\ \bibinfo {author}
  {\bibfnamefont {L.~M.}\ \bibnamefont {Ses\'e}},\ }\href {\doibase
  10.1063/1.3298879} {\bibfield  {journal} {\bibinfo  {journal} {J. Chem.
  Phys.}\ }\textbf {\bibinfo {volume} {132}},\ \bibinfo {pages} {046101}
  (\bibinfo {year} {2010})}\BibitemShut {NoStop}%
\bibitem [{\citenamefont {Layfield}\ and\ \citenamefont
  {Hammes-Schiffer}(2014)}]{SHS_bio_tunel_rev}%
  \BibitemOpen
  \bibfield  {author} {\bibinfo {author} {\bibfnamefont {J.~P.}\ \bibnamefont
  {Layfield}}\ and\ \bibinfo {author} {\bibfnamefont {S.}~\bibnamefont
  {Hammes-Schiffer}},\ }\href {\doibase 10.1021/cr400400p} {\bibfield
  {journal} {\bibinfo  {journal} {Chem. Rev.}\ }\textbf {\bibinfo {volume}
  {114}},\ \bibinfo {pages} {3466} (\bibinfo {year} {2014})}\BibitemShut
  {NoStop}%
\bibitem [{\citenamefont {Pu}\ \emph {et~al.}(2006)\citenamefont {Pu},
  \citenamefont {Gao},\ and\ \citenamefont {Truhlar}}]{bio_rev}%
  \BibitemOpen
  \bibfield  {author} {\bibinfo {author} {\bibfnamefont {J.}~\bibnamefont
  {Pu}}, \bibinfo {author} {\bibfnamefont {J.}~\bibnamefont {Gao}}, \ and\
  \bibinfo {author} {\bibfnamefont {D.~G.}\ \bibnamefont {Truhlar}},\ }\href
  {\doibase 10.1021/cr050308e} {\bibfield  {journal} {\bibinfo  {journal}
  {Chem. Rev.}\ }\textbf {\bibinfo {volume} {106}},\ \bibinfo {pages} {3140}
  (\bibinfo {year} {2006})}\BibitemShut {NoStop}%
\bibitem [{\citenamefont {Jancso}\ and\ \citenamefont
  {Van~Hook}(1974)}]{doi:10.1021/cr60292a004}%
  \BibitemOpen
  \bibfield  {author} {\bibinfo {author} {\bibfnamefont {G.}~\bibnamefont
  {Jancso}}\ and\ \bibinfo {author} {\bibfnamefont {W.~A.}\ \bibnamefont
  {Van~Hook}},\ }\href {\doibase 10.1021/cr60292a004} {\bibfield  {journal}
  {\bibinfo  {journal} {Chem. Rev.}\ }\textbf {\bibinfo {volume} {74}},\
  \bibinfo {pages} {689} (\bibinfo {year} {1974})}\BibitemShut {NoStop}%
\bibitem [{\citenamefont {Ubbelohde}\ and\ \citenamefont
  {Gallagher}(1955)}]{Ubbelohde}%
  \BibitemOpen
  \bibfield  {author} {\bibinfo {author} {\bibfnamefont {A.~R.}\ \bibnamefont
  {Ubbelohde}}\ and\ \bibinfo {author} {\bibfnamefont {K.~J.}\ \bibnamefont
  {Gallagher}},\ }\href@noop {} {\bibfield  {journal} {\bibinfo  {journal}
  {Acta. Crystallogr.}\ }\textbf {\bibinfo {volume} {8}},\ \bibinfo {pages}
  {71} (\bibinfo {year} {1955})}\BibitemShut {NoStop}%
\bibitem [{\citenamefont {Dalal}\ \emph {et~al.}(1998)\citenamefont {Dalal},
  \citenamefont {Klymachyov},\ and\ \citenamefont
  {Bussmann-Holder}}]{PhysRevLett.81.5924}%
  \BibitemOpen
  \bibfield  {author} {\bibinfo {author} {\bibfnamefont {N.}~\bibnamefont
  {Dalal}}, \bibinfo {author} {\bibfnamefont {A.}~\bibnamefont {Klymachyov}}, \
  and\ \bibinfo {author} {\bibfnamefont {A.}~\bibnamefont {Bussmann-Holder}},\
  }\href {\doibase 10.1103/PhysRevLett.81.5924} {\bibfield  {journal} {\bibinfo
   {journal} {Phys. Rev. Lett.}\ }\textbf {\bibinfo {volume} {81}},\ \bibinfo
  {pages} {5924} (\bibinfo {year} {1998})}\BibitemShut {NoStop}%
\bibitem [{\citenamefont {Wikfeldt}\ and\ \citenamefont
  {Michaelides}(2014)}]{doi:10.1063/1.4862740}%
  \BibitemOpen
  \bibfield  {author} {\bibinfo {author} {\bibfnamefont {K.~T.}\ \bibnamefont
  {Wikfeldt}}\ and\ \bibinfo {author} {\bibfnamefont {A.}~\bibnamefont
  {Michaelides}},\ }\href {\doibase 10.1063/1.4862740} {\bibfield  {journal}
  {\bibinfo  {journal} {J. Chem. Phys.}\ }\textbf {\bibinfo {volume} {140}},\
  \bibinfo {pages} {041103} (\bibinfo {year} {2014})}\BibitemShut {NoStop}%
\bibitem [{\citenamefont {G\'omez-Gallego}\ and\ \citenamefont
  {Sierra}(2011)}]{doi:10.1021/cr100436k}%
  \BibitemOpen
  \bibfield  {author} {\bibinfo {author} {\bibfnamefont {M.}~\bibnamefont
  {G\'omez-Gallego}}\ and\ \bibinfo {author} {\bibfnamefont {M.~A.}\
  \bibnamefont {Sierra}},\ }\href {\doibase 10.1021/cr100436k} {\bibfield
  {journal} {\bibinfo  {journal} {Chem. Rev.}\ }\textbf {\bibinfo {volume}
  {111}},\ \bibinfo {pages} {4857} (\bibinfo {year} {2011})}\BibitemShut
  {NoStop}%
\bibitem [{\citenamefont {Chatzidimitriou-Dreismann}\ \emph
  {et~al.}(1997)\citenamefont {Chatzidimitriou-Dreismann}, \citenamefont
  {Abdul~Redah}, \citenamefont {Streffer},\ and\ \citenamefont
  {Mayers}}]{DINS_1}%
  \BibitemOpen
  \bibfield  {author} {\bibinfo {author} {\bibfnamefont {C.~A.}\ \bibnamefont
  {Chatzidimitriou-Dreismann}}, \bibinfo {author} {\bibfnamefont
  {T.}~\bibnamefont {Abdul~Redah}}, \bibinfo {author} {\bibfnamefont
  {R.~M.~F.}\ \bibnamefont {Streffer}}, \ and\ \bibinfo {author} {\bibfnamefont
  {J.}~\bibnamefont {Mayers}},\ }\href {\doibase 10.1103/PhysRevLett.79.2839}
  {\bibfield  {journal} {\bibinfo  {journal} {Phys. Rev. Lett.}\ }\textbf
  {\bibinfo {volume} {79}},\ \bibinfo {pages} {2839} (\bibinfo {year}
  {1997})}\BibitemShut {NoStop}%
\bibitem [{\citenamefont {Romanelli}\ \emph {et~al.}(2013)\citenamefont
  {Romanelli}, \citenamefont {Ceriotti}, \citenamefont {Manolopoulos},
  \citenamefont {Pantalei}, \citenamefont {Senesi},\ and\ \citenamefont
  {Andreani}}]{Romanelli_2013}%
  \BibitemOpen
  \bibfield  {author} {\bibinfo {author} {\bibfnamefont {G.}~\bibnamefont
  {Romanelli}}, \bibinfo {author} {\bibfnamefont {M.}~\bibnamefont {Ceriotti}},
  \bibinfo {author} {\bibfnamefont {D.~E.}\ \bibnamefont {Manolopoulos}},
  \bibinfo {author} {\bibfnamefont {C.}~\bibnamefont {Pantalei}}, \bibinfo
  {author} {\bibfnamefont {R.}~\bibnamefont {Senesi}}, \ and\ \bibinfo {author}
  {\bibfnamefont {C.}~\bibnamefont {Andreani}},\ }\href {\doibase
  10.1021/jz401538r} {\bibfield  {journal} {\bibinfo  {journal} {J. Chem. Phys.
  Lett.}\ }\textbf {\bibinfo {volume} {4}},\ \bibinfo {pages} {3251} (\bibinfo
  {year} {2013})}\BibitemShut {NoStop}%
\bibitem [{\citenamefont {Parmentier}\ \emph {et~al.}(2015)\citenamefont
  {Parmentier}, \citenamefont {Shephard}, \citenamefont {Romanelli},
  \citenamefont {Senesi}, \citenamefont {Salzmann},\ and\ \citenamefont
  {Andreani}}]{DINS_2}%
  \BibitemOpen
  \bibfield  {author} {\bibinfo {author} {\bibfnamefont {A.}~\bibnamefont
  {Parmentier}}, \bibinfo {author} {\bibfnamefont {J.~J.}\ \bibnamefont
  {Shephard}}, \bibinfo {author} {\bibfnamefont {G.}~\bibnamefont {Romanelli}},
  \bibinfo {author} {\bibfnamefont {R.}~\bibnamefont {Senesi}}, \bibinfo
  {author} {\bibfnamefont {C.~G.}\ \bibnamefont {Salzmann}}, \ and\ \bibinfo
  {author} {\bibfnamefont {C.}~\bibnamefont {Andreani}},\ }\href {\doibase
  10.1021/acs.jpclett.5b00711} {\bibfield  {journal} {\bibinfo  {journal} {J.
  Phys. Chem. Lett.}\ }\textbf {\bibinfo {volume} {6}},\ \bibinfo {pages}
  {2038} (\bibinfo {year} {2015})}\BibitemShut {NoStop}%
\bibitem [{\citenamefont {Andreani}\ \emph {et~al.}(2016)\citenamefont
  {Andreani}, \citenamefont {Romanelli},\ and\ \citenamefont
  {Senesi}}]{DINS_3}%
  \BibitemOpen
  \bibfield  {author} {\bibinfo {author} {\bibfnamefont {C.}~\bibnamefont
  {Andreani}}, \bibinfo {author} {\bibfnamefont {G.}~\bibnamefont {Romanelli}},
  \ and\ \bibinfo {author} {\bibfnamefont {R.}~\bibnamefont {Senesi}},\ }\href
  {\doibase 10.1021/acs.jpclett.6b00926} {\bibfield  {journal} {\bibinfo
  {journal} {J. Phys. Chem. Lett.}\ }\textbf {\bibinfo {volume} {7}},\ \bibinfo
  {pages} {2216} (\bibinfo {year} {2016})}\BibitemShut {NoStop}%
\bibitem [{\citenamefont {Binnig}\ and\ \citenamefont
  {Rohrer}(1986)}]{STM_1986}%
  \BibitemOpen
  \bibfield  {author} {\bibinfo {author} {\bibfnamefont {G.}~\bibnamefont
  {Binnig}}\ and\ \bibinfo {author} {\bibfnamefont {H.}~\bibnamefont
  {Rohrer}},\ }\href@noop {} {\bibfield  {journal} {\bibinfo  {journal} {IBM J.
  Res. Dev.}\ }\textbf {\bibinfo {volume} {30}},\ \bibinfo {pages} {355}
  (\bibinfo {year} {1986})}\BibitemShut {NoStop}%
\bibitem [{\citenamefont {Lauhon}\ and\ \citenamefont
  {Ho}(2000)}]{lauhon_direct_2000}%
  \BibitemOpen
  \bibfield  {author} {\bibinfo {author} {\bibfnamefont {L.~J.}\ \bibnamefont
  {Lauhon}}\ and\ \bibinfo {author} {\bibfnamefont {W.}~\bibnamefont {Ho}},\
  }\href {\doibase 10.1103/PhysRevLett.85.4566} {\bibfield  {journal} {\bibinfo
   {journal} {Phys. Rev. Lett.}\ }\textbf {\bibinfo {volume} {85}},\ \bibinfo
  {pages} {4566} (\bibinfo {year} {2000})}\BibitemShut {NoStop}%
\bibitem [{\citenamefont {Kumagai}\ \emph {et~al.}(2010)\citenamefont
  {Kumagai}, \citenamefont {Kaizu}, \citenamefont {Okuyama}, \citenamefont
  {Hatta}, \citenamefont {Aruga}, \citenamefont {Hamada},\ and\ \citenamefont
  {Morikawa}}]{PhysRevB.81.045402}%
  \BibitemOpen
  \bibfield  {author} {\bibinfo {author} {\bibfnamefont {T.}~\bibnamefont
  {Kumagai}}, \bibinfo {author} {\bibfnamefont {M.}~\bibnamefont {Kaizu}},
  \bibinfo {author} {\bibfnamefont {H.}~\bibnamefont {Okuyama}}, \bibinfo
  {author} {\bibfnamefont {S.}~\bibnamefont {Hatta}}, \bibinfo {author}
  {\bibfnamefont {T.}~\bibnamefont {Aruga}}, \bibinfo {author} {\bibfnamefont
  {I.}~\bibnamefont {Hamada}}, \ and\ \bibinfo {author} {\bibfnamefont
  {Y.}~\bibnamefont {Morikawa}},\ }\href {\doibase 10.1103/PhysRevB.81.045402}
  {\bibfield  {journal} {\bibinfo  {journal} {Phys. Rev. B}\ }\textbf {\bibinfo
  {volume} {81}},\ \bibinfo {pages} {045402} (\bibinfo {year}
  {2010})}\BibitemShut {NoStop}%
\bibitem [{\citenamefont {Jewell}\ \emph {et~al.}(2012)\citenamefont {Jewell},
  \citenamefont {Peng}, \citenamefont {Mattera}, \citenamefont {Lewis},
  \citenamefont {Murphy}, \citenamefont {Kyriakou}, \citenamefont
  {Mavrikakis},\ and\ \citenamefont {Sykes}}]{Sykes_Quantum_2012}%
  \BibitemOpen
  \bibfield  {author} {\bibinfo {author} {\bibfnamefont {A.~D.}\ \bibnamefont
  {Jewell}}, \bibinfo {author} {\bibfnamefont {G.}~\bibnamefont {Peng}},
  \bibinfo {author} {\bibfnamefont {M.~F.~G.}\ \bibnamefont {Mattera}},
  \bibinfo {author} {\bibfnamefont {E.~A.}\ \bibnamefont {Lewis}}, \bibinfo
  {author} {\bibfnamefont {C.~J.}\ \bibnamefont {Murphy}}, \bibinfo {author}
  {\bibfnamefont {G.}~\bibnamefont {Kyriakou}}, \bibinfo {author}
  {\bibfnamefont {M.}~\bibnamefont {Mavrikakis}}, \ and\ \bibinfo {author}
  {\bibfnamefont {E.~C.~H.}\ \bibnamefont {Sykes}},\ }\href {\doibase
  10.1021/nn3038463} {\bibfield  {journal} {\bibinfo  {journal} {ACS Nano}\
  }\textbf {\bibinfo {volume} {6}},\ \bibinfo {pages} {10115} (\bibinfo {year}
  {2012})}\BibitemShut {NoStop}%
\bibitem [{\citenamefont {Kyriakou}\ \emph {et~al.}(2014)\citenamefont
  {Kyriakou}, \citenamefont {Davidson}, \citenamefont {Peng}, \citenamefont
  {Roling}, \citenamefont {Singh}, \citenamefont {Boucher}, \citenamefont
  {Marcinkowski}, \citenamefont {Mavrikakis}, \citenamefont {Michaelides},\
  and\ \citenamefont {Sykes}}]{Davidson_2014_2}%
  \BibitemOpen
  \bibfield  {author} {\bibinfo {author} {\bibfnamefont {G.}~\bibnamefont
  {Kyriakou}}, \bibinfo {author} {\bibfnamefont {E.~R.~M.}\ \bibnamefont
  {Davidson}}, \bibinfo {author} {\bibfnamefont {G.}~\bibnamefont {Peng}},
  \bibinfo {author} {\bibfnamefont {L.~T.}\ \bibnamefont {Roling}}, \bibinfo
  {author} {\bibfnamefont {S.}~\bibnamefont {Singh}}, \bibinfo {author}
  {\bibfnamefont {M.~B.}\ \bibnamefont {Boucher}}, \bibinfo {author}
  {\bibfnamefont {M.~D.}\ \bibnamefont {Marcinkowski}}, \bibinfo {author}
  {\bibfnamefont {M.}~\bibnamefont {Mavrikakis}}, \bibinfo {author}
  {\bibfnamefont {A.}~\bibnamefont {Michaelides}}, \ and\ \bibinfo {author}
  {\bibfnamefont {E.~C.~H.}\ \bibnamefont {Sykes}},\ }\href {\doibase
  10.1021/nn500703k} {\bibfield  {journal} {\bibinfo  {journal} {ACS Nano}\
  }\textbf {\bibinfo {volume} {8}},\ \bibinfo {pages} {4827} (\bibinfo {year}
  {2014})}\BibitemShut {NoStop}%
\bibitem [{\citenamefont {Kumagai}\ \emph {et~al.}(2008)\citenamefont
  {Kumagai}, \citenamefont {Kaizu}, \citenamefont {Hatta}, \citenamefont
  {Okuyama}, \citenamefont {Aruga}, \citenamefont {Hamada},\ and\ \citenamefont
  {Morikawa}}]{PhysRevLett.100.166101}%
  \BibitemOpen
  \bibfield  {author} {\bibinfo {author} {\bibfnamefont {T.}~\bibnamefont
  {Kumagai}}, \bibinfo {author} {\bibfnamefont {M.}~\bibnamefont {Kaizu}},
  \bibinfo {author} {\bibfnamefont {S.}~\bibnamefont {Hatta}}, \bibinfo
  {author} {\bibfnamefont {H.}~\bibnamefont {Okuyama}}, \bibinfo {author}
  {\bibfnamefont {T.}~\bibnamefont {Aruga}}, \bibinfo {author} {\bibfnamefont
  {I.}~\bibnamefont {Hamada}}, \ and\ \bibinfo {author} {\bibfnamefont
  {Y.}~\bibnamefont {Morikawa}},\ }\href {\doibase
  10.1103/PhysRevLett.100.166101} {\bibfield  {journal} {\bibinfo  {journal}
  {Phys. Rev. Lett.}\ }\textbf {\bibinfo {volume} {100}},\ \bibinfo {pages}
  {166101} (\bibinfo {year} {2008})}\BibitemShut {NoStop}%
\bibitem [{\citenamefont {Meng}\ \emph {et~al.}(2015)\citenamefont {Meng},
  \citenamefont {Guo}, \citenamefont {Peng}, \citenamefont {Chen},
  \citenamefont {Wang}, \citenamefont {Shi}, \citenamefont {Li}, \citenamefont
  {Wang},\ and\ \citenamefont {Jiang}}]{tetramer_2015}%
  \BibitemOpen
  \bibfield  {author} {\bibinfo {author} {\bibfnamefont {X.}~\bibnamefont
  {Meng}}, \bibinfo {author} {\bibfnamefont {J.}~\bibnamefont {Guo}}, \bibinfo
  {author} {\bibfnamefont {J.}~\bibnamefont {Peng}}, \bibinfo {author}
  {\bibfnamefont {J.}~\bibnamefont {Chen}}, \bibinfo {author} {\bibfnamefont
  {Z.}~\bibnamefont {Wang}}, \bibinfo {author} {\bibfnamefont {J.-R.}\
  \bibnamefont {Shi}}, \bibinfo {author} {\bibfnamefont {X.-Z.}\ \bibnamefont
  {Li}}, \bibinfo {author} {\bibfnamefont {E.-G.}\ \bibnamefont {Wang}}, \ and\
  \bibinfo {author} {\bibfnamefont {Y.}~\bibnamefont {Jiang}},\ }\href
  {\doibase 10.1038/nphys3225} {\bibfield  {journal} {\bibinfo  {journal} {Nat.
  Phys.}\ }\textbf {\bibinfo {volume} {11}},\ \bibinfo {pages} {235} (\bibinfo
  {year} {2015})}\BibitemShut {NoStop}%
\bibitem [{\citenamefont {Jardine}\ \emph {et~al.}(2009)\citenamefont
  {Jardine}, \citenamefont {Hedgeland}, \citenamefont {Alexandrowicz},
  \citenamefont {Allison},\ and\ \citenamefont {Ellis}}]{Jardine_HeSE_2009}%
  \BibitemOpen
  \bibfield  {author} {\bibinfo {author} {\bibfnamefont {A.}~\bibnamefont
  {Jardine}}, \bibinfo {author} {\bibfnamefont {H.}~\bibnamefont {Hedgeland}},
  \bibinfo {author} {\bibfnamefont {G.}~\bibnamefont {Alexandrowicz}}, \bibinfo
  {author} {\bibfnamefont {W.}~\bibnamefont {Allison}}, \ and\ \bibinfo
  {author} {\bibfnamefont {J.}~\bibnamefont {Ellis}},\ }\href {\doibase
  http://dx.doi.org/10.1016/j.progsurf.2009.07.001} {\bibfield  {journal}
  {\bibinfo  {journal} {Prog. Surf. Sci.}\ }\textbf {\bibinfo {volume} {84}},\
  \bibinfo {pages} {323} (\bibinfo {year} {2009})}\BibitemShut {NoStop}%
\bibitem [{\citenamefont {Jardine}\ \emph {et~al.}(2010)\citenamefont
  {Jardine}, \citenamefont {Lee}, \citenamefont {Ward}, \citenamefont
  {Alexandrowicz}, \citenamefont {Hedgeland}, \citenamefont {Allison},
  \citenamefont {Ellis},\ and\ \citenamefont
  {Pollak}}]{jardine_determination_2010}%
  \BibitemOpen
  \bibfield  {author} {\bibinfo {author} {\bibfnamefont {A.~P.}\ \bibnamefont
  {Jardine}}, \bibinfo {author} {\bibfnamefont {E.~Y.~M.}\ \bibnamefont {Lee}},
  \bibinfo {author} {\bibfnamefont {D.~J.}\ \bibnamefont {Ward}}, \bibinfo
  {author} {\bibfnamefont {G.}~\bibnamefont {Alexandrowicz}}, \bibinfo {author}
  {\bibfnamefont {H.}~\bibnamefont {Hedgeland}}, \bibinfo {author}
  {\bibfnamefont {W.}~\bibnamefont {Allison}}, \bibinfo {author} {\bibfnamefont
  {J.}~\bibnamefont {Ellis}}, \ and\ \bibinfo {author} {\bibfnamefont
  {E.}~\bibnamefont {Pollak}},\ }\href {\doibase
  10.1103/PhysRevLett.105.136101} {\bibfield  {journal} {\bibinfo  {journal}
  {Phys. Rev. Lett.}\ }\textbf {\bibinfo {volume} {105}},\ \bibinfo {pages}
  {136101} (\bibinfo {year} {2010})}\BibitemShut {NoStop}%
\bibitem [{\citenamefont {Wang}\ \emph {et~al.}(1998)\citenamefont {Wang},
  \citenamefont {Sun},\ and\ \citenamefont {Miller}}]{LSCIVR}%
  \BibitemOpen
  \bibfield  {author} {\bibinfo {author} {\bibfnamefont {H.}~\bibnamefont
  {Wang}}, \bibinfo {author} {\bibfnamefont {X.}~\bibnamefont {Sun}}, \ and\
  \bibinfo {author} {\bibfnamefont {W.~H.}\ \bibnamefont {Miller}},\
  }\href@noop {} {\bibfield  {journal} {\bibinfo  {journal} {J. Chem. Phys.}\
  }\textbf {\bibinfo {volume} {108}},\ \bibinfo {pages} {9726} (\bibinfo {year}
  {1998})}\BibitemShut {NoStop}%
\bibitem [{\citenamefont {Beck}\ \emph {et~al.}(2000)\citenamefont {Beck},
  \citenamefont {Jäckle}, \citenamefont {Worth},\ and\ \citenamefont
  {Meyer}}]{MCTDH}%
  \BibitemOpen
  \bibfield  {author} {\bibinfo {author} {\bibfnamefont {M.}~\bibnamefont
  {Beck}}, \bibinfo {author} {\bibfnamefont {A.}~\bibnamefont {Jäckle}},
  \bibinfo {author} {\bibfnamefont {G.}~\bibnamefont {Worth}}, \ and\ \bibinfo
  {author} {\bibfnamefont {H.-D.}\ \bibnamefont {Meyer}},\ }\href {\doibase
  https://doi.org/10.1016/S0370-1573(99)00047-2} {\bibfield  {journal}
  {\bibinfo  {journal} {Phys. Rep.}\ }\textbf {\bibinfo {volume} {324}},\
  \bibinfo {pages} {1 } (\bibinfo {year} {2000})}\BibitemShut {NoStop}%
\bibitem [{\citenamefont {Marx}\ and\ \citenamefont
  {Parrinello}(1994)}]{Marx-Parr_1994}%
  \BibitemOpen
  \bibfield  {author} {\bibinfo {author} {\bibfnamefont {D.}~\bibnamefont
  {Marx}}\ and\ \bibinfo {author} {\bibfnamefont {M.}~\bibnamefont
  {Parrinello}},\ }\href {\doibase 10.1007/BF01312185} {\bibfield  {journal}
  {\bibinfo  {journal} {Z. Phys. B, Condens. Matter}\ }\textbf {\bibinfo
  {volume} {95}},\ \bibinfo {pages} {143} (\bibinfo {year} {1994})}\BibitemShut
  {NoStop}%
\bibitem [{\citenamefont {Feynman}\ and\ \citenamefont
  {Hibbs}(1965)}]{Feynman}%
  \BibitemOpen
  \bibfield  {author} {\bibinfo {author} {\bibfnamefont {R.}~\bibnamefont
  {Feynman}}\ and\ \bibinfo {author} {\bibfnamefont {A.}~\bibnamefont
  {Hibbs}},\ }\href {https://books.google.co.uk/books?id=14ApAQAAMAAJ} {\emph
  {\bibinfo {title} {Quantum mechanics and path integrals}}},\ International
  series in pure and applied physics\ (\bibinfo  {publisher} {McGraw-Hill, New
  York, USA},\ \bibinfo {year} {1965})\BibitemShut {NoStop}%
\bibitem [{\citenamefont {Tuckerman}(2010)}]{Tuckerman_book}%
  \BibitemOpen
  \bibfield  {author} {\bibinfo {author} {\bibfnamefont {M.~E.}\ \bibnamefont
  {Tuckerman}},\ }\href@noop {} {\emph {\bibinfo {title} {Statistical
  Mechanics: Theroy and Molecualar Simulation}}}\ (\bibinfo  {publisher}
  {Oxford University Press, Oxford, UK},\ \bibinfo {year} {2010})\BibitemShut
  {NoStop}%
\bibitem [{\citenamefont {Tuckerman}\ \emph {et~al.}(1996)\citenamefont
  {Tuckerman}, \citenamefont {Marx}, \citenamefont {Klein},\ and\ \citenamefont
  {Parrinello}}]{Tuckerman-Marx_1996}%
  \BibitemOpen
  \bibfield  {author} {\bibinfo {author} {\bibfnamefont {M.~E.}\ \bibnamefont
  {Tuckerman}}, \bibinfo {author} {\bibfnamefont {D.}~\bibnamefont {Marx}},
  \bibinfo {author} {\bibfnamefont {M.~L.}\ \bibnamefont {Klein}}, \ and\
  \bibinfo {author} {\bibfnamefont {M.}~\bibnamefont {Parrinello}},\ }\href
  {\doibase 10.1063/1.471771} {\bibfield  {journal} {\bibinfo  {journal} {J.
  Chem. Phys.}\ }\textbf {\bibinfo {volume} {104}},\ \bibinfo {pages} {5579}
  (\bibinfo {year} {1996})}\BibitemShut {NoStop}%
\bibitem [{\citenamefont {Chandler}\ and\ \citenamefont
  {Wolynes}(1981)}]{doi:10.1063/1.441588}%
  \BibitemOpen
  \bibfield  {author} {\bibinfo {author} {\bibfnamefont {D.}~\bibnamefont
  {Chandler}}\ and\ \bibinfo {author} {\bibfnamefont {P.~G.}\ \bibnamefont
  {Wolynes}},\ }\href {\doibase 10.1063/1.441588} {\bibfield  {journal}
  {\bibinfo  {journal} {J. Chem. Phys.}\ }\textbf {\bibinfo {volume} {74}},\
  \bibinfo {pages} {4078} (\bibinfo {year} {1981})}\BibitemShut {NoStop}%
\bibitem [{\citenamefont {Herman}\ \emph {et~al.}(1982)\citenamefont {Herman},
  \citenamefont {Bruskin},\ and\ \citenamefont {Berne}}]{Berne_1982}%
  \BibitemOpen
  \bibfield  {author} {\bibinfo {author} {\bibfnamefont {M.~F.}\ \bibnamefont
  {Herman}}, \bibinfo {author} {\bibfnamefont {E.~J.}\ \bibnamefont {Bruskin}},
  \ and\ \bibinfo {author} {\bibfnamefont {B.~J.}\ \bibnamefont {Berne}},\
  }\href {\doibase 10.1063/1.442815} {\bibfield  {journal} {\bibinfo  {journal}
  {J. Chem. Phys.}\ }\textbf {\bibinfo {volume} {76}},\ \bibinfo {pages} {5150}
  (\bibinfo {year} {1982})}\BibitemShut {NoStop}%
\bibitem [{\citenamefont {Pollock}\ and\ \citenamefont
  {Ceperley}(1984)}]{PhysRevB.30.2555}%
  \BibitemOpen
  \bibfield  {author} {\bibinfo {author} {\bibfnamefont {E.~L.}\ \bibnamefont
  {Pollock}}\ and\ \bibinfo {author} {\bibfnamefont {D.~M.}\ \bibnamefont
  {Ceperley}},\ }\href {\doibase 10.1103/PhysRevB.30.2555} {\bibfield
  {journal} {\bibinfo  {journal} {Phys. Rev. B}\ }\textbf {\bibinfo {volume}
  {30}},\ \bibinfo {pages} {2555} (\bibinfo {year} {1984})}\BibitemShut
  {NoStop}%
\bibitem [{\citenamefont {Cao}\ and\ \citenamefont
  {Voth}(1994)}]{Cao_Voth_1994}%
  \BibitemOpen
  \bibfield  {author} {\bibinfo {author} {\bibfnamefont {J.}~\bibnamefont
  {Cao}}\ and\ \bibinfo {author} {\bibfnamefont {G.~A.}\ \bibnamefont {Voth}},\
  }\href {\doibase 10.1063/1.467175} {\bibfield  {journal} {\bibinfo  {journal}
  {J. Chem. Phys.}\ }\textbf {\bibinfo {volume} {100}},\ \bibinfo {pages}
  {5093} (\bibinfo {year} {1994})}\BibitemShut {NoStop}%
\bibitem [{\citenamefont {Lin}\ \emph {et~al.}(2010)\citenamefont {Lin},
  \citenamefont {Morrone}, \citenamefont {Car},\ and\ \citenamefont
  {Parrinello}}]{PhysRevLett.105.110602}%
  \BibitemOpen
  \bibfield  {author} {\bibinfo {author} {\bibfnamefont {L.}~\bibnamefont
  {Lin}}, \bibinfo {author} {\bibfnamefont {J.~A.}\ \bibnamefont {Morrone}},
  \bibinfo {author} {\bibfnamefont {R.}~\bibnamefont {Car}}, \ and\ \bibinfo
  {author} {\bibfnamefont {M.}~\bibnamefont {Parrinello}},\ }\href {\doibase
  10.1103/PhysRevLett.105.110602} {\bibfield  {journal} {\bibinfo  {journal}
  {Phys. Rev. Lett.}\ }\textbf {\bibinfo {volume} {105}},\ \bibinfo {pages}
  {110602} (\bibinfo {year} {2010})}\BibitemShut {NoStop}%
\bibitem [{\citenamefont {Ceriotti}\ \emph {et~al.}(2014)\citenamefont
  {Ceriotti}, \citenamefont {More},\ and\ \citenamefont {Manolopoulos}}]{ipi}%
  \BibitemOpen
  \bibfield  {author} {\bibinfo {author} {\bibfnamefont {M.}~\bibnamefont
  {Ceriotti}}, \bibinfo {author} {\bibfnamefont {J.}~\bibnamefont {More}}, \
  and\ \bibinfo {author} {\bibfnamefont {D.~E.}\ \bibnamefont {Manolopoulos}},\
  }\href {\doibase http://dx.doi.org/10.1016/j.cpc.2013.10.027} {\bibfield
  {journal} {\bibinfo  {journal} {Comput. Phys. Commun.}\ }\textbf {\bibinfo
  {volume} {185}},\ \bibinfo {pages} {1019} (\bibinfo {year}
  {2014})}\BibitemShut {NoStop}%
\bibitem [{\citenamefont {Herrero}\ and\ \citenamefont
  {Ram\'irez}(2014)}]{PI_Ramirez}%
  \BibitemOpen
  \bibfield  {author} {\bibinfo {author} {\bibfnamefont {C.~P.}\ \bibnamefont
  {Herrero}}\ and\ \bibinfo {author} {\bibfnamefont {R.}~\bibnamefont
  {Ram\'irez}},\ }\href {http://stacks.iop.org/0953-8984/26/i=23/a=233201}
  {\bibfield  {journal} {\bibinfo  {journal} {J. Phys.: Condens. Matter}\
  }\textbf {\bibinfo {volume} {26}},\ \bibinfo {pages} {233201} (\bibinfo
  {year} {2014})}\BibitemShut {NoStop}%
\bibitem [{\citenamefont {Ceriotti}\ \emph {et~al.}(2010)\citenamefont
  {Ceriotti}, \citenamefont {Parrinello}, \citenamefont {Markland},\ and\
  \citenamefont {Manolopoulos}}]{PILE}%
  \BibitemOpen
  \bibfield  {author} {\bibinfo {author} {\bibfnamefont {M.}~\bibnamefont
  {Ceriotti}}, \bibinfo {author} {\bibfnamefont {M.}~\bibnamefont
  {Parrinello}}, \bibinfo {author} {\bibfnamefont {T.~E.}\ \bibnamefont
  {Markland}}, \ and\ \bibinfo {author} {\bibfnamefont {D.~E.}\ \bibnamefont
  {Manolopoulos}},\ }\href {\doibase 10.1063/1.3489925} {\bibfield  {journal}
  {\bibinfo  {journal} {J. Chem. Phys.}\ }\textbf {\bibinfo {volume} {133}},\
  \bibinfo {pages} {124104} (\bibinfo {year} {2010})}\BibitemShut {NoStop}%
\bibitem [{\citenamefont {Ceriotti}\ and\ \citenamefont
  {Markland}(2013)}]{Mass_3}%
  \BibitemOpen
  \bibfield  {author} {\bibinfo {author} {\bibfnamefont {M.}~\bibnamefont
  {Ceriotti}}\ and\ \bibinfo {author} {\bibfnamefont {T.~E.}\ \bibnamefont
  {Markland}},\ }\href {\doibase http://dx.doi.org/10.1063/1.4772676}
  {\bibfield  {journal} {\bibinfo  {journal} {J. Chem. Phys.}\ }\textbf
  {\bibinfo {volume} {138}},\ \bibinfo {pages} {014112} (\bibinfo {year}
  {2013})}\BibitemShut {NoStop}%
\bibitem [{\citenamefont {Habershon}\ \emph {et~al.}(2013)\citenamefont
  {Habershon}, \citenamefont {Manolopoulos}, \citenamefont {Markland},\ and\
  \citenamefont {Miller}}]{habershon_ring-polymer_2013}%
  \BibitemOpen
  \bibfield  {author} {\bibinfo {author} {\bibfnamefont {S.}~\bibnamefont
  {Habershon}}, \bibinfo {author} {\bibfnamefont {D.~E.}\ \bibnamefont
  {Manolopoulos}}, \bibinfo {author} {\bibfnamefont {T.~E.}\ \bibnamefont
  {Markland}}, \ and\ \bibinfo {author} {\bibfnamefont {T.~F.}\ \bibnamefont
  {Miller}},\ }\href {\doibase 10.1146/annurev-physchem-040412-110122}
  {\bibfield  {journal} {\bibinfo  {journal} {Annu. Rev. Phys. Chem.}\ }\textbf
  {\bibinfo {volume} {64}},\ \bibinfo {pages} {387} (\bibinfo {year}
  {2013})}\BibitemShut {NoStop}%
\bibitem [{\citenamefont {Markland}\ and\ \citenamefont
  {Ceriotti}(2018)}]{markland_nuclear_2018}%
  \BibitemOpen
  \bibfield  {author} {\bibinfo {author} {\bibfnamefont {T.~E.}\ \bibnamefont
  {Markland}}\ and\ \bibinfo {author} {\bibfnamefont {M.}~\bibnamefont
  {Ceriotti}},\ }\href {http://dx.doi.org/10.1038/s41570-017-0109} {\bibfield
  {journal} {\bibinfo  {journal} {Nat. Rev. Chem.}\ }\textbf {\bibinfo {volume}
  {2}},\ \bibinfo {pages} {0109} (\bibinfo {year} {2018})}\BibitemShut
  {NoStop}%
\bibitem [{\citenamefont {Parrinello}\ and\ \citenamefont
  {Rahman}(1984)}]{doi:10.1063/1.446740}%
  \BibitemOpen
  \bibfield  {author} {\bibinfo {author} {\bibfnamefont {M.}~\bibnamefont
  {Parrinello}}\ and\ \bibinfo {author} {\bibfnamefont {A.}~\bibnamefont
  {Rahman}},\ }\href {\doibase 10.1063/1.446740} {\bibfield  {journal}
  {\bibinfo  {journal} {J. Chem. Phys.}\ }\textbf {\bibinfo {volume} {80}},\
  \bibinfo {pages} {860} (\bibinfo {year} {1984})}\BibitemShut {NoStop}%
\bibitem [{\citenamefont {Gillan}(1987)}]{Gillan_H_qtst}%
  \BibitemOpen
  \bibfield  {author} {\bibinfo {author} {\bibfnamefont {M.~J.}\ \bibnamefont
  {Gillan}},\ }\href {\doibase 10.1103/PhysRevLett.58.563} {\bibfield
  {journal} {\bibinfo  {journal} {Phys. Rev. Lett.}\ }\textbf {\bibinfo
  {volume} {58}},\ \bibinfo {pages} {563} (\bibinfo {year} {1987})}\BibitemShut
  {NoStop}%
\bibitem [{\citenamefont {Sprik}\ \emph {et~al.}(1985)\citenamefont {Sprik},
  \citenamefont {Klein},\ and\ \citenamefont {Chandler}}]{PhysRevB.31.4234}%
  \BibitemOpen
  \bibfield  {author} {\bibinfo {author} {\bibfnamefont {M.}~\bibnamefont
  {Sprik}}, \bibinfo {author} {\bibfnamefont {M.~L.}\ \bibnamefont {Klein}}, \
  and\ \bibinfo {author} {\bibfnamefont {D.}~\bibnamefont {Chandler}},\ }\href
  {\doibase 10.1103/PhysRevB.31.4234} {\bibfield  {journal} {\bibinfo
  {journal} {Phys. Rev. B}\ }\textbf {\bibinfo {volume} {31}},\ \bibinfo
  {pages} {4234} (\bibinfo {year} {1985})}\BibitemShut {NoStop}%
\bibitem [{\citenamefont {Ceperley}\ and\ \citenamefont
  {Jacucci}(1987)}]{PhysRevLett.58.1648}%
  \BibitemOpen
  \bibfield  {author} {\bibinfo {author} {\bibfnamefont {D.~M.}\ \bibnamefont
  {Ceperley}}\ and\ \bibinfo {author} {\bibfnamefont {G.}~\bibnamefont
  {Jacucci}},\ }\href {\doibase 10.1103/PhysRevLett.58.1648} {\bibfield
  {journal} {\bibinfo  {journal} {Phys. Rev. Lett.}\ }\textbf {\bibinfo
  {volume} {58}},\ \bibinfo {pages} {1648} (\bibinfo {year}
  {1987})}\BibitemShut {NoStop}%
\bibitem [{\citenamefont {Benoit}\ \emph {et~al.}(1998)\citenamefont {Benoit},
  \citenamefont {Marx},\ and\ \citenamefont {Parrinello}}]{Marx1998}%
  \BibitemOpen
  \bibfield  {author} {\bibinfo {author} {\bibfnamefont {M.}~\bibnamefont
  {Benoit}}, \bibinfo {author} {\bibfnamefont {D.}~\bibnamefont {Marx}}, \ and\
  \bibinfo {author} {\bibfnamefont {M.}~\bibnamefont {Parrinello}},\ }\href
  {http://dx.doi.org/10.1038/32609} {\bibfield  {journal} {\bibinfo  {journal}
  {Nature}\ }\textbf {\bibinfo {volume} {392}},\ \bibinfo {pages} {258}
  (\bibinfo {year} {1998})}\BibitemShut {NoStop}%
\bibitem [{\citenamefont {Tuckerman}\ \emph {et~al.}(1997)\citenamefont
  {Tuckerman}, \citenamefont {Marx}, \citenamefont {Klein},\ and\ \citenamefont
  {Parrinello}}]{Tuckerman817}%
  \BibitemOpen
  \bibfield  {author} {\bibinfo {author} {\bibfnamefont {M.~E.}\ \bibnamefont
  {Tuckerman}}, \bibinfo {author} {\bibfnamefont {D.}~\bibnamefont {Marx}},
  \bibinfo {author} {\bibfnamefont {M.~L.}\ \bibnamefont {Klein}}, \ and\
  \bibinfo {author} {\bibfnamefont {M.}~\bibnamefont {Parrinello}},\ }\href
  {\doibase 10.1126/science.275.5301.817} {\bibfield  {journal} {\bibinfo
  {journal} {Science}\ }\textbf {\bibinfo {volume} {275}},\ \bibinfo {pages}
  {817} (\bibinfo {year} {1997})}\BibitemShut {NoStop}%
\bibitem [{\citenamefont {Chen}\ \emph {et~al.}(2003)\citenamefont {Chen},
  \citenamefont {Ivanov}, \citenamefont {Klein},\ and\ \citenamefont
  {Parrinello}}]{Parrinello_2003}%
  \BibitemOpen
  \bibfield  {author} {\bibinfo {author} {\bibfnamefont {B.}~\bibnamefont
  {Chen}}, \bibinfo {author} {\bibfnamefont {I.}~\bibnamefont {Ivanov}},
  \bibinfo {author} {\bibfnamefont {M.~L.}\ \bibnamefont {Klein}}, \ and\
  \bibinfo {author} {\bibfnamefont {M.}~\bibnamefont {Parrinello}},\ }\href
  {\doibase 10.1103/PhysRevLett.91.215503} {\bibfield  {journal} {\bibinfo
  {journal} {Phys. Rev. Lett.}\ }\textbf {\bibinfo {volume} {91}},\ \bibinfo
  {pages} {215503} (\bibinfo {year} {2003})}\BibitemShut {NoStop}%
\bibitem [{\citenamefont {Jang}\ and\ \citenamefont {Voth}(1999)}]{CMD}%
  \BibitemOpen
  \bibfield  {author} {\bibinfo {author} {\bibfnamefont {S.}~\bibnamefont
  {Jang}}\ and\ \bibinfo {author} {\bibfnamefont {G.~A.}\ \bibnamefont
  {Voth}},\ }\href {\doibase 10.1063/1.479515} {\bibfield  {journal} {\bibinfo
  {journal} {J. Chem. Phys.}\ }\textbf {\bibinfo {volume} {111}},\ \bibinfo
  {pages} {2371} (\bibinfo {year} {1999})}\BibitemShut {NoStop}%
\bibitem [{\citenamefont {Voth}\ \emph {et~al.}(1989)\citenamefont {Voth},
  \citenamefont {Chandler},\ and\ \citenamefont {Miller}}]{voth_rigorous_1989}%
  \BibitemOpen
  \bibfield  {author} {\bibinfo {author} {\bibfnamefont {G.~A.}\ \bibnamefont
  {Voth}}, \bibinfo {author} {\bibfnamefont {D.}~\bibnamefont {Chandler}}, \
  and\ \bibinfo {author} {\bibfnamefont {W.~H.}\ \bibnamefont {Miller}},\
  }\href {\doibase 10.1063/1.457242} {\bibfield  {journal} {\bibinfo  {journal}
  {J. Chem. Phys.}\ }\textbf {\bibinfo {volume} {91}},\ \bibinfo {pages} {7749}
  (\bibinfo {year} {1989})}\BibitemShut {NoStop}%
\bibitem [{\citenamefont {Marx}\ \emph {et~al.}(1999)\citenamefont {Marx},
  \citenamefont {Tuckerman},\ and\ \citenamefont {Martyna}}]{MARX1999166}%
  \BibitemOpen
  \bibfield  {author} {\bibinfo {author} {\bibfnamefont {D.}~\bibnamefont
  {Marx}}, \bibinfo {author} {\bibfnamefont {M.~E.}\ \bibnamefont {Tuckerman}},
  \ and\ \bibinfo {author} {\bibfnamefont {G.~J.}\ \bibnamefont {Martyna}},\
  }\href {\doibase https://doi.org/10.1016/S0010-4655(99)00208-8} {\bibfield
  {journal} {\bibinfo  {journal} {Comput. Phys. Commun.}\ }\textbf {\bibinfo
  {volume} {118}},\ \bibinfo {pages} {166 } (\bibinfo {year}
  {1999})}\BibitemShut {NoStop}%
\bibitem [{\citenamefont {Craig}\ and\ \citenamefont
  {Manolopoulos}(2005)}]{craig_refined_2005}%
  \BibitemOpen
  \bibfield  {author} {\bibinfo {author} {\bibfnamefont {I.~R.}\ \bibnamefont
  {Craig}}\ and\ \bibinfo {author} {\bibfnamefont {D.~E.}\ \bibnamefont
  {Manolopoulos}},\ }\href {\doibase 10.1063/1.1954769} {\bibfield  {journal}
  {\bibinfo  {journal} {J. Chem. Phys.}\ }\textbf {\bibinfo {volume} {123}},\
  \bibinfo {pages} {034102} (\bibinfo {year} {2005})}\BibitemShut {NoStop}%
\bibitem [{\citenamefont {Rossi}\ \emph {et~al.}(2014)\citenamefont {Rossi},
  \citenamefont {Ceriotti},\ and\ \citenamefont {Manolopoulos}}]{trpmd}%
  \BibitemOpen
  \bibfield  {author} {\bibinfo {author} {\bibfnamefont {M.}~\bibnamefont
  {Rossi}}, \bibinfo {author} {\bibfnamefont {M.}~\bibnamefont {Ceriotti}}, \
  and\ \bibinfo {author} {\bibfnamefont {D.~E.}\ \bibnamefont {Manolopoulos}},\
  }\href {\doibase 10.1063/1.4883861} {\bibfield  {journal} {\bibinfo
  {journal} {J. Chem. Phys.}\ }\textbf {\bibinfo {volume} {140}},\ \bibinfo
  {pages} {234116} (\bibinfo {year} {2014})}\BibitemShut {NoStop}%
\bibitem [{\citenamefont {Mouhat}\ \emph {et~al.}(2017)\citenamefont {Mouhat},
  \citenamefont {Sorella}, \citenamefont {Vuilleumier}, \citenamefont
  {Saitta},\ and\ \citenamefont {Casula}}]{PI_QMC_1}%
  \BibitemOpen
  \bibfield  {author} {\bibinfo {author} {\bibfnamefont {F.}~\bibnamefont
  {Mouhat}}, \bibinfo {author} {\bibfnamefont {S.}~\bibnamefont {Sorella}},
  \bibinfo {author} {\bibfnamefont {R.}~\bibnamefont {Vuilleumier}}, \bibinfo
  {author} {\bibfnamefont {A.~M.}\ \bibnamefont {Saitta}}, \ and\ \bibinfo
  {author} {\bibfnamefont {M.}~\bibnamefont {Casula}},\ }\href {\doibase
  10.1021/acs.jctc.7b00017} {\bibfield  {journal} {\bibinfo  {journal} {J.
  Chem. Theory Comput.}\ }\textbf {\bibinfo {volume} {13}},\ \bibinfo {pages}
  {2400} (\bibinfo {year} {2017})}\BibitemShut {NoStop}%
\bibitem [{\citenamefont {Kapil}\ \emph {et~al.}(2016)\citenamefont {Kapil},
  \citenamefont {VandeVondele},\ and\ \citenamefont
  {Ceriotti}}]{doi:10.1063/1.4941091}%
  \BibitemOpen
  \bibfield  {author} {\bibinfo {author} {\bibfnamefont {V.}~\bibnamefont
  {Kapil}}, \bibinfo {author} {\bibfnamefont {J.}~\bibnamefont {VandeVondele}},
  \ and\ \bibinfo {author} {\bibfnamefont {M.}~\bibnamefont {Ceriotti}},\
  }\href {\doibase 10.1063/1.4941091} {\bibfield  {journal} {\bibinfo
  {journal} {J. Chem. Phys.}\ }\textbf {\bibinfo {volume} {144}},\ \bibinfo
  {pages} {054111} (\bibinfo {year} {2016})}\BibitemShut {NoStop}%
\bibitem [{\citenamefont {Kawashima}\ \emph {et~al.}(2013)\citenamefont
  {Kawashima}, \citenamefont {Suzuki},\ and\ \citenamefont
  {Tachikawa}}]{Tachikawa_2013}%
  \BibitemOpen
  \bibfield  {author} {\bibinfo {author} {\bibfnamefont {Y.}~\bibnamefont
  {Kawashima}}, \bibinfo {author} {\bibfnamefont {K.}~\bibnamefont {Suzuki}}, \
  and\ \bibinfo {author} {\bibfnamefont {M.}~\bibnamefont {Tachikawa}},\ }\href
  {\doibase 10.1021/jp403295h} {\bibfield  {journal} {\bibinfo  {journal} {J.
  Phys. Chem. A}\ }\textbf {\bibinfo {volume} {117}},\ \bibinfo {pages} {5205}
  (\bibinfo {year} {2013})}\BibitemShut {NoStop}%
\bibitem [{\citenamefont {Spura}\ \emph {et~al.}(2015)\citenamefont {Spura},
  \citenamefont {Elgabarty},\ and\ \citenamefont {K\"{u}hne}}]{C4CP05192K}%
  \BibitemOpen
  \bibfield  {author} {\bibinfo {author} {\bibfnamefont {T.}~\bibnamefont
  {Spura}}, \bibinfo {author} {\bibfnamefont {H.}~\bibnamefont {Elgabarty}}, \
  and\ \bibinfo {author} {\bibfnamefont {T.~D.}\ \bibnamefont {K\"{u}hne}},\
  }\href {\doibase 10.1039/C4CP05192K} {\bibfield  {journal} {\bibinfo
  {journal} {Phys. Chem. Chem. Phys.}\ }\textbf {\bibinfo {volume} {17}},\
  \bibinfo {pages} {14355} (\bibinfo {year} {2015})}\BibitemShut {NoStop}%
\bibitem [{\citenamefont {Miller}(1975)}]{miller_semiclassical_1975}%
  \BibitemOpen
  \bibfield  {author} {\bibinfo {author} {\bibfnamefont {W.~H.}\ \bibnamefont
  {Miller}},\ }\href {\doibase 10.1063/1.430676} {\bibfield  {journal}
  {\bibinfo  {journal} {J. Chem. Phys.}\ }\textbf {\bibinfo {volume} {62}},\
  \bibinfo {pages} {1899} (\bibinfo {year} {1975})}\BibitemShut {NoStop}%
\bibitem [{\citenamefont {Richardson}\ and\ \citenamefont
  {Althorpe}(2009)}]{richardson_ring-polymer_2009}%
  \BibitemOpen
  \bibfield  {author} {\bibinfo {author} {\bibfnamefont {J.~O.}\ \bibnamefont
  {Richardson}}\ and\ \bibinfo {author} {\bibfnamefont {S.~C.}\ \bibnamefont
  {Althorpe}},\ }\href {\doibase 10.1063/1.3267318} {\bibfield  {journal}
  {\bibinfo  {journal} {J. Chem. Phys.}\ }\textbf {\bibinfo {volume} {131}},\
  \bibinfo {pages} {214106} (\bibinfo {year} {2009})}\BibitemShut {NoStop}%
\bibitem [{\citenamefont {Andersson}\ \emph {et~al.}(2009)\citenamefont
  {Andersson}, \citenamefont {Nyman}, \citenamefont {Arnaldsson}, \citenamefont
  {Manthe},\ and\ \citenamefont {J\'onsson}}]{Jonsson_2009}%
  \BibitemOpen
  \bibfield  {author} {\bibinfo {author} {\bibfnamefont {S.}~\bibnamefont
  {Andersson}}, \bibinfo {author} {\bibfnamefont {G.}~\bibnamefont {Nyman}},
  \bibinfo {author} {\bibfnamefont {A.}~\bibnamefont {Arnaldsson}}, \bibinfo
  {author} {\bibfnamefont {U.}~\bibnamefont {Manthe}}, \ and\ \bibinfo {author}
  {\bibfnamefont {H.}~\bibnamefont {J\'onsson}},\ }\href {\doibase
  10.1021/jp811070w} {\bibfield  {journal} {\bibinfo  {journal} {J. Phys. Chem.
  A}\ }\textbf {\bibinfo {volume} {113}},\ \bibinfo {pages} {4468} (\bibinfo
  {year} {2009})}\BibitemShut {NoStop}%
\bibitem [{\citenamefont {K\"{a}stner}(2014)}]{Kastner_2014}%
  \BibitemOpen
  \bibfield  {author} {\bibinfo {author} {\bibfnamefont {J.}~\bibnamefont
  {K\"{a}stner}},\ }\href {\doibase 10.1002/wcms.1165} {\bibfield  {journal}
  {\bibinfo  {journal} {WIREs: Comput. Mol. Sci.}\ }\textbf {\bibinfo {volume}
  {4}},\ \bibinfo {pages} {158} (\bibinfo {year} {2014})}\BibitemShut {NoStop}%
\bibitem [{\citenamefont {Richardson}(2018)}]{Inst_persp}%
  \BibitemOpen
  \bibfield  {author} {\bibinfo {author} {\bibfnamefont {J.~O.}\ \bibnamefont
  {Richardson}},\ }\href {\doibase 10.1063/1.5028352} {\bibfield  {journal}
  {\bibinfo  {journal} {J. Chem. Phys.}\ }\textbf {\bibinfo {volume} {148}},\
  \bibinfo {pages} {200901} (\bibinfo {year} {2018})}\BibitemShut {NoStop}%
\bibitem [{\citenamefont {Raugei}\ and\ \citenamefont
  {Klein}(2003)}]{Klein2003}%
  \BibitemOpen
  \bibfield  {author} {\bibinfo {author} {\bibfnamefont {S.}~\bibnamefont
  {Raugei}}\ and\ \bibinfo {author} {\bibfnamefont {M.~L.}\ \bibnamefont
  {Klein}},\ }\href {\doibase 10.1021/ja0351995} {\bibfield  {journal}
  {\bibinfo  {journal} {J. Am. Chem. Soc.}\ }\textbf {\bibinfo {volume}
  {125}},\ \bibinfo {pages} {8992} (\bibinfo {year} {2003})}\BibitemShut
  {NoStop}%
\bibitem [{\citenamefont {Li}\ \emph {et~al.}(2011)\citenamefont {Li},
  \citenamefont {Walker},\ and\ \citenamefont {Michaelides}}]{XZLi_2011}%
  \BibitemOpen
  \bibfield  {author} {\bibinfo {author} {\bibfnamefont {X.~Z.}\ \bibnamefont
  {Li}}, \bibinfo {author} {\bibfnamefont {B.}~\bibnamefont {Walker}}, \ and\
  \bibinfo {author} {\bibfnamefont {A.}~\bibnamefont {Michaelides}},\
  }\href@noop {} {\bibfield  {journal} {\bibinfo  {journal} {Proc. Natl. Acad.
  Sci. U.S.A.}\ }\textbf {\bibinfo {volume} {108}},\ \bibinfo {pages} {6369}
  (\bibinfo {year} {2011})}\BibitemShut {NoStop}%
\bibitem [{\citenamefont {Tuckerman}\ \emph {et~al.}(2002)\citenamefont
  {Tuckerman}, \citenamefont {Marx},\ and\ \citenamefont
  {Parrinello}}]{tuckerman_nature_2002}%
  \BibitemOpen
  \bibfield  {author} {\bibinfo {author} {\bibfnamefont {M.~E.}\ \bibnamefont
  {Tuckerman}}, \bibinfo {author} {\bibfnamefont {D.}~\bibnamefont {Marx}}, \
  and\ \bibinfo {author} {\bibfnamefont {M.}~\bibnamefont {Parrinello}},\
  }\href {\doibase 10.1038/nature00797} {\bibfield  {journal} {\bibinfo
  {journal} {Nature}\ }\textbf {\bibinfo {volume} {417}},\ \bibinfo {pages}
  {925} (\bibinfo {year} {2002})}\BibitemShut {NoStop}%
\bibitem [{\citenamefont {Paesani}\ and\ \citenamefont
  {Voth}(2009)}]{doi:10.1021/jp810590c}%
  \BibitemOpen
  \bibfield  {author} {\bibinfo {author} {\bibfnamefont {F.}~\bibnamefont
  {Paesani}}\ and\ \bibinfo {author} {\bibfnamefont {G.~A.}\ \bibnamefont
  {Voth}},\ }\href {\doibase 10.1021/jp810590c} {\bibfield  {journal} {\bibinfo
   {journal} {J. Phys. Chem. B}\ }\textbf {\bibinfo {volume} {113}},\ \bibinfo
  {pages} {5702} (\bibinfo {year} {2009})}\BibitemShut {NoStop}%
\bibitem [{\citenamefont {Ceriotti}\ \emph {et~al.}(2016)\citenamefont
  {Ceriotti}, \citenamefont {Fang}, \citenamefont {Kusalik}, \citenamefont
  {McKenzie}, \citenamefont {Michaelides}, \citenamefont {Morales},\ and\
  \citenamefont {Markland}}]{Wei_rev_2016}%
  \BibitemOpen
  \bibfield  {author} {\bibinfo {author} {\bibfnamefont {M.}~\bibnamefont
  {Ceriotti}}, \bibinfo {author} {\bibfnamefont {W.}~\bibnamefont {Fang}},
  \bibinfo {author} {\bibfnamefont {P.~G.}\ \bibnamefont {Kusalik}}, \bibinfo
  {author} {\bibfnamefont {R.~H.}\ \bibnamefont {McKenzie}}, \bibinfo {author}
  {\bibfnamefont {A.}~\bibnamefont {Michaelides}}, \bibinfo {author}
  {\bibfnamefont {M.~A.}\ \bibnamefont {Morales}}, \ and\ \bibinfo {author}
  {\bibfnamefont {T.~E.}\ \bibnamefont {Markland}},\ }\href {\doibase
  10.1021/acs.chemrev.5b00674} {\bibfield  {journal} {\bibinfo  {journal}
  {Chem. Rev.}\ }\textbf {\bibinfo {volume} {116}},\ \bibinfo {pages} {7529}
  (\bibinfo {year} {2016})}\BibitemShut {NoStop}%
\bibitem [{\citenamefont {Wilkins}\ \emph {et~al.}(2017)\citenamefont
  {Wilkins}, \citenamefont {Manolopoulos}, \citenamefont {Pipolo},
  \citenamefont {Laage},\ and\ \citenamefont
  {Hynes}}]{doi:10.1021/acs.jpclett.7b00979}%
  \BibitemOpen
  \bibfield  {author} {\bibinfo {author} {\bibfnamefont {D.~M.}\ \bibnamefont
  {Wilkins}}, \bibinfo {author} {\bibfnamefont {D.~E.}\ \bibnamefont
  {Manolopoulos}}, \bibinfo {author} {\bibfnamefont {S.}~\bibnamefont
  {Pipolo}}, \bibinfo {author} {\bibfnamefont {D.}~\bibnamefont {Laage}}, \
  and\ \bibinfo {author} {\bibfnamefont {J.~T.}\ \bibnamefont {Hynes}},\ }\href
  {\doibase 10.1021/acs.jpclett.7b00979} {\bibfield  {journal} {\bibinfo
  {journal} {J. Phys. Chem. Lett.}\ }\textbf {\bibinfo {volume} {8}},\ \bibinfo
  {pages} {2602} (\bibinfo {year} {2017})}\BibitemShut {NoStop}%
\bibitem [{\citenamefont {Cheng}\ \emph {et~al.}(2018)\citenamefont {Cheng},
  \citenamefont {Paxton},\ and\ \citenamefont
  {Ceriotti}}]{PhysRevLett.120.225901}%
  \BibitemOpen
  \bibfield  {author} {\bibinfo {author} {\bibfnamefont {B.}~\bibnamefont
  {Cheng}}, \bibinfo {author} {\bibfnamefont {A.~T.}\ \bibnamefont {Paxton}}, \
  and\ \bibinfo {author} {\bibfnamefont {M.}~\bibnamefont {Ceriotti}},\ }\href
  {\doibase 10.1103/PhysRevLett.120.225901} {\bibfield  {journal} {\bibinfo
  {journal} {Phys. Rev. Lett.}\ }\textbf {\bibinfo {volume} {120}},\ \bibinfo
  {pages} {225901} (\bibinfo {year} {2018})}\BibitemShut {NoStop}%
\bibitem [{\citenamefont {Rossi}\ \emph
  {et~al.}(2016{\natexlab{a}})\citenamefont {Rossi}, \citenamefont {Ceriotti},\
  and\ \citenamefont {Manolopoulos}}]{rossi_nuclear_2016}%
  \BibitemOpen
  \bibfield  {author} {\bibinfo {author} {\bibfnamefont {M.}~\bibnamefont
  {Rossi}}, \bibinfo {author} {\bibfnamefont {M.}~\bibnamefont {Ceriotti}}, \
  and\ \bibinfo {author} {\bibfnamefont {D.~E.}\ \bibnamefont {Manolopoulos}},\
  }\href {\doibase 10.1021/acs.jpclett.6b01093} {\bibfield  {journal} {\bibinfo
   {journal} {J. Phys. Chem. Lett.}\ }\textbf {\bibinfo {volume} {7}},\
  \bibinfo {pages} {3001} (\bibinfo {year} {2016}{\natexlab{a}})}\BibitemShut
  {NoStop}%
\bibitem [{\citenamefont {Marx}\ \emph {et~al.}(2010)\citenamefont {Marx},
  \citenamefont {Chandra},\ and\ \citenamefont
  {Tuckerman}}]{Marx_Tuckerman_2010}%
  \BibitemOpen
  \bibfield  {author} {\bibinfo {author} {\bibfnamefont {D.}~\bibnamefont
  {Marx}}, \bibinfo {author} {\bibfnamefont {A.}~\bibnamefont {Chandra}}, \
  and\ \bibinfo {author} {\bibfnamefont {M.~E.}\ \bibnamefont {Tuckerman}},\
  }\href {\doibase 10.1021/cr900233f} {\bibfield  {journal} {\bibinfo
  {journal} {Chem. Rev.}\ }\textbf {\bibinfo {volume} {110}},\ \bibinfo {pages}
  {2174} (\bibinfo {year} {2010})}\BibitemShut {NoStop}%
\bibitem [{\citenamefont {Chandler}\ and\ \citenamefont
  {Manolopoulos}(2016)}]{sremarks}%
  \BibitemOpen
  \bibfield  {author} {\bibinfo {author} {\bibfnamefont {D.}~\bibnamefont
  {Chandler}}\ and\ \bibinfo {author} {\bibfnamefont {D.~E.}\ \bibnamefont
  {Manolopoulos}},\ }\href {\doibase 10.1039/C6FD00229C} {\bibfield  {journal}
  {\bibinfo  {journal} {Faraday Discuss.}\ }\textbf {\bibinfo {volume} {195}},\
  \bibinfo {pages} {699} (\bibinfo {year} {2016})}\BibitemShut {NoStop}%
\bibitem [{\citenamefont {Dias}\ and\ \citenamefont
  {Silvera}(2017)}]{dias_observation_2017}%
  \BibitemOpen
  \bibfield  {author} {\bibinfo {author} {\bibfnamefont {R.~P.}\ \bibnamefont
  {Dias}}\ and\ \bibinfo {author} {\bibfnamefont {I.~F.}\ \bibnamefont
  {Silvera}},\ }\href {\doibase 10.1126/science.aal1579} {\bibfield  {journal}
  {\bibinfo  {journal} {Science}\ }\textbf {\bibinfo {volume} {355}},\ \bibinfo
  {pages} {715} (\bibinfo {year} {2017})}\BibitemShut {NoStop}%
\bibitem [{\citenamefont {Silvera}\ and\ \citenamefont
  {Dias}(2017)}]{silvera_response_2017}%
  \BibitemOpen
  \bibfield  {author} {\bibinfo {author} {\bibfnamefont {I.~F.}\ \bibnamefont
  {Silvera}}\ and\ \bibinfo {author} {\bibfnamefont {R.}~\bibnamefont {Dias}},\
  }\href {\doibase 10.1126/science.aan2671} {\bibfield  {journal} {\bibinfo
  {journal} {Science}\ }\textbf {\bibinfo {volume} {357}},\ \bibinfo {pages}
  {eaan2671} (\bibinfo {year} {2017})}\BibitemShut {NoStop}%
\bibitem [{\citenamefont {Liu}\ \emph {et~al.}(2017{\natexlab{a}})\citenamefont
  {Liu}, \citenamefont {Dalladay-Simpson}, \citenamefont {Howie}, \citenamefont
  {Li},\ and\ \citenamefont {Gregoryanz}}]{liu_comment_2017}%
  \BibitemOpen
  \bibfield  {author} {\bibinfo {author} {\bibfnamefont {X.-D.}\ \bibnamefont
  {Liu}}, \bibinfo {author} {\bibfnamefont {P.}~\bibnamefont
  {Dalladay-Simpson}}, \bibinfo {author} {\bibfnamefont {R.~T.}\ \bibnamefont
  {Howie}}, \bibinfo {author} {\bibfnamefont {B.}~\bibnamefont {Li}}, \ and\
  \bibinfo {author} {\bibfnamefont {E.}~\bibnamefont {Gregoryanz}},\ }\href
  {\doibase 10.1126/science.aan2286} {\bibfield  {journal} {\bibinfo  {journal}
  {Science}\ }\textbf {\bibinfo {volume} {357}},\ \bibinfo {pages} {eaan2286}
  (\bibinfo {year} {2017}{\natexlab{a}})}\BibitemShut {NoStop}%
\bibitem [{\citenamefont {Goncharov}\ and\ \citenamefont
  {Struzhkin}(2017)}]{goncharov_comment_2017}%
  \BibitemOpen
  \bibfield  {author} {\bibinfo {author} {\bibfnamefont {A.~F.}\ \bibnamefont
  {Goncharov}}\ and\ \bibinfo {author} {\bibfnamefont {V.~V.}\ \bibnamefont
  {Struzhkin}},\ }\href {\doibase 10.1126/science.aam9736} {\bibfield
  {journal} {\bibinfo  {journal} {Science}\ }\textbf {\bibinfo {volume}
  {357}},\ \bibinfo {pages} {eaam9736} (\bibinfo {year} {2017})}\BibitemShut
  {NoStop}%
\bibitem [{\citenamefont {Pickard}\ and\ \citenamefont
  {Needs}(2007)}]{pickard_structure_2007}%
  \BibitemOpen
  \bibfield  {author} {\bibinfo {author} {\bibfnamefont {C.~J.}\ \bibnamefont
  {Pickard}}\ and\ \bibinfo {author} {\bibfnamefont {R.~J.}\ \bibnamefont
  {Needs}},\ }\href {\doibase 10.1038/nphys625} {\bibfield  {journal} {\bibinfo
   {journal} {Nat. Phys.}\ }\textbf {\bibinfo {volume} {3}},\ \bibinfo {pages}
  {473} (\bibinfo {year} {2007})}\BibitemShut {NoStop}%
\bibitem [{\citenamefont {Drummond}\ \emph {et~al.}(2015)\citenamefont
  {Drummond}, \citenamefont {Monserrat}, \citenamefont {Lloyd-Williams},
  \citenamefont {Ríos}, \citenamefont {Pickard},\ and\ \citenamefont
  {Needs}}]{drummond_quantum_2015}%
  \BibitemOpen
  \bibfield  {author} {\bibinfo {author} {\bibfnamefont {N.~D.}\ \bibnamefont
  {Drummond}}, \bibinfo {author} {\bibfnamefont {B.}~\bibnamefont {Monserrat}},
  \bibinfo {author} {\bibfnamefont {J.~H.}\ \bibnamefont {Lloyd-Williams}},
  \bibinfo {author} {\bibfnamefont {P.~L.}\ \bibnamefont {Ríos}}, \bibinfo
  {author} {\bibfnamefont {C.~J.}\ \bibnamefont {Pickard}}, \ and\ \bibinfo
  {author} {\bibfnamefont {R.~J.}\ \bibnamefont {Needs}},\ }\href {\doibase
  10.1038/ncomms8794} {\bibfield  {journal} {\bibinfo  {journal} {Nat.
  Commun.}\ }\textbf {\bibinfo {volume} {6}},\ \bibinfo {pages} {7794}
  (\bibinfo {year} {2015})}\BibitemShut {NoStop}%
\bibitem [{\citenamefont {Silvera}\ and\ \citenamefont
  {Wijngaarden}(1981)}]{silvera_new_1981}%
  \BibitemOpen
  \bibfield  {author} {\bibinfo {author} {\bibfnamefont {I.~F.}\ \bibnamefont
  {Silvera}}\ and\ \bibinfo {author} {\bibfnamefont {R.~J.}\ \bibnamefont
  {Wijngaarden}},\ }\href {\doibase 10.1103/PhysRevLett.47.39} {\bibfield
  {journal} {\bibinfo  {journal} {Phys. Rev. Lett.}\ }\textbf {\bibinfo
  {volume} {47}},\ \bibinfo {pages} {39} (\bibinfo {year} {1981})}\BibitemShut
  {NoStop}%
\bibitem [{\citenamefont {Lorenzana}\ \emph {et~al.}(1989)\citenamefont
  {Lorenzana}, \citenamefont {Silvera},\ and\ \citenamefont
  {Goettel}}]{lorenzana_evidence_1989}%
  \BibitemOpen
  \bibfield  {author} {\bibinfo {author} {\bibfnamefont {H.~E.}\ \bibnamefont
  {Lorenzana}}, \bibinfo {author} {\bibfnamefont {I.~F.}\ \bibnamefont
  {Silvera}}, \ and\ \bibinfo {author} {\bibfnamefont {K.~A.}\ \bibnamefont
  {Goettel}},\ }\href {\doibase 10.1103/PhysRevLett.63.2080} {\bibfield
  {journal} {\bibinfo  {journal} {Phys. Rev. Lett.}\ }\textbf {\bibinfo
  {volume} {63}},\ \bibinfo {pages} {2080} (\bibinfo {year}
  {1989})}\BibitemShut {NoStop}%
\bibitem [{\citenamefont {Lorenzana}\ \emph {et~al.}(1990)\citenamefont
  {Lorenzana}, \citenamefont {Silvera},\ and\ \citenamefont
  {Goettel}}]{lorenzana_orientational_1990}%
  \BibitemOpen
  \bibfield  {author} {\bibinfo {author} {\bibfnamefont {H.~E.}\ \bibnamefont
  {Lorenzana}}, \bibinfo {author} {\bibfnamefont {I.~F.}\ \bibnamefont
  {Silvera}}, \ and\ \bibinfo {author} {\bibfnamefont {K.~A.}\ \bibnamefont
  {Goettel}},\ }\href {\doibase 10.1103/PhysRevLett.64.1939} {\bibfield
  {journal} {\bibinfo  {journal} {Phys. Rev. Lett.}\ }\textbf {\bibinfo
  {volume} {64}},\ \bibinfo {pages} {1939} (\bibinfo {year}
  {1990})}\BibitemShut {NoStop}%
\bibitem [{\citenamefont {Cui}\ \emph {et~al.}(1994)\citenamefont {Cui},
  \citenamefont {Chen}, \citenamefont {Jeon},\ and\ \citenamefont
  {Silvera}}]{cui_megabar_1994}%
  \BibitemOpen
  \bibfield  {author} {\bibinfo {author} {\bibfnamefont {L.}~\bibnamefont
  {Cui}}, \bibinfo {author} {\bibfnamefont {N.~H.}\ \bibnamefont {Chen}},
  \bibinfo {author} {\bibfnamefont {S.~J.}\ \bibnamefont {Jeon}}, \ and\
  \bibinfo {author} {\bibfnamefont {I.~F.}\ \bibnamefont {Silvera}},\ }\href
  {\doibase 10.1103/PhysRevLett.72.3048} {\bibfield  {journal} {\bibinfo
  {journal} {Phys. Rev. Lett.}\ }\textbf {\bibinfo {volume} {72}},\ \bibinfo
  {pages} {3048} (\bibinfo {year} {1994})}\BibitemShut {NoStop}%
\bibitem [{\citenamefont {Mazin}\ \emph {et~al.}(1997)\citenamefont {Mazin},
  \citenamefont {Hemley}, \citenamefont {Goncharov}, \citenamefont {Hanfland},\
  and\ \citenamefont {Mao}}]{mazin_quantum_1997}%
  \BibitemOpen
  \bibfield  {author} {\bibinfo {author} {\bibfnamefont {I.~I.}\ \bibnamefont
  {Mazin}}, \bibinfo {author} {\bibfnamefont {R.~J.}\ \bibnamefont {Hemley}},
  \bibinfo {author} {\bibfnamefont {A.~F.}\ \bibnamefont {Goncharov}}, \bibinfo
  {author} {\bibfnamefont {M.}~\bibnamefont {Hanfland}}, \ and\ \bibinfo
  {author} {\bibfnamefont {H.-k.}\ \bibnamefont {Mao}},\ }\href {\doibase
  10.1103/PhysRevLett.78.1066} {\bibfield  {journal} {\bibinfo  {journal}
  {Phys. Rev. Lett.}\ }\textbf {\bibinfo {volume} {78}},\ \bibinfo {pages}
  {1066} (\bibinfo {year} {1997})}\BibitemShut {NoStop}%
\bibitem [{\citenamefont {Goncharov}\ \emph {et~al.}(1995)\citenamefont
  {Goncharov}, \citenamefont {Mazin}, \citenamefont {Eggert}, \citenamefont
  {Hemley},\ and\ \citenamefont {Mao}}]{goncharov_invariant_1995}%
  \BibitemOpen
  \bibfield  {author} {\bibinfo {author} {\bibfnamefont {A.~F.}\ \bibnamefont
  {Goncharov}}, \bibinfo {author} {\bibfnamefont {I.~I.}\ \bibnamefont
  {Mazin}}, \bibinfo {author} {\bibfnamefont {J.~H.}\ \bibnamefont {Eggert}},
  \bibinfo {author} {\bibfnamefont {R.~J.}\ \bibnamefont {Hemley}}, \ and\
  \bibinfo {author} {\bibfnamefont {H.-k.}\ \bibnamefont {Mao}},\ }\href
  {\doibase 10.1103/PhysRevLett.75.2514} {\bibfield  {journal} {\bibinfo
  {journal} {Phys. Rev. Lett.}\ }\textbf {\bibinfo {volume} {75}},\ \bibinfo
  {pages} {2514} (\bibinfo {year} {1995})}\BibitemShut {NoStop}%
\bibitem [{\citenamefont {Goncharov}\ \emph
  {et~al.}(1996{\natexlab{a}})\citenamefont {Goncharov}, \citenamefont
  {Eggert}, \citenamefont {Mazin}, \citenamefont {Hemley},\ and\ \citenamefont
  {Mao}}]{goncharov_raman_1996}%
  \BibitemOpen
  \bibfield  {author} {\bibinfo {author} {\bibfnamefont {A.~F.}\ \bibnamefont
  {Goncharov}}, \bibinfo {author} {\bibfnamefont {J.~H.}\ \bibnamefont
  {Eggert}}, \bibinfo {author} {\bibfnamefont {I.~I.}\ \bibnamefont {Mazin}},
  \bibinfo {author} {\bibfnamefont {R.~J.}\ \bibnamefont {Hemley}}, \ and\
  \bibinfo {author} {\bibfnamefont {H.-k.}\ \bibnamefont {Mao}},\ }\href
  {\doibase 10.1103/PhysRevB.54.R15590} {\bibfield  {journal} {\bibinfo
  {journal} {Phys. Rev. B}\ }\textbf {\bibinfo {volume} {54}},\ \bibinfo
  {pages} {R15590} (\bibinfo {year} {1996}{\natexlab{a}})}\BibitemShut
  {NoStop}%
\bibitem [{\citenamefont {Liu}\ \emph {et~al.}(2017{\natexlab{b}})\citenamefont
  {Liu}, \citenamefont {Howie}, \citenamefont {Zhang}, \citenamefont {Chen},\
  and\ \citenamefont {Gregoryanz}}]{PhysRevLett.119.065301}%
  \BibitemOpen
  \bibfield  {author} {\bibinfo {author} {\bibfnamefont {X.-D.}\ \bibnamefont
  {Liu}}, \bibinfo {author} {\bibfnamefont {R.~T.}\ \bibnamefont {Howie}},
  \bibinfo {author} {\bibfnamefont {H.-C.}\ \bibnamefont {Zhang}}, \bibinfo
  {author} {\bibfnamefont {X.-J.}\ \bibnamefont {Chen}}, \ and\ \bibinfo
  {author} {\bibfnamefont {E.}~\bibnamefont {Gregoryanz}},\ }\href {\doibase
  10.1103/PhysRevLett.119.065301} {\bibfield  {journal} {\bibinfo  {journal}
  {Phys. Rev. Lett.}\ }\textbf {\bibinfo {volume} {119}},\ \bibinfo {pages}
  {065301} (\bibinfo {year} {2017}{\natexlab{b}})}\BibitemShut {NoStop}%
\bibitem [{\citenamefont {McMahon}\ \emph {et~al.}(2012)\citenamefont
  {McMahon}, \citenamefont {Morales}, \citenamefont {Pierleoni},\ and\
  \citenamefont {Ceperley}}]{mcmahon_properties_2012}%
  \BibitemOpen
  \bibfield  {author} {\bibinfo {author} {\bibfnamefont {J.~M.}\ \bibnamefont
  {McMahon}}, \bibinfo {author} {\bibfnamefont {M.~A.}\ \bibnamefont
  {Morales}}, \bibinfo {author} {\bibfnamefont {C.}~\bibnamefont {Pierleoni}},
  \ and\ \bibinfo {author} {\bibfnamefont {D.~M.}\ \bibnamefont {Ceperley}},\
  }\href {\doibase 10.1103/RevModPhys.84.1607} {\bibfield  {journal} {\bibinfo
  {journal} {Rev. Mod. Phys.}\ }\textbf {\bibinfo {volume} {84}},\ \bibinfo
  {pages} {1607} (\bibinfo {year} {2012})}\BibitemShut {NoStop}%
\bibitem [{\citenamefont {Goncharenko}\ and\ \citenamefont
  {Loubeyre}(2005)}]{goncharenko_neutron_2005}%
  \BibitemOpen
  \bibfield  {author} {\bibinfo {author} {\bibfnamefont {I.}~\bibnamefont
  {Goncharenko}}\ and\ \bibinfo {author} {\bibfnamefont {P.}~\bibnamefont
  {Loubeyre}},\ }\href {\doibase 10.1038/nature03699} {\bibfield  {journal}
  {\bibinfo  {journal} {Nature}\ }\textbf {\bibinfo {volume} {435}},\ \bibinfo
  {pages} {1206} (\bibinfo {year} {2005})}\BibitemShut {NoStop}%
\bibitem [{\citenamefont {Eremets}\ and\ \citenamefont
  {Troyan}(2011)}]{eremets_conductive_2011}%
  \BibitemOpen
  \bibfield  {author} {\bibinfo {author} {\bibfnamefont {M.~I.}\ \bibnamefont
  {Eremets}}\ and\ \bibinfo {author} {\bibfnamefont {I.~A.}\ \bibnamefont
  {Troyan}},\ }\href {\doibase 10.1038/nmat3175} {\bibfield  {journal}
  {\bibinfo  {journal} {Nat. Mater.}\ }\textbf {\bibinfo {volume} {10}},\
  \bibinfo {pages} {927} (\bibinfo {year} {2011})}\BibitemShut {NoStop}%
\bibitem [{\citenamefont {Howie}\ \emph
  {et~al.}(2012{\natexlab{a}})\citenamefont {Howie}, \citenamefont {Guillaume},
  \citenamefont {Scheler}, \citenamefont {Goncharov},\ and\ \citenamefont
  {Gregoryanz}}]{howie_mixed_2012}%
  \BibitemOpen
  \bibfield  {author} {\bibinfo {author} {\bibfnamefont {R.~T.}\ \bibnamefont
  {Howie}}, \bibinfo {author} {\bibfnamefont {C.~L.}\ \bibnamefont
  {Guillaume}}, \bibinfo {author} {\bibfnamefont {T.}~\bibnamefont {Scheler}},
  \bibinfo {author} {\bibfnamefont {A.~F.}\ \bibnamefont {Goncharov}}, \ and\
  \bibinfo {author} {\bibfnamefont {E.}~\bibnamefont {Gregoryanz}},\ }\href
  {\doibase 10.1103/PhysRevLett.108.125501} {\bibfield  {journal} {\bibinfo
  {journal} {Phys. Rev. Lett.}\ }\textbf {\bibinfo {volume} {108}},\ \bibinfo
  {pages} {125501} (\bibinfo {year} {2012}{\natexlab{a}})}\BibitemShut
  {NoStop}%
\bibitem [{\citenamefont {Howie}\ \emph
  {et~al.}(2012{\natexlab{b}})\citenamefont {Howie}, \citenamefont {Scheler},
  \citenamefont {Guillaume},\ and\ \citenamefont
  {Gregoryanz}}]{howie_proton_2012}%
  \BibitemOpen
  \bibfield  {author} {\bibinfo {author} {\bibfnamefont {R.~T.}\ \bibnamefont
  {Howie}}, \bibinfo {author} {\bibfnamefont {T.}~\bibnamefont {Scheler}},
  \bibinfo {author} {\bibfnamefont {C.~L.}\ \bibnamefont {Guillaume}}, \ and\
  \bibinfo {author} {\bibfnamefont {E.}~\bibnamefont {Gregoryanz}},\ }\href
  {\doibase 10.1103/PhysRevB.86.214104} {\bibfield  {journal} {\bibinfo
  {journal} {Phys. Rev. B}\ }\textbf {\bibinfo {volume} {86}},\ \bibinfo
  {pages} {214104} (\bibinfo {year} {2012}{\natexlab{b}})}\BibitemShut
  {NoStop}%
\bibitem [{\citenamefont {Zha}\ \emph {et~al.}(2012)\citenamefont {Zha},
  \citenamefont {Liu},\ and\ \citenamefont {Hemley}}]{zha_synchrotron_2012}%
  \BibitemOpen
  \bibfield  {author} {\bibinfo {author} {\bibfnamefont {C.-S.}\ \bibnamefont
  {Zha}}, \bibinfo {author} {\bibfnamefont {Z.}~\bibnamefont {Liu}}, \ and\
  \bibinfo {author} {\bibfnamefont {R.~J.}\ \bibnamefont {Hemley}},\ }\href
  {\doibase 10.1103/PhysRevLett.108.146402} {\bibfield  {journal} {\bibinfo
  {journal} {Phys. Rev. Lett.}\ }\textbf {\bibinfo {volume} {108}},\ \bibinfo
  {pages} {146402} (\bibinfo {year} {2012})}\BibitemShut {NoStop}%
\bibitem [{\citenamefont {Loubeyre}\ \emph {et~al.}(2013)\citenamefont
  {Loubeyre}, \citenamefont {Occelli},\ and\ \citenamefont
  {Dumas}}]{loubeyre_hydrogen_2013}%
  \BibitemOpen
  \bibfield  {author} {\bibinfo {author} {\bibfnamefont {P.}~\bibnamefont
  {Loubeyre}}, \bibinfo {author} {\bibfnamefont {F.}~\bibnamefont {Occelli}}, \
  and\ \bibinfo {author} {\bibfnamefont {P.}~\bibnamefont {Dumas}},\ }\href
  {\doibase 10.1103/PhysRevB.87.134101} {\bibfield  {journal} {\bibinfo
  {journal} {Phys. Rev. B}\ }\textbf {\bibinfo {volume} {87}},\ \bibinfo
  {pages} {134101} (\bibinfo {year} {2013})}\BibitemShut {NoStop}%
\bibitem [{\citenamefont {Eremets}\ \emph {et~al.}(2013)\citenamefont
  {Eremets}, \citenamefont {Troyan}, \citenamefont {Lerch},\ and\ \citenamefont
  {Drozdov}}]{eremets_infrared_2013}%
  \BibitemOpen
  \bibfield  {author} {\bibinfo {author} {\bibfnamefont {M.~I.}\ \bibnamefont
  {Eremets}}, \bibinfo {author} {\bibfnamefont {I.~A.}\ \bibnamefont {Troyan}},
  \bibinfo {author} {\bibfnamefont {P.}~\bibnamefont {Lerch}}, \ and\ \bibinfo
  {author} {\bibfnamefont {A.}~\bibnamefont {Drozdov}},\ }\href {\doibase
  10.1080/08957959.2013.794229} {\bibfield  {journal} {\bibinfo  {journal}
  {High Press. Res.}\ }\textbf {\bibinfo {volume} {33}},\ \bibinfo {pages}
  {377} (\bibinfo {year} {2013})}\BibitemShut {NoStop}%
\bibitem [{\citenamefont {Zha}\ \emph {et~al.}(2013)\citenamefont {Zha},
  \citenamefont {Liu}, \citenamefont {Ahart}, \citenamefont {Boehler},\ and\
  \citenamefont {Hemley}}]{zha_high-pressure_2013}%
  \BibitemOpen
  \bibfield  {author} {\bibinfo {author} {\bibfnamefont {C.-S.}\ \bibnamefont
  {Zha}}, \bibinfo {author} {\bibfnamefont {Z.}~\bibnamefont {Liu}}, \bibinfo
  {author} {\bibfnamefont {M.}~\bibnamefont {Ahart}}, \bibinfo {author}
  {\bibfnamefont {R.}~\bibnamefont {Boehler}}, \ and\ \bibinfo {author}
  {\bibfnamefont {R.~J.}\ \bibnamefont {Hemley}},\ }\href {\doibase
  10.1103/PhysRevLett.110.217402} {\bibfield  {journal} {\bibinfo  {journal}
  {Phys. Rev. Lett.}\ }\textbf {\bibinfo {volume} {110}},\ \bibinfo {pages}
  {217402} (\bibinfo {year} {2013})}\BibitemShut {NoStop}%
\bibitem [{\citenamefont {Zha}\ \emph {et~al.}(2014)\citenamefont {Zha},
  \citenamefont {Cohen}, \citenamefont {Mao},\ and\ \citenamefont
  {Hemley}}]{zha_raman_2014}%
  \BibitemOpen
  \bibfield  {author} {\bibinfo {author} {\bibfnamefont {C.-S.}\ \bibnamefont
  {Zha}}, \bibinfo {author} {\bibfnamefont {R.~E.}\ \bibnamefont {Cohen}},
  \bibinfo {author} {\bibfnamefont {H.-k.}\ \bibnamefont {Mao}}, \ and\
  \bibinfo {author} {\bibfnamefont {R.~J.}\ \bibnamefont {Hemley}},\ }\href
  {\doibase 10.1073/pnas.1402737111} {\bibfield  {journal} {\bibinfo  {journal}
  {Proc. Natl. Acad. Sci.}\ }\textbf {\bibinfo {volume} {111}},\ \bibinfo
  {pages} {4792} (\bibinfo {year} {2014})}\BibitemShut {NoStop}%
\bibitem [{\citenamefont {Howie}\ \emph {et~al.}(2015)\citenamefont {Howie},
  \citenamefont {Dalladay-Simpson},\ and\ \citenamefont
  {Gregoryanz}}]{howie_raman_2015}%
  \BibitemOpen
  \bibfield  {author} {\bibinfo {author} {\bibfnamefont {R.~T.}\ \bibnamefont
  {Howie}}, \bibinfo {author} {\bibfnamefont {P.}~\bibnamefont
  {Dalladay-Simpson}}, \ and\ \bibinfo {author} {\bibfnamefont
  {E.}~\bibnamefont {Gregoryanz}},\ }\href {\doibase 10.1038/nmat4213}
  {\bibfield  {journal} {\bibinfo  {journal} {Nat. Mater.}\ }\textbf {\bibinfo
  {volume} {14}},\ \bibinfo {pages} {495} (\bibinfo {year} {2015})}\BibitemShut
  {NoStop}%
\bibitem [{\citenamefont {Dalladay-Simpson}\ \emph {et~al.}(2016)\citenamefont
  {Dalladay-Simpson}, \citenamefont {Howie},\ and\ \citenamefont
  {Gregoryanz}}]{dalladay-simpson_evidence_2016}%
  \BibitemOpen
  \bibfield  {author} {\bibinfo {author} {\bibfnamefont {P.}~\bibnamefont
  {Dalladay-Simpson}}, \bibinfo {author} {\bibfnamefont {R.~T.}\ \bibnamefont
  {Howie}}, \ and\ \bibinfo {author} {\bibfnamefont {E.}~\bibnamefont
  {Gregoryanz}},\ }\href {\doibase 10.1038/nature16164} {\bibfield  {journal}
  {\bibinfo  {journal} {Nature}\ }\textbf {\bibinfo {volume} {529}},\ \bibinfo
  {pages} {63} (\bibinfo {year} {2016})}\BibitemShut {NoStop}%
\bibitem [{\citenamefont {Ackland}(2015)}]{ackland_bearing_2015}%
  \BibitemOpen
  \bibfield  {author} {\bibinfo {author} {\bibfnamefont {G.~J.}\ \bibnamefont
  {Ackland}},\ }\href {\doibase 10.1126/science.aac6626} {\bibfield  {journal}
  {\bibinfo  {journal} {Science}\ }\textbf {\bibinfo {volume} {348}},\ \bibinfo
  {pages} {1429} (\bibinfo {year} {2015})}\BibitemShut {NoStop}%
\bibitem [{\citenamefont {Miller~III}\ and\ \citenamefont
  {Manolopoulos}(2005)}]{doi:10.1063/1.1893956}%
  \BibitemOpen
  \bibfield  {author} {\bibinfo {author} {\bibfnamefont {T.~F.}\ \bibnamefont
  {Miller~III}}\ and\ \bibinfo {author} {\bibfnamefont {D.~E.}\ \bibnamefont
  {Manolopoulos}},\ }\href {\doibase 10.1063/1.1893956} {\bibfield  {journal}
  {\bibinfo  {journal} {J. Chem. Phys.}\ }\textbf {\bibinfo {volume} {122}},\
  \bibinfo {pages} {184503} (\bibinfo {year} {2005})}\BibitemShut {NoStop}%
\bibitem [{\citenamefont {Morales}\ \emph {et~al.}(2013)\citenamefont
  {Morales}, \citenamefont {McMahon}, \citenamefont {Pierleoni},\ and\
  \citenamefont {Ceperley}}]{morales_nuclear_2013}%
  \BibitemOpen
  \bibfield  {author} {\bibinfo {author} {\bibfnamefont {M.~A.}\ \bibnamefont
  {Morales}}, \bibinfo {author} {\bibfnamefont {J.~M.}\ \bibnamefont
  {McMahon}}, \bibinfo {author} {\bibfnamefont {C.}~\bibnamefont {Pierleoni}},
  \ and\ \bibinfo {author} {\bibfnamefont {D.~M.}\ \bibnamefont {Ceperley}},\
  }\href {\doibase 10.1103/PhysRevLett.110.065702} {\bibfield  {journal}
  {\bibinfo  {journal} {Phys. Rev. Lett.}\ }\textbf {\bibinfo {volume} {110}},\
  \bibinfo {pages} {065702} (\bibinfo {year} {2013})}\BibitemShut {NoStop}%
\bibitem [{\citenamefont {Kang}\ \emph {et~al.}(2013)\citenamefont {Kang},
  \citenamefont {Sun}, \citenamefont {Dai}, \citenamefont {Zhao}, \citenamefont
  {Hou}, \citenamefont {Zeng},\ and\ \citenamefont
  {Yuan}}]{kang_revealing_2013}%
  \BibitemOpen
  \bibfield  {author} {\bibinfo {author} {\bibfnamefont {D.}~\bibnamefont
  {Kang}}, \bibinfo {author} {\bibfnamefont {H.}~\bibnamefont {Sun}}, \bibinfo
  {author} {\bibfnamefont {J.}~\bibnamefont {Dai}}, \bibinfo {author}
  {\bibfnamefont {Z.}~\bibnamefont {Zhao}}, \bibinfo {author} {\bibfnamefont
  {Y.}~\bibnamefont {Hou}}, \bibinfo {author} {\bibfnamefont {J.}~\bibnamefont
  {Zeng}}, \ and\ \bibinfo {author} {\bibfnamefont {J.}~\bibnamefont {Yuan}},\
  }\href {http://arxiv.org/abs/1304.0953} {\bibfield  {journal} {\bibinfo
  {journal} {arXiv:1304.0953 [cond-mat, physics:physics]}\ } (\bibinfo {year}
  {2013})},\ \bibinfo {note} {arXiv: 1304.0953}\BibitemShut {NoStop}%
\bibitem [{\citenamefont {Kang}\ \emph {et~al.}(2014)\citenamefont {Kang},
  \citenamefont {Sun}, \citenamefont {Dai}, \citenamefont {Chen}, \citenamefont
  {Zhao}, \citenamefont {Hou}, \citenamefont {Zeng},\ and\ \citenamefont
  {Yuan}}]{kang_nuclear_2014}%
  \BibitemOpen
  \bibfield  {author} {\bibinfo {author} {\bibfnamefont {D.}~\bibnamefont
  {Kang}}, \bibinfo {author} {\bibfnamefont {H.}~\bibnamefont {Sun}}, \bibinfo
  {author} {\bibfnamefont {J.}~\bibnamefont {Dai}}, \bibinfo {author}
  {\bibfnamefont {W.}~\bibnamefont {Chen}}, \bibinfo {author} {\bibfnamefont
  {Z.}~\bibnamefont {Zhao}}, \bibinfo {author} {\bibfnamefont {Y.}~\bibnamefont
  {Hou}}, \bibinfo {author} {\bibfnamefont {J.}~\bibnamefont {Zeng}}, \ and\
  \bibinfo {author} {\bibfnamefont {J.}~\bibnamefont {Yuan}},\ }\href {\doibase
  10.1038/srep05484} {\bibfield  {journal} {\bibinfo  {journal} {Sci. Rep.}\
  }\textbf {\bibinfo {volume} {4}},\ \bibinfo {pages} {5484} (\bibinfo {year}
  {2014})}\BibitemShut {NoStop}%
\bibitem [{\citenamefont {Mao}\ and\ \citenamefont
  {Hemley}(1994)}]{mao_ultrahigh-pressure_1994}%
  \BibitemOpen
  \bibfield  {author} {\bibinfo {author} {\bibfnamefont {H.-K.}\ \bibnamefont
  {Mao}}\ and\ \bibinfo {author} {\bibfnamefont {R.~J.}\ \bibnamefont
  {Hemley}},\ }\href {\doibase 10.1103/RevModPhys.66.671} {\bibfield  {journal}
  {\bibinfo  {journal} {Rev. Mod. Phys.}\ }\textbf {\bibinfo {volume} {66}},\
  \bibinfo {pages} {671} (\bibinfo {year} {1994})}\BibitemShut {NoStop}%
\bibitem [{\citenamefont {Biermann}\ \emph
  {et~al.}(1998{\natexlab{a}})\citenamefont {Biermann}, \citenamefont {Hohl},\
  and\ \citenamefont {Marx}}]{biermann_proton_1998}%
  \BibitemOpen
  \bibfield  {author} {\bibinfo {author} {\bibfnamefont {S.}~\bibnamefont
  {Biermann}}, \bibinfo {author} {\bibfnamefont {D.}~\bibnamefont {Hohl}}, \
  and\ \bibinfo {author} {\bibfnamefont {D.}~\bibnamefont {Marx}},\ }\href
  {\doibase 10.1023/A:1022566818119} {\bibfield  {journal} {\bibinfo  {journal}
  {J. Low Temp. Phys.}\ }\textbf {\bibinfo {volume} {110}},\ \bibinfo {pages}
  {97} (\bibinfo {year} {1998}{\natexlab{a}})}\BibitemShut {NoStop}%
\bibitem [{\citenamefont {Biermann}\ \emph
  {et~al.}(1998{\natexlab{b}})\citenamefont {Biermann}, \citenamefont {Hohl},\
  and\ \citenamefont {Marx}}]{biermann_quantum_1998}%
  \BibitemOpen
  \bibfield  {author} {\bibinfo {author} {\bibfnamefont {S.}~\bibnamefont
  {Biermann}}, \bibinfo {author} {\bibfnamefont {D.}~\bibnamefont {Hohl}}, \
  and\ \bibinfo {author} {\bibfnamefont {D.}~\bibnamefont {Marx}},\ }\href
  {\doibase 10.1016/S0038-1098(98)00388-3} {\bibfield  {journal} {\bibinfo
  {journal} {Solid State Commun.}\ }\textbf {\bibinfo {volume} {108}},\
  \bibinfo {pages} {337} (\bibinfo {year} {1998}{\natexlab{b}})}\BibitemShut
  {NoStop}%
\bibitem [{\citenamefont {Kitamura}\ \emph {et~al.}(2000)\citenamefont
  {Kitamura}, \citenamefont {Tsuneyuki}, \citenamefont {Ogitsu},\ and\
  \citenamefont {Miyake}}]{kitamura_quantum_2000}%
  \BibitemOpen
  \bibfield  {author} {\bibinfo {author} {\bibfnamefont {H.}~\bibnamefont
  {Kitamura}}, \bibinfo {author} {\bibfnamefont {S.}~\bibnamefont {Tsuneyuki}},
  \bibinfo {author} {\bibfnamefont {T.}~\bibnamefont {Ogitsu}}, \ and\ \bibinfo
  {author} {\bibfnamefont {T.}~\bibnamefont {Miyake}},\ }\href {\doibase
  10.1038/35005027} {\bibfield  {journal} {\bibinfo  {journal} {Nature}\
  }\textbf {\bibinfo {volume} {404}},\ \bibinfo {pages} {259} (\bibinfo {year}
  {2000})}\BibitemShut {NoStop}%
\bibitem [{\citenamefont {Klime\v{s}}\ \emph {et~al.}(2011)\citenamefont
  {Klime\v{s}}, \citenamefont {Bowler},\ and\ \citenamefont
  {Michaelides}}]{PhysRevB.83.195131}%
  \BibitemOpen
  \bibfield  {author} {\bibinfo {author} {\bibfnamefont {J.}~\bibnamefont
  {Klime\v{s}}}, \bibinfo {author} {\bibfnamefont {D.~R.}\ \bibnamefont
  {Bowler}}, \ and\ \bibinfo {author} {\bibfnamefont {A.}~\bibnamefont
  {Michaelides}},\ }\href {\doibase 10.1103/PhysRevB.83.195131} {\bibfield
  {journal} {\bibinfo  {journal} {Phys. Rev. B}\ }\textbf {\bibinfo {volume}
  {83}},\ \bibinfo {pages} {195131} (\bibinfo {year} {2011})}\BibitemShut
  {NoStop}%
\bibitem [{\citenamefont {Klime\v{s}}\ and\ \citenamefont
  {Michaelides}(2012)}]{doi:10.1063/1.4754130}%
  \BibitemOpen
  \bibfield  {author} {\bibinfo {author} {\bibfnamefont {J.}~\bibnamefont
  {Klime\v{s}}}\ and\ \bibinfo {author} {\bibfnamefont {A.}~\bibnamefont
  {Michaelides}},\ }\href {\doibase 10.1063/1.4754130} {\bibfield  {journal}
  {\bibinfo  {journal} {J. Chem. Phys.}\ }\textbf {\bibinfo {volume} {137}},\
  \bibinfo {pages} {120901} (\bibinfo {year} {2012})}\BibitemShut {NoStop}%
\bibitem [{\citenamefont {Li}\ \emph {et~al.}(2013)\citenamefont {Li},
  \citenamefont {Walker}, \citenamefont {Probert}, \citenamefont {Pickard},
  \citenamefont {Needs},\ and\ \citenamefont
  {Michaelides}}]{li_classical_2013}%
  \BibitemOpen
  \bibfield  {author} {\bibinfo {author} {\bibfnamefont {X.-Z.}\ \bibnamefont
  {Li}}, \bibinfo {author} {\bibfnamefont {B.}~\bibnamefont {Walker}}, \bibinfo
  {author} {\bibfnamefont {M.~I.~J.}\ \bibnamefont {Probert}}, \bibinfo
  {author} {\bibfnamefont {C.~J.}\ \bibnamefont {Pickard}}, \bibinfo {author}
  {\bibfnamefont {R.~J.}\ \bibnamefont {Needs}}, \ and\ \bibinfo {author}
  {\bibfnamefont {A.}~\bibnamefont {Michaelides}},\ }\href {\doibase
  10.1088/0953-8984/25/8/085402} {\bibfield  {journal} {\bibinfo  {journal} {J.
  Phys.: Condens. Matter}\ }\textbf {\bibinfo {volume} {25}},\ \bibinfo {pages}
  {085402} (\bibinfo {year} {2013})}\BibitemShut {NoStop}%
\bibitem [{\citenamefont {Babaev}\ \emph {et~al.}(2004)\citenamefont {Babaev},
  \citenamefont {Sudbø},\ and\ \citenamefont
  {Ashcroft}}]{babaev_superconductor_2004}%
  \BibitemOpen
  \bibfield  {author} {\bibinfo {author} {\bibfnamefont {E.}~\bibnamefont
  {Babaev}}, \bibinfo {author} {\bibfnamefont {A.}~\bibnamefont {Sudbø}}, \
  and\ \bibinfo {author} {\bibfnamefont {N.~W.}\ \bibnamefont {Ashcroft}},\
  }\href {\doibase 10.1038/nature02910} {\bibfield  {journal} {\bibinfo
  {journal} {Nature}\ }\textbf {\bibinfo {volume} {431}},\ \bibinfo {pages}
  {666} (\bibinfo {year} {2004})}\BibitemShut {NoStop}%
\bibitem [{\citenamefont {Bonev}\ \emph {et~al.}(2004)\citenamefont {Bonev},
  \citenamefont {Schwegler}, \citenamefont {Ogitsu},\ and\ \citenamefont
  {Galli}}]{bonev_quantum_2004}%
  \BibitemOpen
  \bibfield  {author} {\bibinfo {author} {\bibfnamefont {S.~A.}\ \bibnamefont
  {Bonev}}, \bibinfo {author} {\bibfnamefont {E.}~\bibnamefont {Schwegler}},
  \bibinfo {author} {\bibfnamefont {T.}~\bibnamefont {Ogitsu}}, \ and\ \bibinfo
  {author} {\bibfnamefont {G.}~\bibnamefont {Galli}},\ }\href {\doibase
  10.1038/nature02968} {\bibfield  {journal} {\bibinfo  {journal} {Nature}\
  }\textbf {\bibinfo {volume} {431}},\ \bibinfo {pages} {669} (\bibinfo {year}
  {2004})}\BibitemShut {NoStop}%
\bibitem [{\citenamefont {Deemyad}\ and\ \citenamefont
  {Silvera}(2008)}]{deemyad_melting_2008}%
  \BibitemOpen
  \bibfield  {author} {\bibinfo {author} {\bibfnamefont {S.}~\bibnamefont
  {Deemyad}}\ and\ \bibinfo {author} {\bibfnamefont {I.~F.}\ \bibnamefont
  {Silvera}},\ }\href {\doibase 10.1103/PhysRevLett.100.155701} {\bibfield
  {journal} {\bibinfo  {journal} {Phys. Rev. Lett.}\ }\textbf {\bibinfo
  {volume} {100}},\ \bibinfo {pages} {155701} (\bibinfo {year}
  {2008})}\BibitemShut {NoStop}%
\bibitem [{\citenamefont {Chen}\ \emph {et~al.}(2013)\citenamefont {Chen},
  \citenamefont {Li}, \citenamefont {Zhang}, \citenamefont {Probert},
  \citenamefont {Pickard}, \citenamefont {Needs}, \citenamefont {Michaelides},\
  and\ \citenamefont {Wang}}]{chen_quantum_2013}%
  \BibitemOpen
  \bibfield  {author} {\bibinfo {author} {\bibfnamefont {J.}~\bibnamefont
  {Chen}}, \bibinfo {author} {\bibfnamefont {X.-Z.}\ \bibnamefont {Li}},
  \bibinfo {author} {\bibfnamefont {Q.}~\bibnamefont {Zhang}}, \bibinfo
  {author} {\bibfnamefont {M.~I.~J.}\ \bibnamefont {Probert}}, \bibinfo
  {author} {\bibfnamefont {C.~J.}\ \bibnamefont {Pickard}}, \bibinfo {author}
  {\bibfnamefont {R.~J.}\ \bibnamefont {Needs}}, \bibinfo {author}
  {\bibfnamefont {A.}~\bibnamefont {Michaelides}}, \ and\ \bibinfo {author}
  {\bibfnamefont {E.}~\bibnamefont {Wang}},\ }\href {\doibase
  10.1038/ncomms3064} {\bibfield  {journal} {\bibinfo  {journal} {Nat.
  Commun.}\ }\textbf {\bibinfo {volume} {4}},\ \bibinfo {pages} {2064}
  (\bibinfo {year} {2013})}\BibitemShut {NoStop}%
\bibitem [{\citenamefont {Geng}\ \emph {et~al.}(2015)\citenamefont {Geng},
  \citenamefont {Hoffmann},\ and\ \citenamefont {Wu}}]{geng_lattice_2015}%
  \BibitemOpen
  \bibfield  {author} {\bibinfo {author} {\bibfnamefont {H.~Y.}\ \bibnamefont
  {Geng}}, \bibinfo {author} {\bibfnamefont {R.}~\bibnamefont {Hoffmann}}, \
  and\ \bibinfo {author} {\bibfnamefont {Q.}~\bibnamefont {Wu}},\ }\href
  {\doibase 10.1103/PhysRevB.92.104103} {\bibfield  {journal} {\bibinfo
  {journal} {Phys. Rev. B}\ }\textbf {\bibinfo {volume} {92}},\ \bibinfo
  {pages} {104103} (\bibinfo {year} {2015})}\BibitemShut {NoStop}%
\bibitem [{\citenamefont {Geng}\ and\ \citenamefont
  {Wu}(2016)}]{geng_predicted_2016}%
  \BibitemOpen
  \bibfield  {author} {\bibinfo {author} {\bibfnamefont {H.~Y.}\ \bibnamefont
  {Geng}}\ and\ \bibinfo {author} {\bibfnamefont {Q.}~\bibnamefont {Wu}},\
  }\href {\doibase 10.1038/srep36745} {\bibfield  {journal} {\bibinfo
  {journal} {Sci. Rep.}\ }\textbf {\bibinfo {volume} {6}},\ \bibinfo {pages}
  {36745} (\bibinfo {year} {2016})}\BibitemShut {NoStop}%
\bibitem [{\citenamefont {Li}\ \emph {et~al.}(2015)\citenamefont {Li},
  \citenamefont {Chen}, \citenamefont {Li}, \citenamefont {Wang},\ and\
  \citenamefont {Xu}}]{li_supercritical_2015}%
  \BibitemOpen
  \bibfield  {author} {\bibinfo {author} {\bibfnamefont {R.}~\bibnamefont
  {Li}}, \bibinfo {author} {\bibfnamefont {J.}~\bibnamefont {Chen}}, \bibinfo
  {author} {\bibfnamefont {X.}~\bibnamefont {Li}}, \bibinfo {author}
  {\bibfnamefont {E.}~\bibnamefont {Wang}}, \ and\ \bibinfo {author}
  {\bibfnamefont {L.}~\bibnamefont {Xu}},\ }\href {\doibase
  10.1088/1367-2630/17/6/063023} {\bibfield  {journal} {\bibinfo  {journal}
  {New J. Phys.}\ }\textbf {\bibinfo {volume} {17}},\ \bibinfo {pages} {063023}
  (\bibinfo {year} {2015})}\BibitemShut {NoStop}%
\bibitem [{\citenamefont {Vidali}(2013)}]{Interstellar_2013}%
  \BibitemOpen
  \bibfield  {author} {\bibinfo {author} {\bibfnamefont {G.}~\bibnamefont
  {Vidali}},\ }\href {\doibase 10.1021/cr400156b} {\bibfield  {journal}
  {\bibinfo  {journal} {Chem. Rev.}\ }\textbf {\bibinfo {volume} {113}},\
  \bibinfo {pages} {8762} (\bibinfo {year} {2013})}\BibitemShut {NoStop}%
\bibitem [{\citenamefont {Hama}\ and\ \citenamefont
  {Watanabe}(2013)}]{Interstellar_2}%
  \BibitemOpen
  \bibfield  {author} {\bibinfo {author} {\bibfnamefont {T.}~\bibnamefont
  {Hama}}\ and\ \bibinfo {author} {\bibfnamefont {N.}~\bibnamefont
  {Watanabe}},\ }\href {\doibase 10.1021/cr4000978} {\bibfield  {journal}
  {\bibinfo  {journal} {Chem. Rev.}\ }\textbf {\bibinfo {volume} {113}},\
  \bibinfo {pages} {8783} (\bibinfo {year} {2013})}\BibitemShut {NoStop}%
\bibitem [{\citenamefont {Lachawiec}\ \emph {et~al.}(2005)\citenamefont
  {Lachawiec}, \citenamefont {Qi},\ and\ \citenamefont
  {Yang}}]{doi:10.1021/la051659r}%
  \BibitemOpen
  \bibfield  {author} {\bibinfo {author} {\bibfnamefont {A.~J.}\ \bibnamefont
  {Lachawiec}}, \bibinfo {author} {\bibfnamefont {G.}~\bibnamefont {Qi}}, \
  and\ \bibinfo {author} {\bibfnamefont {R.~T.}\ \bibnamefont {Yang}},\ }\href
  {\doibase 10.1021/la051659r} {\bibfield  {journal} {\bibinfo  {journal}
  {Langmuir}\ }\textbf {\bibinfo {volume} {21}},\ \bibinfo {pages} {11418}
  (\bibinfo {year} {2005})}\BibitemShut {NoStop}%
\bibitem [{\citenamefont {Shytov}\ \emph {et~al.}(2009)\citenamefont {Shytov},
  \citenamefont {Abanin},\ and\ \citenamefont
  {Levitov}}]{PhysRevLett.103.016806}%
  \BibitemOpen
  \bibfield  {author} {\bibinfo {author} {\bibfnamefont {A.~V.}\ \bibnamefont
  {Shytov}}, \bibinfo {author} {\bibfnamefont {D.~A.}\ \bibnamefont {Abanin}},
  \ and\ \bibinfo {author} {\bibfnamefont {L.~S.}\ \bibnamefont {Levitov}},\
  }\href {\doibase 10.1103/PhysRevLett.103.016806} {\bibfield  {journal}
  {\bibinfo  {journal} {Phys. Rev. Lett.}\ }\textbf {\bibinfo {volume} {103}},\
  \bibinfo {pages} {016806} (\bibinfo {year} {2009})}\BibitemShut {NoStop}%
\bibitem [{\citenamefont {Elias}\ \emph {et~al.}(2009)\citenamefont {Elias},
  \citenamefont {Nair}, \citenamefont {Mohiuddin}, \citenamefont {Morozov},
  \citenamefont {Blake}, \citenamefont {Halsall}, \citenamefont {Ferrari},
  \citenamefont {Boukhvalov}, \citenamefont {Katsnelson}, \citenamefont
  {Geim},\ and\ \citenamefont {Novoselov}}]{Elias610}%
  \BibitemOpen
  \bibfield  {author} {\bibinfo {author} {\bibfnamefont {D.~C.}\ \bibnamefont
  {Elias}}, \bibinfo {author} {\bibfnamefont {R.~R.}\ \bibnamefont {Nair}},
  \bibinfo {author} {\bibfnamefont {T.~M.~G.}\ \bibnamefont {Mohiuddin}},
  \bibinfo {author} {\bibfnamefont {S.~V.}\ \bibnamefont {Morozov}}, \bibinfo
  {author} {\bibfnamefont {P.}~\bibnamefont {Blake}}, \bibinfo {author}
  {\bibfnamefont {M.~P.}\ \bibnamefont {Halsall}}, \bibinfo {author}
  {\bibfnamefont {A.~C.}\ \bibnamefont {Ferrari}}, \bibinfo {author}
  {\bibfnamefont {D.~W.}\ \bibnamefont {Boukhvalov}}, \bibinfo {author}
  {\bibfnamefont {M.~I.}\ \bibnamefont {Katsnelson}}, \bibinfo {author}
  {\bibfnamefont {A.~K.}\ \bibnamefont {Geim}}, \ and\ \bibinfo {author}
  {\bibfnamefont {K.~S.}\ \bibnamefont {Novoselov}},\ }\href {\doibase
  10.1126/science.1167130} {\bibfield  {journal} {\bibinfo  {journal}
  {Science}\ }\textbf {\bibinfo {volume} {323}},\ \bibinfo {pages} {610}
  (\bibinfo {year} {2009})}\BibitemShut {NoStop}%
\bibitem [{\citenamefont {Sint}\ \emph {et~al.}(2008)\citenamefont {Sint},
  \citenamefont {Wang},\ and\ \citenamefont {Kr\'al}}]{doi:10.1021/ja804409f}%
  \BibitemOpen
  \bibfield  {author} {\bibinfo {author} {\bibfnamefont {K.}~\bibnamefont
  {Sint}}, \bibinfo {author} {\bibfnamefont {B.}~\bibnamefont {Wang}}, \ and\
  \bibinfo {author} {\bibfnamefont {P.}~\bibnamefont {Kr\'al}},\ }\href
  {\doibase 10.1021/ja804409f} {\bibfield  {journal} {\bibinfo  {journal} {J.
  Am. Chem. Soc.}\ }\textbf {\bibinfo {volume} {130}},\ \bibinfo {pages}
  {16448} (\bibinfo {year} {2008})}\BibitemShut {NoStop}%
\bibitem [{\citenamefont {Hornek\ae{}r}\ \emph {et~al.}(2006)\citenamefont
  {Hornek\ae{}r}, \citenamefont {Rauls}, \citenamefont {Xu}, \citenamefont
  {\v{Z}. \v{S}ljivan\v{c}anin}, \citenamefont {Otero}, \citenamefont
  {Stensgaard}, \citenamefont {L\ae{}gsgaard}, \citenamefont {Hammer},\ and\
  \citenamefont {Besenbacher}}]{Hgra_PRL}%
  \BibitemOpen
  \bibfield  {author} {\bibinfo {author} {\bibfnamefont {L.}~\bibnamefont
  {Hornek\ae{}r}}, \bibinfo {author} {\bibfnamefont {E.}~\bibnamefont {Rauls}},
  \bibinfo {author} {\bibfnamefont {W.}~\bibnamefont {Xu}}, \bibinfo {author}
  {\bibnamefont {\v{Z}. \v{S}ljivan\v{c}anin}}, \bibinfo {author}
  {\bibfnamefont {R.}~\bibnamefont {Otero}}, \bibinfo {author} {\bibfnamefont
  {I.}~\bibnamefont {Stensgaard}}, \bibinfo {author} {\bibfnamefont
  {E.}~\bibnamefont {L\ae{}gsgaard}}, \bibinfo {author} {\bibfnamefont
  {B.}~\bibnamefont {Hammer}}, \ and\ \bibinfo {author} {\bibfnamefont
  {F.}~\bibnamefont {Besenbacher}},\ }\href {\doibase
  10.1103/PhysRevLett.97.186102} {\bibfield  {journal} {\bibinfo  {journal}
  {Phys. Rev. Lett.}\ }\textbf {\bibinfo {volume} {97}},\ \bibinfo {pages}
  {186102} (\bibinfo {year} {2006})}\BibitemShut {NoStop}%
\bibitem [{\citenamefont {Goumans}\ and\ \citenamefont
  {K\"{a}stner}(2010)}]{Kastner_2010}%
  \BibitemOpen
  \bibfield  {author} {\bibinfo {author} {\bibfnamefont {T.~P.~M.}\
  \bibnamefont {Goumans}}\ and\ \bibinfo {author} {\bibfnamefont
  {J.}~\bibnamefont {K\"{a}stner}},\ }\href {\doibase 10.1002/anie.201001311}
  {\bibfield  {journal} {\bibinfo  {journal} {Angew. Chem. Int. Ed.}\ }\textbf
  {\bibinfo {volume} {49}},\ \bibinfo {pages} {7350} (\bibinfo {year}
  {2010})}\BibitemShut {NoStop}%
\bibitem [{\citenamefont {Jeloaica}\ and\ \citenamefont
  {Sidis}(1999)}]{JELOAICA1999157}%
  \BibitemOpen
  \bibfield  {author} {\bibinfo {author} {\bibfnamefont {L.}~\bibnamefont
  {Jeloaica}}\ and\ \bibinfo {author} {\bibfnamefont {V.}~\bibnamefont
  {Sidis}},\ }\href {\doibase https://doi.org/10.1016/S0009-2614(98)01337-2}
  {\bibfield  {journal} {\bibinfo  {journal} {Chem. Phys. Lett.}\ }\textbf
  {\bibinfo {volume} {300}},\ \bibinfo {pages} {157 } (\bibinfo {year}
  {1999})}\BibitemShut {NoStop}%
\bibitem [{\citenamefont {Sha}\ \emph {et~al.}(2002)\citenamefont {Sha},
  \citenamefont {Jackson},\ and\ \citenamefont
  {Lemoine}}]{doi:10.1063/1.1463399}%
  \BibitemOpen
  \bibfield  {author} {\bibinfo {author} {\bibfnamefont {X.}~\bibnamefont
  {Sha}}, \bibinfo {author} {\bibfnamefont {B.}~\bibnamefont {Jackson}}, \ and\
  \bibinfo {author} {\bibfnamefont {D.}~\bibnamefont {Lemoine}},\ }\href
  {\doibase 10.1063/1.1463399} {\bibfield  {journal} {\bibinfo  {journal} {J.
  Chem. Phys.}\ }\textbf {\bibinfo {volume} {116}},\ \bibinfo {pages} {7158}
  (\bibinfo {year} {2002})}\BibitemShut {NoStop}%
\bibitem [{\citenamefont {Ivanovskaya}\ \emph {et~al.}(2010)\citenamefont
  {Ivanovskaya}, \citenamefont {Zobelli}, \citenamefont {Teillet-Billy},
  \citenamefont {Rougeau}, \citenamefont {Sidis},\ and\ \citenamefont
  {Briddon}}]{EPJB_Hadgra}%
  \BibitemOpen
  \bibfield  {author} {\bibinfo {author} {\bibfnamefont {V.~V.}\ \bibnamefont
  {Ivanovskaya}}, \bibinfo {author} {\bibfnamefont {A.}~\bibnamefont
  {Zobelli}}, \bibinfo {author} {\bibfnamefont {D.}~\bibnamefont
  {Teillet-Billy}}, \bibinfo {author} {\bibfnamefont {N.}~\bibnamefont
  {Rougeau}}, \bibinfo {author} {\bibfnamefont {V.}~\bibnamefont {Sidis}}, \
  and\ \bibinfo {author} {\bibfnamefont {P.~R.}\ \bibnamefont {Briddon}},\
  }\href {\doibase 10.1140/epjb/e2010-00238-7} {\bibfield  {journal} {\bibinfo
  {journal} {Eur. Phys. J. B}\ }\textbf {\bibinfo {volume} {76}},\ \bibinfo
  {pages} {481} (\bibinfo {year} {2010})}\BibitemShut {NoStop}%
\bibitem [{\citenamefont {Hornek{\ae}r}\ \emph {et~al.}(2003)\citenamefont
  {Hornek{\ae}r}, \citenamefont {Baurichter}, \citenamefont {Petrunin},
  \citenamefont {Field},\ and\ \citenamefont {Luntz}}]{Hornekaer1943}%
  \BibitemOpen
  \bibfield  {author} {\bibinfo {author} {\bibfnamefont {L.}~\bibnamefont
  {Hornek{\ae}r}}, \bibinfo {author} {\bibfnamefont {A.}~\bibnamefont
  {Baurichter}}, \bibinfo {author} {\bibfnamefont {V.~V.}\ \bibnamefont
  {Petrunin}}, \bibinfo {author} {\bibfnamefont {D.}~\bibnamefont {Field}}, \
  and\ \bibinfo {author} {\bibfnamefont {A.~C.}\ \bibnamefont {Luntz}},\ }\href
  {\doibase 10.1126/science.1090820} {\bibfield  {journal} {\bibinfo  {journal}
  {Science}\ }\textbf {\bibinfo {volume} {302}},\ \bibinfo {pages} {1943}
  (\bibinfo {year} {2003})}\BibitemShut {NoStop}%
\bibitem [{\citenamefont {Davidson}\ \emph {et~al.}(2014)\citenamefont
  {Davidson}, \citenamefont {Klime\v{s}}, \citenamefont {Alf\`e},\ and\
  \citenamefont {Michaelides}}]{Davidson_2014_1}%
  \BibitemOpen
  \bibfield  {author} {\bibinfo {author} {\bibfnamefont {E.~R.~M.}\
  \bibnamefont {Davidson}}, \bibinfo {author} {\bibfnamefont {J.}~\bibnamefont
  {Klime\v{s}}}, \bibinfo {author} {\bibfnamefont {D.}~\bibnamefont {Alf\`e}},
  \ and\ \bibinfo {author} {\bibfnamefont {A.}~\bibnamefont {Michaelides}},\
  }\href {\doibase 10.1021/nn505578x} {\bibfield  {journal} {\bibinfo
  {journal} {ACS Nano}\ }\textbf {\bibinfo {volume} {8}},\ \bibinfo {pages}
  {9905} (\bibinfo {year} {2014})}\BibitemShut {NoStop}%
\bibitem [{\citenamefont {Bonfanti}\ \emph {et~al.}(2015)\citenamefont
  {Bonfanti}, \citenamefont {Jackson}, \citenamefont {Hughes}, \citenamefont
  {Burghardt},\ and\ \citenamefont {Martinazzo}}]{doi:10.1063/1.4931117}%
  \BibitemOpen
  \bibfield  {author} {\bibinfo {author} {\bibfnamefont {M.}~\bibnamefont
  {Bonfanti}}, \bibinfo {author} {\bibfnamefont {B.}~\bibnamefont {Jackson}},
  \bibinfo {author} {\bibfnamefont {K.~H.}\ \bibnamefont {Hughes}}, \bibinfo
  {author} {\bibfnamefont {I.}~\bibnamefont {Burghardt}}, \ and\ \bibinfo
  {author} {\bibfnamefont {R.}~\bibnamefont {Martinazzo}},\ }\href {\doibase
  10.1063/1.4931117} {\bibfield  {journal} {\bibinfo  {journal} {J. Chem.
  Phys.}\ }\textbf {\bibinfo {volume} {143}},\ \bibinfo {pages} {124704}
  (\bibinfo {year} {2015})}\BibitemShut {NoStop}%
\bibitem [{\citenamefont {Casolo}\ \emph {et~al.}(2016)\citenamefont {Casolo},
  \citenamefont {Tantardini},\ and\ \citenamefont
  {Martinazzo}}]{doi:10.1021/acs.jpca.5b12761}%
  \BibitemOpen
  \bibfield  {author} {\bibinfo {author} {\bibfnamefont {S.}~\bibnamefont
  {Casolo}}, \bibinfo {author} {\bibfnamefont {G.~F.}\ \bibnamefont
  {Tantardini}}, \ and\ \bibinfo {author} {\bibfnamefont {R.}~\bibnamefont
  {Martinazzo}},\ }\href {\doibase 10.1021/acs.jpca.5b12761} {\bibfield
  {journal} {\bibinfo  {journal} {J. Phys. Chem. A}\ }\textbf {\bibinfo
  {volume} {120}},\ \bibinfo {pages} {5032} (\bibinfo {year}
  {2016})}\BibitemShut {NoStop}%
\bibitem [{\citenamefont {Ar\'eou}\ \emph {et~al.}(2011)\citenamefont
  {Ar\'eou}, \citenamefont {Cartry}, \citenamefont {Layet},\ and\ \citenamefont
  {Angot}}]{doi:10.1063/1.3518981}%
  \BibitemOpen
  \bibfield  {author} {\bibinfo {author} {\bibfnamefont {E.}~\bibnamefont
  {Ar\'eou}}, \bibinfo {author} {\bibfnamefont {G.}~\bibnamefont {Cartry}},
  \bibinfo {author} {\bibfnamefont {J.-M.}\ \bibnamefont {Layet}}, \ and\
  \bibinfo {author} {\bibfnamefont {T.}~\bibnamefont {Angot}},\ }\href
  {\doibase 10.1063/1.3518981} {\bibfield  {journal} {\bibinfo  {journal} {J.
  Chem. Phys.}\ }\textbf {\bibinfo {volume} {134}},\ \bibinfo {pages} {014701}
  (\bibinfo {year} {2011})}\BibitemShut {NoStop}%
\bibitem [{\citenamefont {Sims}(2013)}]{sims_low-temperature_2013}%
  \BibitemOpen
  \bibfield  {author} {\bibinfo {author} {\bibfnamefont {I.~R.}\ \bibnamefont
  {Sims}},\ }\href {\doibase 10.1038/nchem.1736} {\bibfield  {journal}
  {\bibinfo  {journal} {Nat. Chem.}\ }\textbf {\bibinfo {volume} {5}},\
  \bibinfo {pages} {734} (\bibinfo {year} {2013})}\BibitemShut {NoStop}%
\bibitem [{\citenamefont {Sims}\ and\ \citenamefont
  {Smith}(1995)}]{doi:10.1146/annurev.pc.46.100195.000545}%
  \BibitemOpen
  \bibfield  {author} {\bibinfo {author} {\bibfnamefont {I.~R.}\ \bibnamefont
  {Sims}}\ and\ \bibinfo {author} {\bibfnamefont {I.~W.~M.}\ \bibnamefont
  {Smith}},\ }\href {\doibase 10.1146/annurev.pc.46.100195.000545} {\bibfield
  {journal} {\bibinfo  {journal} {Annu. Rev. Phys. Chem.}\ }\textbf {\bibinfo
  {volume} {46}},\ \bibinfo {pages} {109} (\bibinfo {year} {1995})}\BibitemShut
  {NoStop}%
\bibitem [{\citenamefont {Graetz}(2009)}]{B718842K}%
  \BibitemOpen
  \bibfield  {author} {\bibinfo {author} {\bibfnamefont {J.}~\bibnamefont
  {Graetz}},\ }\href {\doibase 10.1039/B718842K} {\bibfield  {journal}
  {\bibinfo  {journal} {Chem. Soc. Rev.}\ }\textbf {\bibinfo {volume} {38}},\
  \bibinfo {pages} {73} (\bibinfo {year} {2009})}\BibitemShut {NoStop}%
\bibitem [{\citenamefont {Lopez}\ \emph {et~al.}(2004)\citenamefont {Lopez},
  \citenamefont {Lodziana}, \citenamefont {Illas},\ and\ \citenamefont
  {Salmeron}}]{lopez_when_2004}%
  \BibitemOpen
  \bibfield  {author} {\bibinfo {author} {\bibfnamefont {N.}~\bibnamefont
  {Lopez}}, \bibinfo {author} {\bibfnamefont {Z.}~\bibnamefont {Lodziana}},
  \bibinfo {author} {\bibfnamefont {F.}~\bibnamefont {Illas}}, \ and\ \bibinfo
  {author} {\bibfnamefont {M.}~\bibnamefont {Salmeron}},\ }\href {\doibase
  10.1103/PhysRevLett.93.146103} {\bibfield  {journal} {\bibinfo  {journal}
  {Phys. Rev. Lett.}\ }\textbf {\bibinfo {volume} {93}},\ \bibinfo {pages}
  {146103} (\bibinfo {year} {2004})}\BibitemShut {NoStop}%
\bibitem [{\citenamefont {Teschner}\ \emph {et~al.}(2008)\citenamefont
  {Teschner}, \citenamefont {Borsodi}, \citenamefont {Wootsch}, \citenamefont
  {Révay}, \citenamefont {Hävecker}, \citenamefont {Knop-Gericke},
  \citenamefont {Jackson},\ and\ \citenamefont
  {Schl\'ogl}}]{teschner_roles_2008}%
  \BibitemOpen
  \bibfield  {author} {\bibinfo {author} {\bibfnamefont {D.}~\bibnamefont
  {Teschner}}, \bibinfo {author} {\bibfnamefont {J.}~\bibnamefont {Borsodi}},
  \bibinfo {author} {\bibfnamefont {A.}~\bibnamefont {Wootsch}}, \bibinfo
  {author} {\bibfnamefont {Z.}~\bibnamefont {Révay}}, \bibinfo {author}
  {\bibfnamefont {M.}~\bibnamefont {Hävecker}}, \bibinfo {author}
  {\bibfnamefont {A.}~\bibnamefont {Knop-Gericke}}, \bibinfo {author}
  {\bibfnamefont {S.~D.}\ \bibnamefont {Jackson}}, \ and\ \bibinfo {author}
  {\bibfnamefont {R.}~\bibnamefont {Schl\'ogl}},\ }\href {\doibase
  10.1126/science.1155200} {\bibfield  {journal} {\bibinfo  {journal}
  {Science}\ }\textbf {\bibinfo {volume} {320}},\ \bibinfo {pages} {86}
  (\bibinfo {year} {2008})}\BibitemShut {NoStop}%
\bibitem [{\citenamefont {Sakintuna}\ \emph {et~al.}(2007)\citenamefont
  {Sakintuna}, \citenamefont {Lamari-Darkrim},\ and\ \citenamefont
  {Hirscher}}]{SAKINTUNA20071121}%
  \BibitemOpen
  \bibfield  {author} {\bibinfo {author} {\bibfnamefont {B.}~\bibnamefont
  {Sakintuna}}, \bibinfo {author} {\bibfnamefont {F.}~\bibnamefont
  {Lamari-Darkrim}}, \ and\ \bibinfo {author} {\bibfnamefont {M.}~\bibnamefont
  {Hirscher}},\ }\href {\doibase
  https://doi.org/10.1016/j.ijhydene.2006.11.022} {\bibfield  {journal}
  {\bibinfo  {journal} {Int. J. Hydro. Ener.}\ }\textbf {\bibinfo {volume}
  {32}},\ \bibinfo {pages} {1121 } (\bibinfo {year} {2007})}\BibitemShut
  {NoStop}%
\bibitem [{\citenamefont {Tierney}\ \emph {et~al.}(2009)\citenamefont
  {Tierney}, \citenamefont {Baber}, \citenamefont {Kitchin},\ and\
  \citenamefont {Sykes}}]{PhysRevLett.103.246102}%
  \BibitemOpen
  \bibfield  {author} {\bibinfo {author} {\bibfnamefont {H.~L.}\ \bibnamefont
  {Tierney}}, \bibinfo {author} {\bibfnamefont {A.~E.}\ \bibnamefont {Baber}},
  \bibinfo {author} {\bibfnamefont {J.~R.}\ \bibnamefont {Kitchin}}, \ and\
  \bibinfo {author} {\bibfnamefont {E.~C.~H.}\ \bibnamefont {Sykes}},\ }\href
  {\doibase 10.1103/PhysRevLett.103.246102} {\bibfield  {journal} {\bibinfo
  {journal} {Phys. Rev. Lett.}\ }\textbf {\bibinfo {volume} {103}},\ \bibinfo
  {pages} {246102} (\bibinfo {year} {2009})}\BibitemShut {NoStop}%
\bibitem [{\citenamefont {Darby}\ \emph
  {et~al.}(2018{\natexlab{a}})\citenamefont {Darby}, \citenamefont
  {Stamatakis}, \citenamefont {Michaelides},\ and\ \citenamefont
  {Sykes}}]{doi:10.1021/acs.jpclett.8b01888}%
  \BibitemOpen
  \bibfield  {author} {\bibinfo {author} {\bibfnamefont {M.~T.}\ \bibnamefont
  {Darby}}, \bibinfo {author} {\bibfnamefont {M.}~\bibnamefont {Stamatakis}},
  \bibinfo {author} {\bibfnamefont {A.}~\bibnamefont {Michaelides}}, \ and\
  \bibinfo {author} {\bibfnamefont {E.~C.~H.}\ \bibnamefont {Sykes}},\ }\href
  {\doibase 10.1021/acs.jpclett.8b01888} {\bibfield  {journal} {\bibinfo
  {journal} {J. Phys. Chem. Lett.}\ }\textbf {\bibinfo {volume} {9}},\ \bibinfo
  {pages} {5636} (\bibinfo {year} {2018}{\natexlab{a}})}\BibitemShut {NoStop}%
\bibitem [{\citenamefont {Darby}\ \emph
  {et~al.}(2018{\natexlab{b}})\citenamefont {Darby}, \citenamefont
  {R\'eocreux}, \citenamefont {Sykes}, \citenamefont {Michaelides},\ and\
  \citenamefont {Stamatakis}}]{doi:10.1021/acscatal.8b00881}%
  \BibitemOpen
  \bibfield  {author} {\bibinfo {author} {\bibfnamefont {M.~T.}\ \bibnamefont
  {Darby}}, \bibinfo {author} {\bibfnamefont {R.}~\bibnamefont {R\'eocreux}},
  \bibinfo {author} {\bibfnamefont {E.~C.~H.}\ \bibnamefont {Sykes}}, \bibinfo
  {author} {\bibfnamefont {A.}~\bibnamefont {Michaelides}}, \ and\ \bibinfo
  {author} {\bibfnamefont {M.}~\bibnamefont {Stamatakis}},\ }\href {\doibase
  10.1021/acscatal.8b00881} {\bibfield  {journal} {\bibinfo  {journal} {ACS
  Catal.}\ }\textbf {\bibinfo {volume} {8}},\ \bibinfo {pages} {5038} (\bibinfo
  {year} {2018}{\natexlab{b}})}\BibitemShut {NoStop}%
\bibitem [{\citenamefont {Darby}\ \emph
  {et~al.}(2018{\natexlab{c}})\citenamefont {Darby}, \citenamefont {Liu},
  \citenamefont {Wimble}, \citenamefont {Lucci}, \citenamefont {Lee},
  \citenamefont {Michaelides}, \citenamefont {Flytzani-Stephanopoulos},
  \citenamefont {Stamatakis},\ and\ \citenamefont
  {Sykes}}]{doi:10.1038/nchem.2915}%
  \BibitemOpen
  \bibfield  {author} {\bibinfo {author} {\bibfnamefont {M.~T.}\ \bibnamefont
  {Darby}}, \bibinfo {author} {\bibfnamefont {J.}~\bibnamefont {Liu}}, \bibinfo
  {author} {\bibfnamefont {J.~M.}\ \bibnamefont {Wimble}}, \bibinfo {author}
  {\bibfnamefont {F.~R.}\ \bibnamefont {Lucci}}, \bibinfo {author}
  {\bibfnamefont {S.}~\bibnamefont {Lee}}, \bibinfo {author} {\bibfnamefont
  {A.}~\bibnamefont {Michaelides}}, \bibinfo {author} {\bibfnamefont
  {M.}~\bibnamefont {Flytzani-Stephanopoulos}}, \bibinfo {author}
  {\bibfnamefont {M.}~\bibnamefont {Stamatakis}}, \ and\ \bibinfo {author}
  {\bibfnamefont {E.~C.~H.}\ \bibnamefont {Sykes}},\ }\href {\doibase
  10.1038/nchem.2915} {\bibfield  {journal} {\bibinfo  {journal} {Nat. Chem.}\
  }\textbf {\bibinfo {volume} {10}},\ \bibinfo {pages} {325} (\bibinfo {year}
  {2018}{\natexlab{c}})}\BibitemShut {NoStop}%
\bibitem [{\citenamefont {Martinazzo}\ \emph {et~al.}(2013)\citenamefont
  {Martinazzo}, \citenamefont {Casolo},\ and\ \citenamefont
  {Hornek{\ae}r}}]{martinazzo_hydrogen_2013}%
  \BibitemOpen
  \bibfield  {author} {\bibinfo {author} {\bibfnamefont {R.}~\bibnamefont
  {Martinazzo}}, \bibinfo {author} {\bibfnamefont {S.}~\bibnamefont {Casolo}},
  \ and\ \bibinfo {author} {\bibfnamefont {L.~H.}\ \bibnamefont
  {Hornek{\ae}r}},\ }\href@noop {} {\emph {\bibinfo {title} {Dynamics of
  {Gas}-{Surface} {Interactions}}}}\ (\bibinfo  {publisher} {Springer, Berlin,
  Heidelberg},\ \bibinfo {year} {2013})\BibitemShut {NoStop}%
\bibitem [{\citenamefont {D\"urr}\ and\ \citenamefont
  {H\"ofer}(2013)}]{DURR201361}%
  \BibitemOpen
  \bibfield  {author} {\bibinfo {author} {\bibfnamefont {M.}~\bibnamefont
  {D\"urr}}\ and\ \bibinfo {author} {\bibfnamefont {U.}~\bibnamefont
  {H\"ofer}},\ }\href {\doibase https://doi.org/10.1016/j.progsurf.2013.01.001}
  {\bibfield  {journal} {\bibinfo  {journal} {Prog. Surf. Sci.}\ }\textbf
  {\bibinfo {volume} {88}},\ \bibinfo {pages} {61 } (\bibinfo {year}
  {2013})}\BibitemShut {NoStop}%
\bibitem [{\citenamefont {Huang}\ \emph {et~al.}(2010)\citenamefont {Huang},
  \citenamefont {Ni}, \citenamefont {Zheng}, \citenamefont {Zhou},
  \citenamefont {Li},\ and\ \citenamefont {Zeng}}]{JPCC_Hdiff_gra}%
  \BibitemOpen
  \bibfield  {author} {\bibinfo {author} {\bibfnamefont {L.~F.}\ \bibnamefont
  {Huang}}, \bibinfo {author} {\bibfnamefont {M.~Y.}\ \bibnamefont {Ni}},
  \bibinfo {author} {\bibfnamefont {X.~H.}\ \bibnamefont {Zheng}}, \bibinfo
  {author} {\bibfnamefont {W.~H.}\ \bibnamefont {Zhou}}, \bibinfo {author}
  {\bibfnamefont {Y.~G.}\ \bibnamefont {Li}}, \ and\ \bibinfo {author}
  {\bibfnamefont {Z.}~\bibnamefont {Zeng}},\ }\href {\doibase
  10.1021/jp109160c} {\bibfield  {journal} {\bibinfo  {journal} {J. Phys. Chem.
  C}\ }\textbf {\bibinfo {volume} {114}},\ \bibinfo {pages} {22636} (\bibinfo
  {year} {2010})}\BibitemShut {NoStop}%
\bibitem [{\citenamefont {Herrero}\ and\ \citenamefont
  {Ram\'{\i}rez}(2009)}]{PhysRevB.79.115429}%
  \BibitemOpen
  \bibfield  {author} {\bibinfo {author} {\bibfnamefont {C.~P.}\ \bibnamefont
  {Herrero}}\ and\ \bibinfo {author} {\bibfnamefont {R.}~\bibnamefont
  {Ram\'{\i}rez}},\ }\href {\doibase 10.1103/PhysRevB.79.115429} {\bibfield
  {journal} {\bibinfo  {journal} {Phys. Rev. B}\ }\textbf {\bibinfo {volume}
  {79}},\ \bibinfo {pages} {115429} (\bibinfo {year} {2009})}\BibitemShut
  {NoStop}%
\bibitem [{\citenamefont {Castelli}\ \emph {et~al.}(2018)\citenamefont
  {Castelli}, \citenamefont {Soriga},\ and\ \citenamefont
  {Man}}]{doi:10.1063/1.5029329}%
  \BibitemOpen
  \bibfield  {author} {\bibinfo {author} {\bibfnamefont {I.~E.}\ \bibnamefont
  {Castelli}}, \bibinfo {author} {\bibfnamefont {S.~G.}\ \bibnamefont
  {Soriga}}, \ and\ \bibinfo {author} {\bibfnamefont {I.~C.}\ \bibnamefont
  {Man}},\ }\href {\doibase 10.1063/1.5029329} {\bibfield  {journal} {\bibinfo
  {journal} {J. Chem. Phys.}\ }\textbf {\bibinfo {volume} {149}},\ \bibinfo
  {pages} {034704} (\bibinfo {year} {2018})}\BibitemShut {NoStop}%
\bibitem [{\citenamefont {Kristinsd\'ottir}\ and\ \citenamefont
  {Sk\'ulason}(2012)}]{Skulason20121400}%
  \BibitemOpen
  \bibfield  {author} {\bibinfo {author} {\bibfnamefont {L.}~\bibnamefont
  {Kristinsd\'ottir}}\ and\ \bibinfo {author} {\bibfnamefont {E.}~\bibnamefont
  {Sk\'ulason}},\ }\href {\doibase
  http://dx.doi.org/10.1016/j.susc.2012.04.028} {\bibfield  {journal} {\bibinfo
   {journal} {Surf. Sci.}\ }\textbf {\bibinfo {volume} {606}},\ \bibinfo
  {pages} {1400} (\bibinfo {year} {2012})}\BibitemShut {NoStop}%
\bibitem [{\citenamefont {Lin}\ and\ \citenamefont
  {Gomer}(1991)}]{lin_diffusion_1991}%
  \BibitemOpen
  \bibfield  {author} {\bibinfo {author} {\bibfnamefont {T.~S.}\ \bibnamefont
  {Lin}}\ and\ \bibinfo {author} {\bibfnamefont {R.}~\bibnamefont {Gomer}},\
  }\href {\doibase 10.1016/0039-6028(91)90010-P} {\bibfield  {journal}
  {\bibinfo  {journal} {Surf. Sci.}\ }\textbf {\bibinfo {volume} {255}},\
  \bibinfo {pages} {41} (\bibinfo {year} {1991})}\BibitemShut {NoStop}%
\bibitem [{\citenamefont {McIntosh}\ \emph {et~al.}(2013)\citenamefont
  {McIntosh}, \citenamefont {Wikfeldt}, \citenamefont {Ellis}, \citenamefont
  {Michaelides},\ and\ \citenamefont {Allison}}]{Ru_2013}%
  \BibitemOpen
  \bibfield  {author} {\bibinfo {author} {\bibfnamefont {E.~M.}\ \bibnamefont
  {McIntosh}}, \bibinfo {author} {\bibfnamefont {K.~T.}\ \bibnamefont
  {Wikfeldt}}, \bibinfo {author} {\bibfnamefont {J.}~\bibnamefont {Ellis}},
  \bibinfo {author} {\bibfnamefont {A.}~\bibnamefont {Michaelides}}, \ and\
  \bibinfo {author} {\bibfnamefont {W.}~\bibnamefont {Allison}},\ }\href
  {\doibase 10.1021/jz400622v} {\bibfield  {journal} {\bibinfo  {journal} {J.
  Phys. Chem. Lett.}\ }\textbf {\bibinfo {volume} {4}},\ \bibinfo {pages}
  {1565} (\bibinfo {year} {2013})}\BibitemShut {NoStop}%
\bibitem [{\citenamefont {Michaelides}\ \emph {et~al.}(1999)\citenamefont
  {Michaelides}, \citenamefont {Hu},\ and\ \citenamefont
  {Alavi}}]{doi:10.1063/1.479392}%
  \BibitemOpen
  \bibfield  {author} {\bibinfo {author} {\bibfnamefont {A.}~\bibnamefont
  {Michaelides}}, \bibinfo {author} {\bibfnamefont {P.}~\bibnamefont {Hu}}, \
  and\ \bibinfo {author} {\bibfnamefont {A.}~\bibnamefont {Alavi}},\ }\href
  {\doibase 10.1063/1.479392} {\bibfield  {journal} {\bibinfo  {journal} {J.
  Chem. Phys.}\ }\textbf {\bibinfo {volume} {111}},\ \bibinfo {pages} {1343}
  (\bibinfo {year} {1999})}\BibitemShut {NoStop}%
\bibitem [{\citenamefont {Zhang}\ and\ \citenamefont
  {Michaelides}(2011)}]{ZHANG2011689}%
  \BibitemOpen
  \bibfield  {author} {\bibinfo {author} {\bibfnamefont {C.}~\bibnamefont
  {Zhang}}\ and\ \bibinfo {author} {\bibfnamefont {A.}~\bibnamefont
  {Michaelides}},\ }\href {\doibase https://doi.org/10.1016/j.susc.2011.01.004}
  {\bibfield  {journal} {\bibinfo  {journal} {Surf. Sci.}\ }\textbf {\bibinfo
  {volume} {605}},\ \bibinfo {pages} {689 } (\bibinfo {year}
  {2011})}\BibitemShut {NoStop}%
\bibitem [{\citenamefont {Mattsson}\ \emph {et~al.}(1993)\citenamefont
  {Mattsson}, \citenamefont {Engberg},\ and\ \citenamefont
  {Wahnstr\"{o}m}}]{mattsson_h_1993}%
  \BibitemOpen
  \bibfield  {author} {\bibinfo {author} {\bibfnamefont {T.~R.}\ \bibnamefont
  {Mattsson}}, \bibinfo {author} {\bibfnamefont {U.}~\bibnamefont {Engberg}}, \
  and\ \bibinfo {author} {\bibfnamefont {G.}~\bibnamefont {Wahnstr\"{o}m}},\
  }\href {\doibase 10.1103/PhysRevLett.71.2615} {\bibfield  {journal} {\bibinfo
   {journal} {Phys. Rev. Lett.}\ }\textbf {\bibinfo {volume} {71}},\ \bibinfo
  {pages} {2615} (\bibinfo {year} {1993})}\BibitemShut {NoStop}%
\bibitem [{\citenamefont {Mattsson}\ and\ \citenamefont
  {Wahnstr\"{o}m}(1997)}]{mattsson_isotope_1997}%
  \BibitemOpen
  \bibfield  {author} {\bibinfo {author} {\bibfnamefont {T.~R.}\ \bibnamefont
  {Mattsson}}\ and\ \bibinfo {author} {\bibfnamefont {G.}~\bibnamefont
  {Wahnstr\"{o}m}},\ }\href {\doibase 10.1103/PhysRevB.56.14944} {\bibfield
  {journal} {\bibinfo  {journal} {Phys. Rev. B}\ }\textbf {\bibinfo {volume}
  {56}},\ \bibinfo {pages} {14944} (\bibinfo {year} {1997})}\BibitemShut
  {NoStop}%
\bibitem [{\citenamefont {Sundell}\ and\ \citenamefont
  {Wahnstr\"{o}m}(2004)}]{sundell_quantum_2004}%
  \BibitemOpen
  \bibfield  {author} {\bibinfo {author} {\bibfnamefont {P.~G.}\ \bibnamefont
  {Sundell}}\ and\ \bibinfo {author} {\bibfnamefont {G.}~\bibnamefont
  {Wahnstr\"{o}m}},\ }\href {\doibase 10.1103/PhysRevB.70.081403} {\bibfield
  {journal} {\bibinfo  {journal} {Phys. Rev. B}\ }\textbf {\bibinfo {volume}
  {70}},\ \bibinfo {pages} {081403} (\bibinfo {year} {2004})}\BibitemShut
  {NoStop}%
\bibitem [{\citenamefont {Sundell}\ and\ \citenamefont
  {Wahnstr\"{o}m}(2005)}]{sundell_hydrogen_2005}%
  \BibitemOpen
  \bibfield  {author} {\bibinfo {author} {\bibfnamefont {P.~G.}\ \bibnamefont
  {Sundell}}\ and\ \bibinfo {author} {\bibfnamefont {G.}~\bibnamefont
  {Wahnstr\"{o}m}},\ }\href {\doibase 10.1016/j.susc.2005.06.051} {\bibfield
  {journal} {\bibinfo  {journal} {Surf. Sci.}\ }\textbf {\bibinfo {volume}
  {593}},\ \bibinfo {pages} {102} (\bibinfo {year} {2005})}\BibitemShut
  {NoStop}%
\bibitem [{\citenamefont {Sundell}\ and\ \citenamefont
  {Wahnstr\"om}(2004)}]{Sundell_2004_2}%
  \BibitemOpen
  \bibfield  {author} {\bibinfo {author} {\bibfnamefont {P.~G.}\ \bibnamefont
  {Sundell}}\ and\ \bibinfo {author} {\bibfnamefont {G.}~\bibnamefont
  {Wahnstr\"om}},\ }\href {\doibase 10.1103/PhysRevLett.92.155901} {\bibfield
  {journal} {\bibinfo  {journal} {Phys. Rev. Lett.}\ }\textbf {\bibinfo
  {volume} {92}},\ \bibinfo {pages} {155901} (\bibinfo {year}
  {2004})}\BibitemShut {NoStop}%
\bibitem [{\citenamefont {Fang}\ \emph {et~al.}(2017)\citenamefont {Fang},
  \citenamefont {Richardson}, \citenamefont {Chen}, \citenamefont {Li},\ and\
  \citenamefont {Michaelides}}]{Wei_HD_2017}%
  \BibitemOpen
  \bibfield  {author} {\bibinfo {author} {\bibfnamefont {W.}~\bibnamefont
  {Fang}}, \bibinfo {author} {\bibfnamefont {J.~O.}\ \bibnamefont
  {Richardson}}, \bibinfo {author} {\bibfnamefont {J.}~\bibnamefont {Chen}},
  \bibinfo {author} {\bibfnamefont {X.-Z.}\ \bibnamefont {Li}}, \ and\ \bibinfo
  {author} {\bibfnamefont {A.}~\bibnamefont {Michaelides}},\ }\href {\doibase
  10.1103/PhysRevLett.119.126001} {\bibfield  {journal} {\bibinfo  {journal}
  {Phys. Rev. Lett.}\ }\textbf {\bibinfo {volume} {119}},\ \bibinfo {pages}
  {126001} (\bibinfo {year} {2017})}\BibitemShut {NoStop}%
\bibitem [{\citenamefont {Vlasov}\ \emph {et~al.}(2016)\citenamefont {Vlasov},
  \citenamefont {Bessarab}, \citenamefont {Uzdin},\ and\ \citenamefont
  {J\'onsson}}]{C6FD00136J}%
  \BibitemOpen
  \bibfield  {author} {\bibinfo {author} {\bibfnamefont {S.}~\bibnamefont
  {Vlasov}}, \bibinfo {author} {\bibfnamefont {P.~F.}\ \bibnamefont
  {Bessarab}}, \bibinfo {author} {\bibfnamefont {V.~M.}\ \bibnamefont {Uzdin}},
  \ and\ \bibinfo {author} {\bibfnamefont {H.}~\bibnamefont {J\'onsson}},\
  }\href {\doibase 10.1039/C6FD00136J} {\bibfield  {journal} {\bibinfo
  {journal} {Faraday Discuss.}\ }\textbf {\bibinfo {volume} {195}},\ \bibinfo
  {pages} {93} (\bibinfo {year} {2016})}\BibitemShut {NoStop}%
\bibitem [{\citenamefont {Miao}\ \emph {et~al.}(2013)\citenamefont {Miao},
  \citenamefont {Nardelli}, \citenamefont {Wang},\ and\ \citenamefont
  {Liu}}]{C3CP52318G}%
  \BibitemOpen
  \bibfield  {author} {\bibinfo {author} {\bibfnamefont {M.}~\bibnamefont
  {Miao}}, \bibinfo {author} {\bibfnamefont {M.~B.}\ \bibnamefont {Nardelli}},
  \bibinfo {author} {\bibfnamefont {Q.}~\bibnamefont {Wang}}, \ and\ \bibinfo
  {author} {\bibfnamefont {Y.}~\bibnamefont {Liu}},\ }\href {\doibase
  10.1039/C3CP52318G} {\bibfield  {journal} {\bibinfo  {journal} {Phys. Chem.
  Chem. Phys.}\ }\textbf {\bibinfo {volume} {15}},\ \bibinfo {pages} {16132}
  (\bibinfo {year} {2013})}\BibitemShut {NoStop}%
\bibitem [{\citenamefont {Zhang}\ \emph {et~al.}(2016)\citenamefont {Zhang},
  \citenamefont {Ju}, \citenamefont {Chen},\ and\ \citenamefont
  {Zeng}}]{doi:10.1021/acs.jpclett.6b01507}%
  \BibitemOpen
  \bibfield  {author} {\bibinfo {author} {\bibfnamefont {Q.}~\bibnamefont
  {Zhang}}, \bibinfo {author} {\bibfnamefont {M.}~\bibnamefont {Ju}}, \bibinfo
  {author} {\bibfnamefont {L.}~\bibnamefont {Chen}}, \ and\ \bibinfo {author}
  {\bibfnamefont {X.~C.}\ \bibnamefont {Zeng}},\ }\href {\doibase
  10.1021/acs.jpclett.6b01507} {\bibfield  {journal} {\bibinfo  {journal} {J.
  Phys. Chem. Lett.}\ }\textbf {\bibinfo {volume} {7}},\ \bibinfo {pages}
  {3395} (\bibinfo {year} {2016})}\BibitemShut {NoStop}%
\bibitem [{\citenamefont {Kroes}\ \emph {et~al.}(2017)\citenamefont {Kroes},
  \citenamefont {Fasolino},\ and\ \citenamefont {Katsnelson}}]{C6CP08923B}%
  \BibitemOpen
  \bibfield  {author} {\bibinfo {author} {\bibfnamefont {J.~M.~H.}\
  \bibnamefont {Kroes}}, \bibinfo {author} {\bibfnamefont {A.}~\bibnamefont
  {Fasolino}}, \ and\ \bibinfo {author} {\bibfnamefont {M.~I.}\ \bibnamefont
  {Katsnelson}},\ }\href {\doibase 10.1039/C6CP08923B} {\bibfield  {journal}
  {\bibinfo  {journal} {Phys. Chem. Chem. Phys.}\ }\textbf {\bibinfo {volume}
  {19}},\ \bibinfo {pages} {5813} (\bibinfo {year} {2017})}\BibitemShut
  {NoStop}%
\bibitem [{\citenamefont {Hu}\ \emph {et~al.}(2014)\citenamefont {Hu},
  \citenamefont {Lozada-Hidalgo}, \citenamefont {Wang}, \citenamefont
  {Mishchenko}, \citenamefont {Schedin}, \citenamefont {Nair}, \citenamefont
  {Hill}, \citenamefont {Boukhvalov}, \citenamefont {Katsnelson}, \citenamefont
  {Dryfe}, \citenamefont {Grigorieva}, \citenamefont {Wu},\ and\ \citenamefont
  {Geim}}]{hu_proton_2014}%
  \BibitemOpen
  \bibfield  {author} {\bibinfo {author} {\bibfnamefont {S.}~\bibnamefont
  {Hu}}, \bibinfo {author} {\bibfnamefont {M.}~\bibnamefont {Lozada-Hidalgo}},
  \bibinfo {author} {\bibfnamefont {F.~C.}\ \bibnamefont {Wang}}, \bibinfo
  {author} {\bibfnamefont {A.}~\bibnamefont {Mishchenko}}, \bibinfo {author}
  {\bibfnamefont {F.}~\bibnamefont {Schedin}}, \bibinfo {author} {\bibfnamefont
  {R.~R.}\ \bibnamefont {Nair}}, \bibinfo {author} {\bibfnamefont {E.~W.}\
  \bibnamefont {Hill}}, \bibinfo {author} {\bibfnamefont {D.~W.}\ \bibnamefont
  {Boukhvalov}}, \bibinfo {author} {\bibfnamefont {M.~I.}\ \bibnamefont
  {Katsnelson}}, \bibinfo {author} {\bibfnamefont {R.~a.~W.}\ \bibnamefont
  {Dryfe}}, \bibinfo {author} {\bibfnamefont {I.~V.}\ \bibnamefont
  {Grigorieva}}, \bibinfo {author} {\bibfnamefont {H.~A.}\ \bibnamefont {Wu}},
  \ and\ \bibinfo {author} {\bibfnamefont {A.~K.}\ \bibnamefont {Geim}},\
  }\href {\doibase 10.1038/nature14015} {\bibfield  {journal} {\bibinfo
  {journal} {Nature}\ }\textbf {\bibinfo {volume} {516}},\ \bibinfo {pages}
  {227} (\bibinfo {year} {2014})}\BibitemShut {NoStop}%
\bibitem [{\citenamefont {Lozada-Hidalgo}\ \emph {et~al.}(2016)\citenamefont
  {Lozada-Hidalgo}, \citenamefont {Hu}, \citenamefont {Marshall}, \citenamefont
  {Mishchenko}, \citenamefont {Grigorenko}, \citenamefont {Dryfe},
  \citenamefont {Radha}, \citenamefont {Grigorieva},\ and\ \citenamefont
  {Geim}}]{Lozada-Hidalgo68}%
  \BibitemOpen
  \bibfield  {author} {\bibinfo {author} {\bibfnamefont {M.}~\bibnamefont
  {Lozada-Hidalgo}}, \bibinfo {author} {\bibfnamefont {S.}~\bibnamefont {Hu}},
  \bibinfo {author} {\bibfnamefont {O.}~\bibnamefont {Marshall}}, \bibinfo
  {author} {\bibfnamefont {A.}~\bibnamefont {Mishchenko}}, \bibinfo {author}
  {\bibfnamefont {A.~N.}\ \bibnamefont {Grigorenko}}, \bibinfo {author}
  {\bibfnamefont {R.~A.~W.}\ \bibnamefont {Dryfe}}, \bibinfo {author}
  {\bibfnamefont {B.}~\bibnamefont {Radha}}, \bibinfo {author} {\bibfnamefont
  {I.~V.}\ \bibnamefont {Grigorieva}}, \ and\ \bibinfo {author} {\bibfnamefont
  {A.~K.}\ \bibnamefont {Geim}},\ }\href {\doibase 10.1126/science.aac9726}
  {\bibfield  {journal} {\bibinfo  {journal} {Science}\ }\textbf {\bibinfo
  {volume} {351}},\ \bibinfo {pages} {68} (\bibinfo {year} {2016})}\BibitemShut
  {NoStop}%
\bibitem [{\citenamefont {Ekanayake}\ \emph {et~al.}(2017)\citenamefont
  {Ekanayake}, \citenamefont {Huang}, \citenamefont {Jakowski}, \citenamefont
  {Sumpter},\ and\ \citenamefont {Garashchuk}}]{doi:10.1021/acs.jpcc.7b08152}%
  \BibitemOpen
  \bibfield  {author} {\bibinfo {author} {\bibfnamefont {N.~T.}\ \bibnamefont
  {Ekanayake}}, \bibinfo {author} {\bibfnamefont {J.}~\bibnamefont {Huang}},
  \bibinfo {author} {\bibfnamefont {J.}~\bibnamefont {Jakowski}}, \bibinfo
  {author} {\bibfnamefont {B.~G.}\ \bibnamefont {Sumpter}}, \ and\ \bibinfo
  {author} {\bibfnamefont {S.}~\bibnamefont {Garashchuk}},\ }\href {\doibase
  10.1021/acs.jpcc.7b08152} {\bibfield  {journal} {\bibinfo  {journal} {J.
  Phys. Chem. C}\ }\textbf {\bibinfo {volume} {121}},\ \bibinfo {pages} {24335}
  (\bibinfo {year} {2017})}\BibitemShut {NoStop}%
\bibitem [{\citenamefont {Poltavsky}\ \emph {et~al.}(2018)\citenamefont
  {Poltavsky}, \citenamefont {Zheng},\ and\ \citenamefont
  {Majid~Mortazavi}}]{Tka_tunneling_2016}%
  \BibitemOpen
  \bibfield  {author} {\bibinfo {author} {\bibfnamefont {I.}~\bibnamefont
  {Poltavsky}}, \bibinfo {author} {\bibfnamefont {L.}~\bibnamefont {Zheng}}, \
  and\ \bibinfo {author} {\bibfnamefont {A.~T.}\ \bibnamefont
  {Majid~Mortazavi}},\ }\href
  {https://aip.scitation.org/doi/abs/10.1063/1.5024317} {\bibfield  {journal}
  {\bibinfo  {journal} {J. Chem. Phys.}\ }\textbf {\bibinfo {volume} {148}},\
  \bibinfo {pages} {204707} (\bibinfo {year} {2018})}\BibitemShut {NoStop}%
\bibitem [{\citenamefont {Feng}\ \emph {et~al.}(2017)\citenamefont {Feng},
  \citenamefont {Chen}, \citenamefont {Fang}, \citenamefont {Wang},
  \citenamefont {Michaelides},\ and\ \citenamefont
  {Li}}]{doi:10.1021/acs.jpclett.7b02820}%
  \BibitemOpen
  \bibfield  {author} {\bibinfo {author} {\bibfnamefont {Y.}~\bibnamefont
  {Feng}}, \bibinfo {author} {\bibfnamefont {J.}~\bibnamefont {Chen}}, \bibinfo
  {author} {\bibfnamefont {W.}~\bibnamefont {Fang}}, \bibinfo {author}
  {\bibfnamefont {E.-G.}\ \bibnamefont {Wang}}, \bibinfo {author}
  {\bibfnamefont {A.}~\bibnamefont {Michaelides}}, \ and\ \bibinfo {author}
  {\bibfnamefont {X.-Z.}\ \bibnamefont {Li}},\ }\href {\doibase
  10.1021/acs.jpclett.7b02820} {\bibfield  {journal} {\bibinfo  {journal} {J.
  Phys. Chem. Lett.}\ }\textbf {\bibinfo {volume} {8}},\ \bibinfo {pages}
  {6009} (\bibinfo {year} {2017})}\BibitemShut {NoStop}%
\bibitem [{\citenamefont {Marx}(2006)}]{Marx_waterrev_2006}%
  \BibitemOpen
  \bibfield  {author} {\bibinfo {author} {\bibfnamefont {D.}~\bibnamefont
  {Marx}},\ }\href@noop {} {\bibfield  {journal} {\bibinfo  {journal} {Chem.
  Phys. Chem.}\ }\textbf {\bibinfo {volume} {7}},\ \bibinfo {pages} {1848}
  (\bibinfo {year} {2006})}\BibitemShut {NoStop}%
\bibitem [{\citenamefont {Li}\ \emph {et~al.}(2010)\citenamefont {Li},
  \citenamefont {Probert}, \citenamefont {Alavi},\ and\ \citenamefont
  {Michaelides}}]{water_metal_NQEs}%
  \BibitemOpen
  \bibfield  {author} {\bibinfo {author} {\bibfnamefont {X.-Z.}\ \bibnamefont
  {Li}}, \bibinfo {author} {\bibfnamefont {M.~I.~J.}\ \bibnamefont {Probert}},
  \bibinfo {author} {\bibfnamefont {A.}~\bibnamefont {Alavi}}, \ and\ \bibinfo
  {author} {\bibfnamefont {A.}~\bibnamefont {Michaelides}},\ }\href {\doibase
  10.1103/PhysRevLett.104.066102} {\bibfield  {journal} {\bibinfo  {journal}
  {Phys. Rev. Lett.}\ }\textbf {\bibinfo {volume} {104}},\ \bibinfo {pages}
  {066102} (\bibinfo {year} {2010})}\BibitemShut {NoStop}%
\bibitem [{\citenamefont {Ram\'irez}\ \emph {et~al.}(2013)\citenamefont
  {Ram\'irez}, \citenamefont {Neuerburg},\ and\ \citenamefont
  {Herrero}}]{doi:10.1063/1.4818875}%
  \BibitemOpen
  \bibfield  {author} {\bibinfo {author} {\bibfnamefont {R.}~\bibnamefont
  {Ram\'irez}}, \bibinfo {author} {\bibfnamefont {N.}~\bibnamefont
  {Neuerburg}}, \ and\ \bibinfo {author} {\bibfnamefont {C.~P.}\ \bibnamefont
  {Herrero}},\ }\href {\doibase 10.1063/1.4818875} {\bibfield  {journal}
  {\bibinfo  {journal} {J. Chem. Phys.}\ }\textbf {\bibinfo {volume} {139}},\
  \bibinfo {pages} {084503} (\bibinfo {year} {2013})}\BibitemShut {NoStop}%
\bibitem [{\citenamefont {Pruzan}\ \emph {et~al.}(1993)\citenamefont {Pruzan},
  \citenamefont {Chervin},\ and\ \citenamefont {Canny}}]{doi:10.1063/1.465467}%
  \BibitemOpen
  \bibfield  {author} {\bibinfo {author} {\bibfnamefont {P.}~\bibnamefont
  {Pruzan}}, \bibinfo {author} {\bibfnamefont {J.~C.}\ \bibnamefont {Chervin}},
  \ and\ \bibinfo {author} {\bibfnamefont {B.}~\bibnamefont {Canny}},\ }\href
  {\doibase 10.1063/1.465467} {\bibfield  {journal} {\bibinfo  {journal} {J.
  Chem. Phys.}\ }\textbf {\bibinfo {volume} {99}},\ \bibinfo {pages} {9842}
  (\bibinfo {year} {1993})}\BibitemShut {NoStop}%
\bibitem [{\citenamefont {Holzapfel}(1972)}]{doi:10.1063/1.1677221}%
  \BibitemOpen
  \bibfield  {author} {\bibinfo {author} {\bibfnamefont {W.~B.}\ \bibnamefont
  {Holzapfel}},\ }\href {\doibase 10.1063/1.1677221} {\bibfield  {journal}
  {\bibinfo  {journal} {J. Chem. Phys.}\ }\textbf {\bibinfo {volume} {56}},\
  \bibinfo {pages} {712} (\bibinfo {year} {1972})}\BibitemShut {NoStop}%
\bibitem [{\citenamefont {Goncharov}\ \emph
  {et~al.}(1996{\natexlab{b}})\citenamefont {Goncharov}, \citenamefont
  {Struzhkin}, \citenamefont {Somayazulu}, \citenamefont {Hemley},\ and\
  \citenamefont {Mao}}]{Goncharov218}%
  \BibitemOpen
  \bibfield  {author} {\bibinfo {author} {\bibfnamefont {A.~F.}\ \bibnamefont
  {Goncharov}}, \bibinfo {author} {\bibfnamefont {V.~V.}\ \bibnamefont
  {Struzhkin}}, \bibinfo {author} {\bibfnamefont {M.~S.}\ \bibnamefont
  {Somayazulu}}, \bibinfo {author} {\bibfnamefont {R.~J.}\ \bibnamefont
  {Hemley}}, \ and\ \bibinfo {author} {\bibfnamefont {H.~K.}\ \bibnamefont
  {Mao}},\ }\href {\doibase 10.1126/science.273.5272.218} {\bibfield  {journal}
  {\bibinfo  {journal} {Science}\ }\textbf {\bibinfo {volume} {273}},\ \bibinfo
  {pages} {218} (\bibinfo {year} {1996}{\natexlab{b}})}\BibitemShut {NoStop}%
\bibitem [{\citenamefont {Michaelides}\ and\ \citenamefont
  {Morgenstern}(2007)}]{Morgenstern_2007_1}%
  \BibitemOpen
  \bibfield  {author} {\bibinfo {author} {\bibfnamefont {A.}~\bibnamefont
  {Michaelides}}\ and\ \bibinfo {author} {\bibfnamefont {K.}~\bibnamefont
  {Morgenstern}},\ }\href@noop {} {\bibfield  {journal} {\bibinfo  {journal}
  {Nat. Mater.}\ }\textbf {\bibinfo {volume} {6}},\ \bibinfo {pages} {597}
  (\bibinfo {year} {2007})}\BibitemShut {NoStop}%
\bibitem [{\citenamefont {Carrasco}\ \emph {et~al.}(2012)\citenamefont
  {Carrasco}, \citenamefont {Hodgson},\ and\ \citenamefont
  {Michaelides}}]{Carrasco_2012_1}%
  \BibitemOpen
  \bibfield  {author} {\bibinfo {author} {\bibfnamefont {J.}~\bibnamefont
  {Carrasco}}, \bibinfo {author} {\bibfnamefont {A.}~\bibnamefont {Hodgson}}, \
  and\ \bibinfo {author} {\bibfnamefont {A.}~\bibnamefont {Michaelides}},\
  }\href {\doibase 10.1038/nmat3354} {\bibfield  {journal} {\bibinfo  {journal}
  {Nat. Mater.}\ }\textbf {\bibinfo {volume} {11}},\ \bibinfo {pages} {667 }
  (\bibinfo {year} {2012})}\BibitemShut {NoStop}%
\bibitem [{\citenamefont {Maier}\ and\ \citenamefont
  {Salmeron}(2015)}]{Salmeron_2015}%
  \BibitemOpen
  \bibfield  {author} {\bibinfo {author} {\bibfnamefont {S.}~\bibnamefont
  {Maier}}\ and\ \bibinfo {author} {\bibfnamefont {M.}~\bibnamefont
  {Salmeron}},\ }\href {\doibase 10.1021/acs.accounts.5b00214} {\bibfield
  {journal} {\bibinfo  {journal} {Acc. Chem. Res.}\ }\textbf {\bibinfo {volume}
  {48}},\ \bibinfo {pages} {2783} (\bibinfo {year} {2015})}\BibitemShut
  {NoStop}%
\bibitem [{\citenamefont {Bj\"orneholm}\ \emph {et~al.}(2016)\citenamefont
  {Bj\"orneholm}, \citenamefont {Hansen}, \citenamefont {Hodgson},
  \citenamefont {Liu}, \citenamefont {Limmer}, \citenamefont {Michaelides},
  \citenamefont {Pedevilla}, \citenamefont {Rossmeisl}, \citenamefont {Shen},
  \citenamefont {Tocci}, \citenamefont {Tyrode}, \citenamefont {Walz},
  \citenamefont {Werner},\ and\ \citenamefont
  {Bluhm}}]{doi:10.1021/acs.chemrev.6b00045}%
  \BibitemOpen
  \bibfield  {author} {\bibinfo {author} {\bibfnamefont {O.}~\bibnamefont
  {Bj\"orneholm}}, \bibinfo {author} {\bibfnamefont {M.~H.}\ \bibnamefont
  {Hansen}}, \bibinfo {author} {\bibfnamefont {A.}~\bibnamefont {Hodgson}},
  \bibinfo {author} {\bibfnamefont {L.-M.}\ \bibnamefont {Liu}}, \bibinfo
  {author} {\bibfnamefont {D.~T.}\ \bibnamefont {Limmer}}, \bibinfo {author}
  {\bibfnamefont {A.}~\bibnamefont {Michaelides}}, \bibinfo {author}
  {\bibfnamefont {P.}~\bibnamefont {Pedevilla}}, \bibinfo {author}
  {\bibfnamefont {J.}~\bibnamefont {Rossmeisl}}, \bibinfo {author}
  {\bibfnamefont {H.}~\bibnamefont {Shen}}, \bibinfo {author} {\bibfnamefont
  {G.}~\bibnamefont {Tocci}}, \bibinfo {author} {\bibfnamefont
  {E.}~\bibnamefont {Tyrode}}, \bibinfo {author} {\bibfnamefont {M.-M.}\
  \bibnamefont {Walz}}, \bibinfo {author} {\bibfnamefont {J.}~\bibnamefont
  {Werner}}, \ and\ \bibinfo {author} {\bibfnamefont {H.}~\bibnamefont
  {Bluhm}},\ }\href {\doibase 10.1021/acs.chemrev.6b00045} {\bibfield
  {journal} {\bibinfo  {journal} {Chem. Rev.}\ }\textbf {\bibinfo {volume}
  {116}},\ \bibinfo {pages} {7698} (\bibinfo {year} {2016})}\BibitemShut
  {NoStop}%
\bibitem [{\citenamefont {Michaelides}\ and\ \citenamefont
  {Hu}(2001)}]{doi:10.1021/ja003576x}%
  \BibitemOpen
  \bibfield  {author} {\bibinfo {author} {\bibfnamefont {A.}~\bibnamefont
  {Michaelides}}\ and\ \bibinfo {author} {\bibfnamefont {P.}~\bibnamefont
  {Hu}},\ }\href {\doibase 10.1021/ja003576x} {\bibfield  {journal} {\bibinfo
  {journal} {J. Am. Chem. Soc.}\ }\textbf {\bibinfo {volume} {123}},\ \bibinfo
  {pages} {4235} (\bibinfo {year} {2001})}\BibitemShut {NoStop}%
\bibitem [{\citenamefont {Michaelides}\ \emph {et~al.}(2004)\citenamefont
  {Michaelides}, \citenamefont {Alavi},\ and\ \citenamefont
  {King}}]{PhysRevB.69.113404}%
  \BibitemOpen
  \bibfield  {author} {\bibinfo {author} {\bibfnamefont {A.}~\bibnamefont
  {Michaelides}}, \bibinfo {author} {\bibfnamefont {A.}~\bibnamefont {Alavi}},
  \ and\ \bibinfo {author} {\bibfnamefont {D.~A.}\ \bibnamefont {King}},\
  }\href {\doibase 10.1103/PhysRevB.69.113404} {\bibfield  {journal} {\bibinfo
  {journal} {Phys. Rev. B}\ }\textbf {\bibinfo {volume} {69}},\ \bibinfo
  {pages} {113404} (\bibinfo {year} {2004})}\BibitemShut {NoStop}%
\bibitem [{\citenamefont {Herrero}\ and\ \citenamefont
  {Ram\'irez}(2015)}]{HERRERO2015125}%
  \BibitemOpen
  \bibfield  {author} {\bibinfo {author} {\bibfnamefont {C.~P.}\ \bibnamefont
  {Herrero}}\ and\ \bibinfo {author} {\bibfnamefont {R.}~\bibnamefont
  {Ram\'irez}},\ }\href {\doibase
  https://doi.org/10.1016/j.chemphys.2015.09.011} {\bibfield  {journal}
  {\bibinfo  {journal} {Chem. Phys.}\ }\textbf {\bibinfo {volume} {461}},\
  \bibinfo {pages} {125 } (\bibinfo {year} {2015})}\BibitemShut {NoStop}%
\bibitem [{\citenamefont {Errea}\ \emph {et~al.}(2016)\citenamefont {Errea},
  \citenamefont {Calandra}, \citenamefont {Pickard}, \citenamefont {Nelson},
  \citenamefont {Needs}, \citenamefont {Li}, \citenamefont {Liu}, \citenamefont
  {Zhang}, \citenamefont {Ma},\ and\ \citenamefont
  {Mauri}}]{errea_quantum_2016}%
  \BibitemOpen
  \bibfield  {author} {\bibinfo {author} {\bibfnamefont {I.}~\bibnamefont
  {Errea}}, \bibinfo {author} {\bibfnamefont {M.}~\bibnamefont {Calandra}},
  \bibinfo {author} {\bibfnamefont {C.~J.}\ \bibnamefont {Pickard}}, \bibinfo
  {author} {\bibfnamefont {J.~R.}\ \bibnamefont {Nelson}}, \bibinfo {author}
  {\bibfnamefont {R.~J.}\ \bibnamefont {Needs}}, \bibinfo {author}
  {\bibfnamefont {Y.}~\bibnamefont {Li}}, \bibinfo {author} {\bibfnamefont
  {H.}~\bibnamefont {Liu}}, \bibinfo {author} {\bibfnamefont {Y.}~\bibnamefont
  {Zhang}}, \bibinfo {author} {\bibfnamefont {Y.}~\bibnamefont {Ma}}, \ and\
  \bibinfo {author} {\bibfnamefont {F.}~\bibnamefont {Mauri}},\ }\href
  {\doibase 10.1038/nature17175} {\bibfield  {journal} {\bibinfo  {journal}
  {Nature}\ }\textbf {\bibinfo {volume} {532}},\ \bibinfo {pages} {81}
  (\bibinfo {year} {2016})}\BibitemShut {NoStop}%
\bibitem [{\citenamefont {McKenzie}(2012)}]{McKenzie2012}%
  \BibitemOpen
  \bibfield  {author} {\bibinfo {author} {\bibfnamefont {R.~H.}\ \bibnamefont
  {McKenzie}},\ }\href {\doibase
  http://dx.doi.org/10.1016/j.cplett.2012.03.064} {\bibfield  {journal}
  {\bibinfo  {journal} {Chem. Phys. Lett.}\ }\textbf {\bibinfo {volume}
  {535}},\ \bibinfo {pages} {196} (\bibinfo {year} {2012})}\BibitemShut
  {NoStop}%
\bibitem [{\citenamefont {McKenzie}\ \emph {et~al.}(2014)\citenamefont
  {McKenzie}, \citenamefont {Bekker}, \citenamefont {Athokpam},\ and\
  \citenamefont {Ramesh}}]{McKenzie_2014}%
  \BibitemOpen
  \bibfield  {author} {\bibinfo {author} {\bibfnamefont {R.~H.}\ \bibnamefont
  {McKenzie}}, \bibinfo {author} {\bibfnamefont {C.}~\bibnamefont {Bekker}},
  \bibinfo {author} {\bibfnamefont {B.}~\bibnamefont {Athokpam}}, \ and\
  \bibinfo {author} {\bibfnamefont {S.~G.}\ \bibnamefont {Ramesh}},\ }\href
  {\doibase http://dx.doi.org/10.1063/1.4873352} {\bibfield  {journal}
  {\bibinfo  {journal} {J. Chem. Phys.}\ }\textbf {\bibinfo {volume} {140}},\
  \bibinfo {pages} {174508} (\bibinfo {year} {2014})}\BibitemShut {NoStop}%
\bibitem [{\citenamefont {Habershon}\ \emph {et~al.}(2009)\citenamefont
  {Habershon}, \citenamefont {Markland},\ and\ \citenamefont
  {Manolopoulos}}]{Habershon_2009}%
  \BibitemOpen
  \bibfield  {author} {\bibinfo {author} {\bibfnamefont {S.}~\bibnamefont
  {Habershon}}, \bibinfo {author} {\bibfnamefont {T.~E.}\ \bibnamefont
  {Markland}}, \ and\ \bibinfo {author} {\bibfnamefont {D.~E.}\ \bibnamefont
  {Manolopoulos}},\ }\href {\doibase http://dx.doi.org/10.1063/1.3167790}
  {\bibfield  {journal} {\bibinfo  {journal} {J. Chem. Phys.}\ }\textbf
  {\bibinfo {volume} {131}},\ \bibinfo {pages} {024501} (\bibinfo {year}
  {2009})}\BibitemShut {NoStop}%
\bibitem [{\citenamefont {Habershon}\ and\ \citenamefont
  {Manolopoulos}(2011{\natexlab{a}})}]{C1CP21520E}%
  \BibitemOpen
  \bibfield  {author} {\bibinfo {author} {\bibfnamefont {S.}~\bibnamefont
  {Habershon}}\ and\ \bibinfo {author} {\bibfnamefont {D.~E.}\ \bibnamefont
  {Manolopoulos}},\ }\href {\doibase 10.1039/C1CP21520E} {\bibfield  {journal}
  {\bibinfo  {journal} {Phys. Chem. Chem. Phys.}\ }\textbf {\bibinfo {volume}
  {13}},\ \bibinfo {pages} {19714} (\bibinfo {year}
  {2011}{\natexlab{a}})}\BibitemShut {NoStop}%
\bibitem [{\citenamefont {Markland}\ and\ \citenamefont
  {Berne}(2012)}]{Markland_2012}%
  \BibitemOpen
  \bibfield  {author} {\bibinfo {author} {\bibfnamefont {T.~E.}\ \bibnamefont
  {Markland}}\ and\ \bibinfo {author} {\bibfnamefont {B.~J.}\ \bibnamefont
  {Berne}},\ }\href@noop {} {\bibfield  {journal} {\bibinfo  {journal} {Proc.
  Natl. Acad. Sci. U.S.A.}\ }\textbf {\bibinfo {volume} {109}},\ \bibinfo
  {pages} {7988} (\bibinfo {year} {2012})}\BibitemShut {NoStop}%
\bibitem [{\citenamefont {Liu}\ \emph {et~al.}(2013)\citenamefont {Liu},
  \citenamefont {Andino}, \citenamefont {Miller}, \citenamefont {Chen},
  \citenamefont {Wilkins}, \citenamefont {Ceriotti},\ and\ \citenamefont
  {Manolopoulos}}]{Ceriotti_2013_1}%
  \BibitemOpen
  \bibfield  {author} {\bibinfo {author} {\bibfnamefont {J.}~\bibnamefont
  {Liu}}, \bibinfo {author} {\bibfnamefont {R.~S.}\ \bibnamefont {Andino}},
  \bibinfo {author} {\bibfnamefont {C.~M.}\ \bibnamefont {Miller}}, \bibinfo
  {author} {\bibfnamefont {X.}~\bibnamefont {Chen}}, \bibinfo {author}
  {\bibfnamefont {D.~M.}\ \bibnamefont {Wilkins}}, \bibinfo {author}
  {\bibfnamefont {M.}~\bibnamefont {Ceriotti}}, \ and\ \bibinfo {author}
  {\bibfnamefont {D.~E.}\ \bibnamefont {Manolopoulos}},\ }\href
  {http://dx.doi.org/10.1021/jp311986m} {\bibfield  {journal} {\bibinfo
  {journal} {J. Phys. Chem. C}\ }\textbf {\bibinfo {volume} {117}},\ \bibinfo
  {pages} {2944} (\bibinfo {year} {2013})}\BibitemShut {NoStop}%
\bibitem [{\citenamefont {Wang}\ \emph {et~al.}(2014)\citenamefont {Wang},
  \citenamefont {Ceriotti},\ and\ \citenamefont {Markland}}]{LWang_2014}%
  \BibitemOpen
  \bibfield  {author} {\bibinfo {author} {\bibfnamefont {L.}~\bibnamefont
  {Wang}}, \bibinfo {author} {\bibfnamefont {M.}~\bibnamefont {Ceriotti}}, \
  and\ \bibinfo {author} {\bibfnamefont {T.~E.}\ \bibnamefont {Markland}},\
  }\href {\doibase http://dx.doi.org/10.1063/1.4894287} {\bibfield  {journal}
  {\bibinfo  {journal} {J. Chem. Phys.}\ }\textbf {\bibinfo {volume} {141}},\
  \bibinfo {pages} {104502} (\bibinfo {year} {2014})}\BibitemShut {NoStop}%
\bibitem [{\citenamefont {Wilkins}\ \emph {et~al.}(2015)\citenamefont
  {Wilkins}, \citenamefont {Manolopoulos},\ and\ \citenamefont
  {Dang}}]{doi:10.1063/1.4907554}%
  \BibitemOpen
  \bibfield  {author} {\bibinfo {author} {\bibfnamefont {D.~M.}\ \bibnamefont
  {Wilkins}}, \bibinfo {author} {\bibfnamefont {D.~E.}\ \bibnamefont
  {Manolopoulos}}, \ and\ \bibinfo {author} {\bibfnamefont {L.~X.}\
  \bibnamefont {Dang}},\ }\href {\doibase 10.1063/1.4907554} {\bibfield
  {journal} {\bibinfo  {journal} {J. Chem. Phys.}\ }\textbf {\bibinfo {volume}
  {142}},\ \bibinfo {pages} {064509} (\bibinfo {year} {2015})}\BibitemShut
  {NoStop}%
\bibitem [{\citenamefont {Morrone}\ and\ \citenamefont
  {Car}(2008)}]{Morrone_2008}%
  \BibitemOpen
  \bibfield  {author} {\bibinfo {author} {\bibfnamefont {J.~A.}\ \bibnamefont
  {Morrone}}\ and\ \bibinfo {author} {\bibfnamefont {R.}~\bibnamefont {Car}},\
  }\href {\doibase 10.1103/PhysRevLett.101.017801} {\bibfield  {journal}
  {\bibinfo  {journal} {Phys. Rev. Lett.}\ }\textbf {\bibinfo {volume} {101}},\
  \bibinfo {pages} {017801} (\bibinfo {year} {2008})}\BibitemShut {NoStop}%
\bibitem [{\citenamefont {Medders}\ \emph {et~al.}(2014)\citenamefont
  {Medders}, \citenamefont {Babin},\ and\ \citenamefont
  {Paesani}}]{water_ordf}%
  \BibitemOpen
  \bibfield  {author} {\bibinfo {author} {\bibfnamefont {G.~R.}\ \bibnamefont
  {Medders}}, \bibinfo {author} {\bibfnamefont {V.}~\bibnamefont {Babin}}, \
  and\ \bibinfo {author} {\bibfnamefont {F.}~\bibnamefont {Paesani}},\ }\href
  {\doibase 10.1021/ct5004115} {\bibfield  {journal} {\bibinfo  {journal} {J.
  Chem. Theory Comput.}\ }\textbf {\bibinfo {volume} {10}},\ \bibinfo {pages}
  {2906} (\bibinfo {year} {2014})}\BibitemShut {NoStop}%
\bibitem [{\citenamefont {Fang}\ \emph {et~al.}(2016)\citenamefont {Fang},
  \citenamefont {Chen}, \citenamefont {Rossi}, \citenamefont {Feng},
  \citenamefont {Li},\ and\ \citenamefont {Michaelides}}]{Wei_BP_BFE}%
  \BibitemOpen
  \bibfield  {author} {\bibinfo {author} {\bibfnamefont {W.}~\bibnamefont
  {Fang}}, \bibinfo {author} {\bibfnamefont {J.}~\bibnamefont {Chen}}, \bibinfo
  {author} {\bibfnamefont {M.}~\bibnamefont {Rossi}}, \bibinfo {author}
  {\bibfnamefont {Y.}~\bibnamefont {Feng}}, \bibinfo {author} {\bibfnamefont
  {X.-Z.}\ \bibnamefont {Li}}, \ and\ \bibinfo {author} {\bibfnamefont
  {A.}~\bibnamefont {Michaelides}},\ }\href {\doibase
  10.1021/acs.jpclett.6b00777} {\bibfield  {journal} {\bibinfo  {journal} {J.
  Phys. Chem. Lett.}\ }\textbf {\bibinfo {volume} {7}},\ \bibinfo {pages}
  {2125} (\bibinfo {year} {2016})}\BibitemShut {NoStop}%
\bibitem [{\citenamefont {Cheng}\ \emph {et~al.}(2016)\citenamefont {Cheng},
  \citenamefont {Behler},\ and\ \citenamefont {Ceriotti}}]{Cheng_2016}%
  \BibitemOpen
  \bibfield  {author} {\bibinfo {author} {\bibfnamefont {B.}~\bibnamefont
  {Cheng}}, \bibinfo {author} {\bibfnamefont {J.}~\bibnamefont {Behler}}, \
  and\ \bibinfo {author} {\bibfnamefont {M.}~\bibnamefont {Ceriotti}},\ }\href
  {\doibase 10.1021/acs.jpclett.6b00729} {\bibfield  {journal} {\bibinfo
  {journal} {J. Phys. Chem. Lett.}\ }\textbf {\bibinfo {volume} {7}},\ \bibinfo
  {pages} {2210} (\bibinfo {year} {2016})}\BibitemShut {NoStop}%
\bibitem [{\citenamefont {Guo}\ \emph {et~al.}(2016)\citenamefont {Guo},
  \citenamefont {L{\"u}}, \citenamefont {Feng}, \citenamefont {Chen},
  \citenamefont {Peng}, \citenamefont {Lin}, \citenamefont {Meng},
  \citenamefont {Wang}, \citenamefont {Li}, \citenamefont {Wang},\ and\
  \citenamefont {Jiang}}]{Guo321}%
  \BibitemOpen
  \bibfield  {author} {\bibinfo {author} {\bibfnamefont {J.}~\bibnamefont
  {Guo}}, \bibinfo {author} {\bibfnamefont {J.-T.}\ \bibnamefont {L{\"u}}},
  \bibinfo {author} {\bibfnamefont {Y.}~\bibnamefont {Feng}}, \bibinfo {author}
  {\bibfnamefont {J.}~\bibnamefont {Chen}}, \bibinfo {author} {\bibfnamefont
  {J.}~\bibnamefont {Peng}}, \bibinfo {author} {\bibfnamefont {Z.}~\bibnamefont
  {Lin}}, \bibinfo {author} {\bibfnamefont {X.}~\bibnamefont {Meng}}, \bibinfo
  {author} {\bibfnamefont {Z.}~\bibnamefont {Wang}}, \bibinfo {author}
  {\bibfnamefont {X.-Z.}\ \bibnamefont {Li}}, \bibinfo {author} {\bibfnamefont
  {E.-G.}\ \bibnamefont {Wang}}, \ and\ \bibinfo {author} {\bibfnamefont
  {Y.}~\bibnamefont {Jiang}},\ }\href {\doibase 10.1126/science.aaf2042}
  {\bibfield  {journal} {\bibinfo  {journal} {Science}\ }\textbf {\bibinfo
  {volume} {352}},\ \bibinfo {pages} {321} (\bibinfo {year}
  {2016})}\BibitemShut {NoStop}%
\bibitem [{\citenamefont {P\'erez}\ and\ \citenamefont {von
  Lilienfeld}(2011)}]{Mass_2}%
  \BibitemOpen
  \bibfield  {author} {\bibinfo {author} {\bibfnamefont {A.}~\bibnamefont
  {P\'erez}}\ and\ \bibinfo {author} {\bibfnamefont {O.~A.}\ \bibnamefont {von
  Lilienfeld}},\ }\href {\doibase 10.1021/ct2000556} {\bibfield  {journal}
  {\bibinfo  {journal} {J. Chem. Theory Comput.}\ }\textbf {\bibinfo {volume}
  {7}},\ \bibinfo {pages} {2358} (\bibinfo {year} {2011})}\BibitemShut
  {NoStop}%
\bibitem [{\citenamefont {Rossi}\ \emph {et~al.}(2015)\citenamefont {Rossi},
  \citenamefont {Fang},\ and\ \citenamefont {Michaelides}}]{Rossi_2015}%
  \BibitemOpen
  \bibfield  {author} {\bibinfo {author} {\bibfnamefont {M.}~\bibnamefont
  {Rossi}}, \bibinfo {author} {\bibfnamefont {W.}~\bibnamefont {Fang}}, \ and\
  \bibinfo {author} {\bibfnamefont {A.}~\bibnamefont {Michaelides}},\ }\href
  {http://dx.doi.org/10.1021/acs.jpclett.5b01899} {\bibfield  {journal}
  {\bibinfo  {journal} {J. Phys. Chem. Lett.}\ }\textbf {\bibinfo {volume}
  {6}},\ \bibinfo {pages} {4233} (\bibinfo {year} {2015})}\BibitemShut
  {NoStop}%
\bibitem [{\citenamefont {Morales}\ and\ \citenamefont
  {Singer}(1991)}]{Morales_Singer}%
  \BibitemOpen
  \bibfield  {author} {\bibinfo {author} {\bibfnamefont {J.}~\bibnamefont
  {Morales}}\ and\ \bibinfo {author} {\bibfnamefont {K.}~\bibnamefont
  {Singer}},\ }\href {\doibase 10.1080/00268979100101621} {\bibfield  {journal}
  {\bibinfo  {journal} {Mol. Phys.}\ }\textbf {\bibinfo {volume} {73}},\
  \bibinfo {pages} {873} (\bibinfo {year} {1991})}\BibitemShut {NoStop}%
\bibitem [{\citenamefont {Frenkel}\ and\ \citenamefont
  {Smit}(2002)}]{DaanFrenkel_book}%
  \BibitemOpen
  \bibfield  {author} {\bibinfo {author} {\bibfnamefont {D.}~\bibnamefont
  {Frenkel}}\ and\ \bibinfo {author} {\bibfnamefont {B.}~\bibnamefont {Smit}},\
  }\href@noop {} {\emph {\bibinfo {title} {Understanding Molecular Simulation:
  From Algorithms to Applications}}}\ (\bibinfo  {publisher} {Academic Press,
  San Diego, USA},\ \bibinfo {year} {2002})\BibitemShut {NoStop}%
\bibitem [{\citenamefont {Ram\'{i}rez}\ \emph {et~al.}(2008)\citenamefont
  {Ram\'{i}rez}, \citenamefont {Herrero}, \citenamefont {Antonelli},\ and\
  \citenamefont {Hern\'{a}ndez}}]{doi:10.1063/1.2966006}%
  \BibitemOpen
  \bibfield  {author} {\bibinfo {author} {\bibfnamefont {R.}~\bibnamefont
  {Ram\'{i}rez}}, \bibinfo {author} {\bibfnamefont {C.~P.}\ \bibnamefont
  {Herrero}}, \bibinfo {author} {\bibfnamefont {A.}~\bibnamefont {Antonelli}},
  \ and\ \bibinfo {author} {\bibfnamefont {E.~R.}\ \bibnamefont
  {Hern\'{a}ndez}},\ }\href {\doibase 10.1063/1.2966006} {\bibfield  {journal}
  {\bibinfo  {journal} {J. Chem. Phys.}\ }\textbf {\bibinfo {volume} {129}},\
  \bibinfo {pages} {064110} (\bibinfo {year} {2008})}\BibitemShut {NoStop}%
\bibitem [{\citenamefont {Geerke}\ \emph {et~al.}(2008)\citenamefont {Geerke},
  \citenamefont {Luber}, \citenamefont {Marti},\ and\ \citenamefont
  {Van~Gunsteren}}]{doi:10.1002/jcc.21070}%
  \BibitemOpen
  \bibfield  {author} {\bibinfo {author} {\bibfnamefont {D.~P.}\ \bibnamefont
  {Geerke}}, \bibinfo {author} {\bibfnamefont {S.}~\bibnamefont {Luber}},
  \bibinfo {author} {\bibfnamefont {K.~H.}\ \bibnamefont {Marti}}, \ and\
  \bibinfo {author} {\bibfnamefont {W.~F.}\ \bibnamefont {Van~Gunsteren}},\
  }\href {\doibase 10.1002/jcc.21070} {\bibfield  {journal} {\bibinfo
  {journal} {J. Chem. Phys.}\ }\textbf {\bibinfo {volume} {30}},\ \bibinfo
  {pages} {514} (\bibinfo {year} {2008})}\BibitemShut {NoStop}%
\bibitem [{\citenamefont {Habershon}\ and\ \citenamefont
  {Manolopoulos}(2011{\natexlab{b}})}]{scaled_coordinate}%
  \BibitemOpen
  \bibfield  {author} {\bibinfo {author} {\bibfnamefont {S.}~\bibnamefont
  {Habershon}}\ and\ \bibinfo {author} {\bibfnamefont {D.~E.}\ \bibnamefont
  {Manolopoulos}},\ }\href {\doibase 10.1063/1.3666011} {\bibfield  {journal}
  {\bibinfo  {journal} {J. Chem. Phys.}\ }\textbf {\bibinfo {volume} {135}},\
  \bibinfo {pages} {224111} (\bibinfo {year} {2011}{\natexlab{b}})}\BibitemShut
  {NoStop}%
\bibitem [{\citenamefont {Rossi}\ \emph
  {et~al.}(2016{\natexlab{b}})\citenamefont {Rossi}, \citenamefont
  {Gasparotto},\ and\ \citenamefont {Ceriotti}}]{Rossi_2016_2}%
  \BibitemOpen
  \bibfield  {author} {\bibinfo {author} {\bibfnamefont {M.}~\bibnamefont
  {Rossi}}, \bibinfo {author} {\bibfnamefont {P.}~\bibnamefont {Gasparotto}}, \
  and\ \bibinfo {author} {\bibfnamefont {M.}~\bibnamefont {Ceriotti}},\ }\href
  {\doibase 10.1103/PhysRevLett.117.115702} {\bibfield  {journal} {\bibinfo
  {journal} {Phys. Rev. Lett.}\ }\textbf {\bibinfo {volume} {117}},\ \bibinfo
  {pages} {115702} (\bibinfo {year} {2016}{\natexlab{b}})}\BibitemShut
  {NoStop}%
\bibitem [{\citenamefont {Gillan}\ \emph {et~al.}(2016)\citenamefont {Gillan},
  \citenamefont {Alf\`e},\ and\ \citenamefont
  {Michaelides}}]{Gillan_DFT_water}%
  \BibitemOpen
  \bibfield  {author} {\bibinfo {author} {\bibfnamefont {M.~J.}\ \bibnamefont
  {Gillan}}, \bibinfo {author} {\bibfnamefont {D.}~\bibnamefont {Alf\`e}}, \
  and\ \bibinfo {author} {\bibfnamefont {A.}~\bibnamefont {Michaelides}},\
  }\href {\doibase 10.1063/1.4944633} {\bibfield  {journal} {\bibinfo
  {journal} {J. Chem. Phys.}\ }\textbf {\bibinfo {volume} {144}},\ \bibinfo
  {pages} {130901} (\bibinfo {year} {2016})}\BibitemShut {NoStop}%
\bibitem [{\citenamefont {Burke}(2012)}]{doi:10.1063/1.4704546}%
  \BibitemOpen
  \bibfield  {author} {\bibinfo {author} {\bibfnamefont {K.}~\bibnamefont
  {Burke}},\ }\href {\doibase 10.1063/1.4704546} {\bibfield  {journal}
  {\bibinfo  {journal} {J. Chem. Phys.}\ }\textbf {\bibinfo {volume} {136}},\
  \bibinfo {pages} {150901} (\bibinfo {year} {2012})}\BibitemShut {NoStop}%
\bibitem [{\citenamefont {Becke}(2014)}]{doi:10.1063/1.4869598}%
  \BibitemOpen
  \bibfield  {author} {\bibinfo {author} {\bibfnamefont {A.~D.}\ \bibnamefont
  {Becke}},\ }\href {\doibase 10.1063/1.4869598} {\bibfield  {journal}
  {\bibinfo  {journal} {J. Chem. Phys.}\ }\textbf {\bibinfo {volume} {140}},\
  \bibinfo {pages} {18A301} (\bibinfo {year} {2014})}\BibitemShut {NoStop}%
\bibitem [{\citenamefont {Brandenburg}\ \emph {et~al.}(2016)\citenamefont
  {Brandenburg}, \citenamefont {Bates}, \citenamefont {Sun},\ and\
  \citenamefont {Perdew}}]{PhysRevB.94.115144}%
  \BibitemOpen
  \bibfield  {author} {\bibinfo {author} {\bibfnamefont {J.~G.}\ \bibnamefont
  {Brandenburg}}, \bibinfo {author} {\bibfnamefont {J.~E.}\ \bibnamefont
  {Bates}}, \bibinfo {author} {\bibfnamefont {J.}~\bibnamefont {Sun}}, \ and\
  \bibinfo {author} {\bibfnamefont {J.~P.}\ \bibnamefont {Perdew}},\ }\href
  {\doibase 10.1103/PhysRevB.94.115144} {\bibfield  {journal} {\bibinfo
  {journal} {Phys. Rev. B}\ }\textbf {\bibinfo {volume} {94}},\ \bibinfo
  {pages} {115144} (\bibinfo {year} {2016})}\BibitemShut {NoStop}%
\bibitem [{\citenamefont {Chen}\ \emph {et~al.}(2014)\citenamefont {Chen},
  \citenamefont {Ren}, \citenamefont {Li}, \citenamefont {Alf\`e},\ and\
  \citenamefont {Wang}}]{chen_room-temperature_2014}%
  \BibitemOpen
  \bibfield  {author} {\bibinfo {author} {\bibfnamefont {J.}~\bibnamefont
  {Chen}}, \bibinfo {author} {\bibfnamefont {X.}~\bibnamefont {Ren}}, \bibinfo
  {author} {\bibfnamefont {X.-Z.}\ \bibnamefont {Li}}, \bibinfo {author}
  {\bibfnamefont {D.}~\bibnamefont {Alf\`e}}, \ and\ \bibinfo {author}
  {\bibfnamefont {E.}~\bibnamefont {Wang}},\ }\href {\doibase
  10.1063/1.4886075} {\bibfield  {journal} {\bibinfo  {journal} {J. Chem.
  Phys.}\ }\textbf {\bibinfo {volume} {141}},\ \bibinfo {pages} {024501}
  (\bibinfo {year} {2014})}\BibitemShut {NoStop}%
\bibitem [{\citenamefont {Foulkes}\ \emph {et~al.}(2001)\citenamefont
  {Foulkes}, \citenamefont {Mitas}, \citenamefont {Needs},\ and\ \citenamefont
  {Rajagopal}}]{RevModPhys.73.33}%
  \BibitemOpen
  \bibfield  {author} {\bibinfo {author} {\bibfnamefont {W.~M.~C.}\
  \bibnamefont {Foulkes}}, \bibinfo {author} {\bibfnamefont {L.}~\bibnamefont
  {Mitas}}, \bibinfo {author} {\bibfnamefont {R.~J.}\ \bibnamefont {Needs}}, \
  and\ \bibinfo {author} {\bibfnamefont {G.}~\bibnamefont {Rajagopal}},\ }\href
  {\doibase 10.1103/RevModPhys.73.33} {\bibfield  {journal} {\bibinfo
  {journal} {Rev. Mod. Phys.}\ }\textbf {\bibinfo {volume} {73}},\ \bibinfo
  {pages} {33} (\bibinfo {year} {2001})}\BibitemShut {NoStop}%
\bibitem [{\citenamefont {Booth}\ \emph {et~al.}(2012)\citenamefont {Booth},
  \citenamefont {Gr\"{u}neis}, \citenamefont {Kresse},\ and\ \citenamefont
  {Alavi}}]{AlaviFCIQMC}%
  \BibitemOpen
  \bibfield  {author} {\bibinfo {author} {\bibfnamefont {G.~H.}\ \bibnamefont
  {Booth}}, \bibinfo {author} {\bibfnamefont {A.}~\bibnamefont {Gr\"{u}neis}},
  \bibinfo {author} {\bibfnamefont {G.}~\bibnamefont {Kresse}}, \ and\ \bibinfo
  {author} {\bibfnamefont {A.}~\bibnamefont {Alavi}},\ }\href {\doibase
  10.1038/nature11770} {\bibfield  {journal} {\bibinfo  {journal} {Nature}\
  }\textbf {\bibinfo {volume} {493}},\ \bibinfo {pages} {365} (\bibinfo {year}
  {2012})}\BibitemShut {NoStop}%
\bibitem [{\citenamefont {Zen}\ \emph {et~al.}(2015)\citenamefont {Zen},
  \citenamefont {Luo}, \citenamefont {Mazzola}, \citenamefont {Guidoni},\ and\
  \citenamefont {Sorella}}]{doi:10.1063/1.4917171}%
  \BibitemOpen
  \bibfield  {author} {\bibinfo {author} {\bibfnamefont {A.}~\bibnamefont
  {Zen}}, \bibinfo {author} {\bibfnamefont {Y.}~\bibnamefont {Luo}}, \bibinfo
  {author} {\bibfnamefont {G.}~\bibnamefont {Mazzola}}, \bibinfo {author}
  {\bibfnamefont {L.}~\bibnamefont {Guidoni}}, \ and\ \bibinfo {author}
  {\bibfnamefont {S.}~\bibnamefont {Sorella}},\ }\href {\doibase
  10.1063/1.4917171} {\bibfield  {journal} {\bibinfo  {journal} {J. Chem.
  Phys.}\ }\textbf {\bibinfo {volume} {142}},\ \bibinfo {pages} {144111}
  (\bibinfo {year} {2015})}\BibitemShut {NoStop}%
\bibitem [{\citenamefont {Zen}\ \emph {et~al.}(2018)\citenamefont {Zen},
  \citenamefont {Brandenburg}, \citenamefont {Klime{\v s}}, \citenamefont
  {Tkatchenko}, \citenamefont {Alf{\`e}},\ and\ \citenamefont
  {Michaelides}}]{Zen201715434}%
  \BibitemOpen
  \bibfield  {author} {\bibinfo {author} {\bibfnamefont {A.}~\bibnamefont
  {Zen}}, \bibinfo {author} {\bibfnamefont {J.~G.}\ \bibnamefont
  {Brandenburg}}, \bibinfo {author} {\bibfnamefont {J.}~\bibnamefont {Klime{\v
  s}}}, \bibinfo {author} {\bibfnamefont {A.}~\bibnamefont {Tkatchenko}},
  \bibinfo {author} {\bibfnamefont {D.}~\bibnamefont {Alf{\`e}}}, \ and\
  \bibinfo {author} {\bibfnamefont {A.}~\bibnamefont {Michaelides}},\ }\href
  {\doibase 10.1073/pnas.1715434115} {\bibfield  {journal} {\bibinfo  {journal}
  {Proc. Natl. Acad. Sci. U.S.A.}\ }\textbf {\bibinfo {volume} {115}},\
  \bibinfo {pages} {1724} (\bibinfo {year} {2018})}\BibitemShut {NoStop}%
\bibitem [{\citenamefont {Zen}\ \emph {et~al.}(2016)\citenamefont {Zen},
  \citenamefont {Sorella}, \citenamefont {Gillan}, \citenamefont
  {Michaelides},\ and\ \citenamefont {Alf\`e}}]{PhysRevB.93.241118}%
  \BibitemOpen
  \bibfield  {author} {\bibinfo {author} {\bibfnamefont {A.}~\bibnamefont
  {Zen}}, \bibinfo {author} {\bibfnamefont {S.}~\bibnamefont {Sorella}},
  \bibinfo {author} {\bibfnamefont {M.~J.}\ \bibnamefont {Gillan}}, \bibinfo
  {author} {\bibfnamefont {A.}~\bibnamefont {Michaelides}}, \ and\ \bibinfo
  {author} {\bibfnamefont {D.}~\bibnamefont {Alf\`e}},\ }\href {\doibase
  10.1103/PhysRevB.93.241118} {\bibfield  {journal} {\bibinfo  {journal} {Phys.
  Rev. B}\ }\textbf {\bibinfo {volume} {93}},\ \bibinfo {pages} {241118}
  (\bibinfo {year} {2016})}\BibitemShut {NoStop}%
\bibitem [{\citenamefont {Richardson}\ \emph {et~al.}(2017)\citenamefont
  {Richardson}, \citenamefont {Meyer}, \citenamefont {Pleinert},\ and\
  \citenamefont {Thoss}}]{RICHARDSON2017124}%
  \BibitemOpen
  \bibfield  {author} {\bibinfo {author} {\bibfnamefont {J.~O.}\ \bibnamefont
  {Richardson}}, \bibinfo {author} {\bibfnamefont {P.}~\bibnamefont {Meyer}},
  \bibinfo {author} {\bibfnamefont {M.~O.}\ \bibnamefont {Pleinert}}, \ and\
  \bibinfo {author} {\bibfnamefont {M.}~\bibnamefont {Thoss}},\ }\href
  {\doibase https://doi.org/10.1016/j.chemphys.2016.09.036} {\bibfield
  {journal} {\bibinfo  {journal} {Chem. Phys.}\ }\textbf {\bibinfo {volume}
  {482}},\ \bibinfo {pages} {124 } (\bibinfo {year} {2017})}\BibitemShut
  {NoStop}%
\bibitem [{\citenamefont {Richardson}(2015)}]{doi:10.1063/1.4932362}%
  \BibitemOpen
  \bibfield  {author} {\bibinfo {author} {\bibfnamefont {J.~O.}\ \bibnamefont
  {Richardson}},\ }\href {\doibase 10.1063/1.4932362} {\bibfield  {journal}
  {\bibinfo  {journal} {J. Chem. Phys.}\ }\textbf {\bibinfo {volume} {143}},\
  \bibinfo {pages} {134116} (\bibinfo {year} {2015})}\BibitemShut {NoStop}%
\bibitem [{\citenamefont {Min}\ \emph {et~al.}(2017)\citenamefont {Min},
  \citenamefont {Agostini}, \citenamefont {Tavernelli},\ and\ \citenamefont
  {Gross}}]{doi:10.1021/acs.jpclett.7b01249}%
  \BibitemOpen
  \bibfield  {author} {\bibinfo {author} {\bibfnamefont {S.~K.}\ \bibnamefont
  {Min}}, \bibinfo {author} {\bibfnamefont {F.}~\bibnamefont {Agostini}},
  \bibinfo {author} {\bibfnamefont {I.}~\bibnamefont {Tavernelli}}, \ and\
  \bibinfo {author} {\bibfnamefont {E.~K.~U.}\ \bibnamefont {Gross}},\ }\href
  {\doibase 10.1021/acs.jpclett.7b01249} {\bibfield  {journal} {\bibinfo
  {journal} {J. Phys. Chem. Lett.}\ }\textbf {\bibinfo {volume} {8}},\ \bibinfo
  {pages} {3048} (\bibinfo {year} {2017})}\BibitemShut {NoStop}%
\bibitem [{\citenamefont {Wolynes}(1987)}]{doi:10.1063/1.453440}%
  \BibitemOpen
  \bibfield  {author} {\bibinfo {author} {\bibfnamefont {P.~G.}\ \bibnamefont
  {Wolynes}},\ }\href {\doibase 10.1063/1.453440} {\bibfield  {journal}
  {\bibinfo  {journal} {J. Chem. Phys.}\ }\textbf {\bibinfo {volume} {87}},\
  \bibinfo {pages} {6559} (\bibinfo {year} {1987})}\BibitemShut {NoStop}%
\bibitem [{\citenamefont {Menzeleev}\ \emph {et~al.}(2014)\citenamefont
  {Menzeleev}, \citenamefont {Bell},\ and\ \citenamefont
  {Miller}}]{doi:10.1063/1.4863919}%
  \BibitemOpen
  \bibfield  {author} {\bibinfo {author} {\bibfnamefont {A.~R.}\ \bibnamefont
  {Menzeleev}}, \bibinfo {author} {\bibfnamefont {F.}~\bibnamefont {Bell}}, \
  and\ \bibinfo {author} {\bibfnamefont {T.~F.}\ \bibnamefont {Miller}},\
  }\href {\doibase 10.1063/1.4863919} {\bibfield  {journal} {\bibinfo
  {journal} {J. Chem. Phys.}\ }\textbf {\bibinfo {volume} {140}},\ \bibinfo
  {pages} {064103} (\bibinfo {year} {2014})}\BibitemShut {NoStop}%
\bibitem [{\citenamefont {Curchod}\ \emph {et~al.}(2012)\citenamefont
  {Curchod}, \citenamefont {Rothlisberger},\ and\ \citenamefont
  {Tavernelli}}]{doi:10.1002/cphc.201200941}%
  \BibitemOpen
  \bibfield  {author} {\bibinfo {author} {\bibfnamefont {B.~F.~E.}\
  \bibnamefont {Curchod}}, \bibinfo {author} {\bibfnamefont {U.}~\bibnamefont
  {Rothlisberger}}, \ and\ \bibinfo {author} {\bibfnamefont {I.}~\bibnamefont
  {Tavernelli}},\ }\href {\doibase 10.1002/cphc.201200941} {\bibfield
  {journal} {\bibinfo  {journal} {Chem. Phys. Chem.}\ }\textbf {\bibinfo
  {volume} {14}},\ \bibinfo {pages} {1314} (\bibinfo {year}
  {2012})}\BibitemShut {NoStop}%
\bibitem [{\citenamefont {Rittmeyer}\ \emph {et~al.}(2018)\citenamefont
  {Rittmeyer}, \citenamefont {Bukas},\ and\ \citenamefont
  {Reuter}}]{doi:10.1080/23746149.2017.1381574}%
  \BibitemOpen
  \bibfield  {author} {\bibinfo {author} {\bibfnamefont {S.~P.}\ \bibnamefont
  {Rittmeyer}}, \bibinfo {author} {\bibfnamefont {V.~J.}\ \bibnamefont
  {Bukas}}, \ and\ \bibinfo {author} {\bibfnamefont {K.}~\bibnamefont
  {Reuter}},\ }\href {\doibase 10.1080/23746149.2017.1381574} {\bibfield
  {journal} {\bibinfo  {journal} {Adv. Phys. X}\ }\textbf {\bibinfo {volume}
  {3}},\ \bibinfo {pages} {1381574} (\bibinfo {year} {2018})}\BibitemShut
  {NoStop}%
\bibitem [{\citenamefont {Li}\ and\ \citenamefont
  {Wahnstr\"om}(1992)}]{PhysRevLett.68.3444}%
  \BibitemOpen
  \bibfield  {author} {\bibinfo {author} {\bibfnamefont {Y.}~\bibnamefont
  {Li}}\ and\ \bibinfo {author} {\bibfnamefont {G.}~\bibnamefont
  {Wahnstr\"om}},\ }\href {\doibase 10.1103/PhysRevLett.68.3444} {\bibfield
  {journal} {\bibinfo  {journal} {Phys. Rev. Lett.}\ }\textbf {\bibinfo
  {volume} {68}},\ \bibinfo {pages} {3444} (\bibinfo {year}
  {1992})}\BibitemShut {NoStop}%
\bibitem [{\citenamefont {Head‐Gordon}\ and\ \citenamefont
  {Tully}(1995)}]{doi:10.1063/1.469915}%
  \BibitemOpen
  \bibfield  {author} {\bibinfo {author} {\bibfnamefont {M.}~\bibnamefont
  {Head‐Gordon}}\ and\ \bibinfo {author} {\bibfnamefont {J.~C.}\ \bibnamefont
  {Tully}},\ }\href {\doibase 10.1063/1.469915} {\bibfield  {journal} {\bibinfo
   {journal} {The Journal of Chemical Physics}\ }\textbf {\bibinfo {volume}
  {103}},\ \bibinfo {pages} {10137} (\bibinfo {year} {1995})}\BibitemShut
  {NoStop}%
\bibitem [{\citenamefont {Askerka}\ \emph {et~al.}(2016)\citenamefont
  {Askerka}, \citenamefont {Maurer}, \citenamefont {Batista},\ and\
  \citenamefont {Tully}}]{PhysRevLett.116.217601}%
  \BibitemOpen
  \bibfield  {author} {\bibinfo {author} {\bibfnamefont {M.}~\bibnamefont
  {Askerka}}, \bibinfo {author} {\bibfnamefont {R.~J.}\ \bibnamefont {Maurer}},
  \bibinfo {author} {\bibfnamefont {V.~S.}\ \bibnamefont {Batista}}, \ and\
  \bibinfo {author} {\bibfnamefont {J.~C.}\ \bibnamefont {Tully}},\ }\href
  {\doibase 10.1103/PhysRevLett.116.217601} {\bibfield  {journal} {\bibinfo
  {journal} {Phys. Rev. Lett.}\ }\textbf {\bibinfo {volume} {116}},\ \bibinfo
  {pages} {217601} (\bibinfo {year} {2016})}\BibitemShut {NoStop}%
\bibitem [{\citenamefont {Wang}\ \emph {et~al.}(2002)\citenamefont {Wang},
  \citenamefont {Darling},\ and\ \citenamefont {Holloway}}]{WANG200266}%
  \BibitemOpen
  \bibfield  {author} {\bibinfo {author} {\bibfnamefont {Z.}~\bibnamefont
  {Wang}}, \bibinfo {author} {\bibfnamefont {G.}~\bibnamefont {Darling}}, \
  and\ \bibinfo {author} {\bibfnamefont {S.}~\bibnamefont {Holloway}},\ }\href
  {\doibase https://doi.org/10.1016/S0039-6028(01)01853-2} {\bibfield
  {journal} {\bibinfo  {journal} {Surf. Sci.}\ }\textbf {\bibinfo {volume}
  {504}},\ \bibinfo {pages} {66 } (\bibinfo {year} {2002})}\BibitemShut
  {NoStop}%
\bibitem [{\citenamefont {Ceriotti}\ and\ \citenamefont
  {Manolopoulos}(2012)}]{PIGLET}%
  \BibitemOpen
  \bibfield  {author} {\bibinfo {author} {\bibfnamefont {M.}~\bibnamefont
  {Ceriotti}}\ and\ \bibinfo {author} {\bibfnamefont {D.~E.}\ \bibnamefont
  {Manolopoulos}},\ }\href {\doibase 10.1103/PhysRevLett.109.100604} {\bibfield
   {journal} {\bibinfo  {journal} {Phys. Rev. Lett.}\ }\textbf {\bibinfo
  {volume} {109}},\ \bibinfo {pages} {100604} (\bibinfo {year}
  {2012})}\BibitemShut {NoStop}%
\bibitem [{\citenamefont {Flammini}\ \emph {et~al.}(2012)\citenamefont
  {Flammini}, \citenamefont {Pietropaolo}, \citenamefont {Senesi},
  \citenamefont {Andreani}, \citenamefont {McBride}, \citenamefont {Hodgson},
  \citenamefont {Adams}, \citenamefont {Lin},\ and\ \citenamefont
  {Car}}]{doi:10.1063/1.3675838}%
  \BibitemOpen
  \bibfield  {author} {\bibinfo {author} {\bibfnamefont {D.}~\bibnamefont
  {Flammini}}, \bibinfo {author} {\bibfnamefont {A.}~\bibnamefont
  {Pietropaolo}}, \bibinfo {author} {\bibfnamefont {R.}~\bibnamefont {Senesi}},
  \bibinfo {author} {\bibfnamefont {C.}~\bibnamefont {Andreani}}, \bibinfo
  {author} {\bibfnamefont {F.}~\bibnamefont {McBride}}, \bibinfo {author}
  {\bibfnamefont {A.}~\bibnamefont {Hodgson}}, \bibinfo {author} {\bibfnamefont
  {M.~A.}\ \bibnamefont {Adams}}, \bibinfo {author} {\bibfnamefont
  {L.}~\bibnamefont {Lin}}, \ and\ \bibinfo {author} {\bibfnamefont
  {R.}~\bibnamefont {Car}},\ }\href {\doibase 10.1063/1.3675838} {\bibfield
  {journal} {\bibinfo  {journal} {J. Chem. Phys.}\ }\textbf {\bibinfo {volume}
  {136}},\ \bibinfo {pages} {024504} (\bibinfo {year} {2012})}\BibitemShut
  {NoStop}%
\bibitem [{\citenamefont {Kapil}\ \emph {et~al.}(2018)\citenamefont {Kapil},
  \citenamefont {Cuzzocrea},\ and\ \citenamefont
  {Ceriotti}}]{doi:10.1021/acs.jpcb.8b03896}%
  \BibitemOpen
  \bibfield  {author} {\bibinfo {author} {\bibfnamefont {V.}~\bibnamefont
  {Kapil}}, \bibinfo {author} {\bibfnamefont {A.}~\bibnamefont {Cuzzocrea}}, \
  and\ \bibinfo {author} {\bibfnamefont {M.}~\bibnamefont {Ceriotti}},\ }\href
  {\doibase 10.1021/acs.jpcb.8b03896} {\bibfield  {journal} {\bibinfo
  {journal} {J. Phys. Chem. B}\ }\textbf {\bibinfo {volume} {122}},\ \bibinfo
  {pages} {6048} (\bibinfo {year} {2018})}\BibitemShut {NoStop}%
\bibitem [{\citenamefont {Bart\'ok}\ \emph {et~al.}(2010)\citenamefont
  {Bart\'ok}, \citenamefont {Payne}, \citenamefont {Kondor},\ and\
  \citenamefont {Cs\'anyi}}]{PhysRevLett.104.136403}%
  \BibitemOpen
  \bibfield  {author} {\bibinfo {author} {\bibfnamefont {A.~P.}\ \bibnamefont
  {Bart\'ok}}, \bibinfo {author} {\bibfnamefont {M.~C.}\ \bibnamefont {Payne}},
  \bibinfo {author} {\bibfnamefont {R.}~\bibnamefont {Kondor}}, \ and\ \bibinfo
  {author} {\bibfnamefont {G.}~\bibnamefont {Cs\'anyi}},\ }\href {\doibase
  10.1103/PhysRevLett.104.136403} {\bibfield  {journal} {\bibinfo  {journal}
  {Phys. Rev. Lett.}\ }\textbf {\bibinfo {volume} {104}},\ \bibinfo {pages}
  {136403} (\bibinfo {year} {2010})}\BibitemShut {NoStop}%
\bibitem [{\citenamefont {Behler}(2015)}]{doi:10.1002/qua.24890}%
  \BibitemOpen
  \bibfield  {author} {\bibinfo {author} {\bibfnamefont {J.}~\bibnamefont
  {Behler}},\ }\href {\doibase 10.1002/qua.24890} {\bibfield  {journal}
  {\bibinfo  {journal} {Int. J. Quant. Chem.}\ }\textbf {\bibinfo {volume}
  {115}},\ \bibinfo {pages} {1032} (\bibinfo {year} {2015})}\BibitemShut
  {NoStop}%
\bibitem [{\citenamefont {Laude}\ \emph {et~al.}(2018)\citenamefont {Laude},
  \citenamefont {Calderini}, \citenamefont {Tew},\ and\ \citenamefont
  {Richardson}}]{Laude_2018}%
  \BibitemOpen
  \bibfield  {author} {\bibinfo {author} {\bibfnamefont {G.}~\bibnamefont
  {Laude}}, \bibinfo {author} {\bibfnamefont {D.}~\bibnamefont {Calderini}},
  \bibinfo {author} {\bibfnamefont {D.~P.}\ \bibnamefont {Tew}}, \ and\
  \bibinfo {author} {\bibfnamefont {J.~O.}\ \bibnamefont {Richardson}},\ }\href
  {\doibase 10.1039/C8FD00085A} {\bibfield  {journal} {\bibinfo  {journal}
  {Faraday Discuss.}\ ,\ } (\bibinfo {year} {2018})}\BibitemShut {NoStop}%
\bibitem [{\citenamefont {Rowe}\ \emph {et~al.}(2018)\citenamefont {Rowe},
  \citenamefont {Cs\'anyi}, \citenamefont {Alf\`e},\ and\ \citenamefont
  {Michaelides}}]{PhysRevB.97.054303}%
  \BibitemOpen
  \bibfield  {author} {\bibinfo {author} {\bibfnamefont {P.}~\bibnamefont
  {Rowe}}, \bibinfo {author} {\bibfnamefont {G.}~\bibnamefont {Cs\'anyi}},
  \bibinfo {author} {\bibfnamefont {D.}~\bibnamefont {Alf\`e}}, \ and\ \bibinfo
  {author} {\bibfnamefont {A.}~\bibnamefont {Michaelides}},\ }\href {\doibase
  10.1103/PhysRevB.97.054303} {\bibfield  {journal} {\bibinfo  {journal} {Phys.
  Rev. B}\ }\textbf {\bibinfo {volume} {97}},\ \bibinfo {pages} {054303}
  (\bibinfo {year} {2018})}\BibitemShut {NoStop}%
\end{thebibliography}%

\end{document}